\documentclass[12pt,oneside,english]{book}
\usepackage[T1]{fontenc}
\usepackage[latin9]{inputenc}
\usepackage{geometry}
\usepackage{color}
\usepackage{float}
\usepackage{amsthm}
\usepackage{amsmath}
\usepackage{amssymb}
\usepackage{graphicx}
\usepackage{setspace}
\usepackage{esint}
\makeatletter
\numberwithin{equation}{section}
\numberwithin{figure}{section}

\usepackage{pdfpages}

\makeatother

\usepackage{babel}
\begin{document}


# Characteristics of Black Hole Entropy

**Thesis submitted in partial fulfillment
of the requirements for the degree of
"DOCTOR OF PHILOSOPHY"**

**by**

*Judith Kupferman*

**Submitted to the Senate of Ben-Gurion University
of the Negev**

**November 18, 2013**

**Beer-Sheva**

# Characteristics of Black Hole Entropy

Thesis submitted in partial fulfillment
of the requirements for the degree of
"DOCTOR OF PHILOSOPHY"

by

**Judith**      **Kupferman**

Submitted to the Senate of Ben-Gurion University
of the Negev

Approved by the advisor

Approved by the Dean of the Kreitman School of Advanced Graduate Studies

November 18, 2013

Beer-Sheva

This work was carried out under the supervision of

**Abstract**

Black hole entropy has been studied for the past forty years, but there is still a lack of clarity as to its meaning, and the various interpretations this term is given in the fields of thermodynamics, quantum information and geometry. In the hope of attaining understanding of the concept of entropy in black holes I study specific characteristics of this entropy, and compare these characteristics with respect to the various fields. In this research I focused on the following definitions of entropy: thermodynamic and statistical entropy, entropy of entanglement, Wald entropy as a Noether charge, and the entropy developed by Carlip from conformal symmetry.

First I studied the question of divergence of entropy on the black hole horizon. It is known that the statistical entropy of a matter field in the black hole metric diverges on the horizon, whereas entanglement entropy does not diverge. Thus it appears that the two entropies differ. I show that it is possible that the divergence is the result of inappropriate treatment of the question, and so it may simply stem from quantum uncertainty between position and momentum, and is not an inherent characteristic of statistical entropy. Thus it is possible that statistical entropy and entanglement entropy are related to the same quantity.

I then turned to the question of observer dependence of entropy. Wald's entropy is defined in terms of intrinsic curvature. I clarify the definition of statistical entropy in curved relativistic space, and I find a transformation which laves the entropy invariant while the curvature changes. There is a subgroup in which the transformation of the number of states preserves invariant curvature, but the counter example presented here proves that it is not possible to assume that a given number of states and the statistical entropy derived from it relate to a space time with a given curvature.

Finally, I examined a geometric variation of the statistical entropy of a matter field near a black hole, and compared this to the variation of Wald's entropy. I find that the statistical entropy contains terms that coincide with the variation of Wald's entropy but can have additional terms as well related to the energy momentum tensor.

## Acknowledgments

First of all I would like to thank my adviser, Prof. Ramy Brustein: for introducing me to the wonderful field of black hole thermodynamics. For showing me that ideas are very important but that you must always look for a calculation to back them up. For the opportunity he granted me to work abroad as well. For his confidence in me right at the start, for his useful feedback and criticism. From Ramy I also learned not only to welcome criticism but to actively seek it out.

Profound thanks to Dr. Merav Hadad, adviser, friend and general guardian angel for her constant input, patience and support in all things theoretical, practical and moral. Thanks also to Dr. Frol Zapolsky for his brilliant mathematical insights and for always being there when needed.

I want to thank other physicists who took time to talk to me and offer their experience and advice: Prof. Aharon Davidson who reminded me at my lowest points that physics is wonderful fun, Dr. Oded Kenneth and Dr. Dmitri Melnikov who spent hours helping me understand things, and Dr. Joey Medved for his helpful comments and for sharing his extensive knowledge of the literature. I'm also grateful to my fellow students at Ben Gurion University and Tel Aviv University, whose company and input were a help and a pleasure, and above all: Nati Aharon, Eran Avraham, Ido Ben Dayan, Amir Erez, Uri Kol and Ben Yellin.

A Ph.D. is not only a scholastic but also a personal adventure. I owe deep thanks to my brother Dr. David Malkiel, and to Sharona, Jenny, Orna, Joe and Tali, for their constant faith in me and excellent practical advice. A special thank you to Prof. Aharon Eviatar who helped me from my very first steps in physics: helped me get into the university despite my lack of background, and for his friendship and help all along the way.

And finally I want to thank my mother, Ann Kupfermann, for her unflagging confidence in me, as well as her explanation of the logic on which this was based. For her energy, enthusiasm and practicality, which have been an inspiration and source of strength.

Dedicated to my father, Ted Kupfermann, who looked at everything in the world with curiosity and thoughtfulness, and who would have been so interested and pleased.



\tableofcontents{}

\part{Aspects of entropy}

\chapter{Introduction}

\section{Overview}

The field of black hole thermodynamics was born forty years ago, when
black holes were shown to have entropy \cite{Hawking Bardeen,Bekenstein}.
However it is not clear just what this entropy is. Entropy appears
in various contexts and meanings. In thermodynamics the change in
entropy for a canonical ensemble is related to the change in energy,
if volume and other factors are constant. Statistical mechanics in
treating the microcanonical ensemble defines entropy as the number
of accessible states. In classical information theory entropy quantifies
the amount of information in a message, while in quantum information
theory, entanglement entropy is a measure of quantum correlations
between parts of a system. Black hole entropy was originally discussed
in the thermodynamic context. However since then it has been explored
in other contexts, and to this day when referring to black hole entropy
there is still a lack of clarity as to what exactly the term refers
to. If, as in statistical mechanics, we are counting microstates,
just what are we counting in the black hole? If quantum correlations,
between what? The vacuum within and without the black hole? Hawking
particles? Matter within and without?

In this thesis I focus on specific aspects of black hole entropy with
the aim of clarifying the similarities and differences in the various
interpretations of the term. After an introductory review of selected
treatments of black hole entropy, I examine the issue of divergence
on the black hole horizon, in an attempt to clarify the distinction
between statistical and entanglement entropy. I then focus on the
treatment of the horizon as boundary, as a partition in space, and
on the distinction between treatment that imposes boundary conditions
and treatment that does not do so. Another aspect is observer dependence:
can entropy be a function of an invariant such as scalar curvature
or is it observer dependent? The third part of the thesis examines
a variation of entropy, in an effort to clarify the distinction between
Wald entropy and statistical mechanical entropy. If the variations
were shown to be identical that could indicate that the two kinds
of entropy originate in a common source. 

In focusing on these specific issues I find that entanglement entropy
and the entropy of statistical mechanics may refer to the same phenomenon,
but that these differ from geometrical concepts such as Wald's Noether
charge entropy and Carlip's derivation of entropy from conformal field
theory. A general conclusion is that in determining the real meaning
of black hole entropy, a careful and unambiguous description of these
aspects is necessary in order to dispel confusion and to distinguish
the particular theory under discussion.

\emph{Remark on notation}: Throughout this work natural units are
used, so that $\hbar,c,$ and the Newton and Boltzmann constants are
taken as unity, except where explicitly quoting previous work. The
personal pronoun varies between ``I'' and ``we'', not the royal
plural but rather due to discomfort at seeing one hundred pages full
of myself alone.

\section{Background}

Black hole entropy has had a rich and varied history over the past
forty years. Below I review only a few examples which are relevant
to the work in this thesis. These include examples of thermodynamic
and statistical mechanical entropy, entanglement entropy, the geometric
formulations of Wald and Carlip, and the mapping of the gravitational
equations of motion to the first law of thermodynamics. In cases where
extensive details of the calculations are relevant to what follows,
these can be found in Part II of the thesis.

\subsection{Thermodynamics }

Black hole entropy was first discussed in the context of thermodynamics.
In 1973 Bardeen, Carter and Hawking \cite{Hawking Bardeen} formulated
the four laws of black hole mechanics, analogous to the laws of thermodynamics,
where the surface gravity played the role of temperature, and the
horizon area that of entropy. They viewed this strictly in terms of
an analogy, taking pains to emphasize that ``$\kappa/8\pi$ and $A$
are distinct from the temperature and entropy of the black hole....In
this sense a black hole can be said to transcend the second law of
thermodynamics.''%
\footnote{\cite{Hawking Bardeen}, p.168.%
} 

Hawking had shown that the area of a black hole cannot decrease \cite{Hawking area}.
Bekenstein \cite{Bekenstein} reasoned that just as area increases
so does thermodynamic entropy, and so the area of a black hole can
be interpreted as its entropy. The generalized second law of thermodynamics
holds that the sum of the entropy of matter outside a black hole together
with the area of the black hole never decreases. 

In \cite{Hawking Bardeen} the authors had taken care to point out
that the parallel between the laws of black hole mechanics and thermodynamics
is purely an analogy, but that the two though similar are distinct.
\textcolor{black}{This approach underwent a startling change in 1974,
when Hawking showed that black holes emit radiation \cite{hawking explosions}.
Using quantum field theory he showed that this radiation is thermal,
and obtained the temperature, which could then be used in the thermodynamic
formulation.}

Gibbons and Hawking \cite{Gibbons hawking} used the path integral
formalism to obtain the partition function for a black hole. The partition
function is taken as
\begin{equation}
Z=\int DgD\phi e^{iI\left[g,\phi\right]}
\end{equation}
where $Dg$ is a measure on the space of metrics, and $D\phi$ on
that of matter fields, and $I\left[g,\phi\right]$ the action. The
result gives the probability of the occurrence in the vacuum state
of a black hole with the relevant parameters that appear in the action.
The gravitational action has a surface term chosen so that variation
of the action gives the Einstein equations. 

In order to avoid singularities the authors define a new coordinate,
$\tau=it$ so that the metric becomes Euclidean. This procedure can
then be extended to non Euclidean systems because the quantities in
the integral, which include the Ricci scalar, the electromagnetic
field tensor and the extrinsic curvature, are holomorphic functions
on the complexified spacetime except at the singularities. Thus the
action integral is in fact a contour integral, and it will have the
same value on any section of the complexified spacetime corresponding
to the Euclidean section even though the induced metric on this section
may be complex. 

In the path integral approach to quantization of a field, going to
a Euclidean system gives the thermodynamic partition function. Thus
\begin{equation}
Tr\left(e^{-\beta H}\right)=\int D\phi e^{iI\left[\phi\right]}
\end{equation}
where the integral is taken over all fields which are periodic with
period $\beta$ in imaginary time. Gibbons and Hawking use this to
obtain thermodynamic quantities from the gravitational and matter
action. Since the dominant contribution to the path integral comes
from metrics $g$ and matter fields $\phi$ which are near background
fields $g_{0}$ and $\phi_{0}$ they expand the action in a Taylor
series around the background fields, neglect higher order terms, and
write the resulting partition function $Z$. From this they obtain
the free energy using the thermodynamic relation $lnZ=-\beta F$.
In a specific example for a rotating black hole they identify the
temperature with the surface gravity, and write the free energy in
terms of the black hole mass, charge and angular rotation, obtaining
\begin{equation}
\frac{1}{2}M=TS+\frac{1}{2}\Phi Q+\Omega J
\end{equation}
and plugging in the generalized Smarr formula \cite{Smarr} 
\begin{equation}
\frac{1}{2}M=\frac{\kappa}{8\pi}A+\frac{1}{2}\Phi Q+\Omega J
\end{equation}
they find 
\begin{equation}
S=\frac{A}{4}.
\end{equation}
Thus the path integral is used in conjunction with a Wick rotation
to give the thermodynamic partition function, and the entropy is derived
from this by plugging in thermodynamic relations.

\subsection{Statistical mechanics}

Statistical and thermodynamic entropy are of course interrelated,
but here we use the term statistical entropy when the computation
is the number of accessible states, while thermodynamics is used in
the context of the canonical ensemble and based on macroscopic properties
of black holes.

A seminal work obtaining black hole entropy from the number of states
was that of 't Hooft \cite{'t Hooft}. He calculates the number of
states of a quantum matter field in the region of a black hole, and
from the number of states he obtains the free energy and the entropy.
To find the number of states for a particle in the black hole metric,
t'Hooft uses a one dimensional WKB approximation. He takes the contribution
to energy of the transverse momenta as an effective radial potential
$V_{eff}=l(l+1)/r^{2}$ , since it behaves as a centrifugal potential.

The one-dimensional WKB approximation gives the number of radial modes
$n$ thus:
\begin{eqnarray}
n\pi & = & \intop_{2M+h}^{R}dr\sqrt{g_{rr}}k(r)
\end{eqnarray}
The integral ought to be taken from the horizon, but to avoid divergence
't Hooft goes a short distance $h$ from the horizon, known as the
``brick wall.'' 

From the wave equation one obtains the square of the radial eigenfunction
\begin{equation}
\left(k(r)\right)^{2}=g_{rr}\left(g^{00}E^{2}-g^{\theta\theta}\left(k\left(\theta\right)\right)^{2}-g^{\phi\phi}\left(k\left(\phi\right)\right)^{2}-m^{2}\right)
\end{equation}
where $k\left(\theta\right),k\left(\phi\right)$ denote the eigenfunctions
of the angular components of the Laplacian. In the Schwarzschild metric
this becomes
\begin{equation}
\left(k(r)\right)^{2}=\frac{1}{1-\frac{2M}{r}}\left(\frac{1}{1-\frac{2M}{r}}E^{2}-\frac{1}{r^{2}}\left(l\left(l+1\right)\right)-m^{2}\right)
\end{equation}
The number of radial modes is then summed over the angular degrees
of freedom, 
\begin{equation}
N\pi=\intop_{2M+h}^{R}dr\frac{1}{1-\frac{2M}{r}}\intop_{0}^{E^{2}r^{2}}dl(2l+1)\sqrt{E^{2}-\left(1-\frac{2M}{r}\right)\left(\frac{l(l+1)}{r^{2}}+m^{2}\right)}
\end{equation}
where the upper limit of the second integral is in order to ensure
a positive root. Integration gives two terms. One is the contribution
from the vacuum surrounding the system at large distances and 't Hooft
discards it. The second term is the horizon contribution, which diverges
as $h\rightarrow0.$ From this he obtains the free energy and the
entropy, which also diverge on the horizon. Further details of the
calculation appear in Sec.\ref{sec:'t-Hooft's-calculation}.

The motivation for this work is to reconcile black hole physics with
quantum mechanics. At the time it was written black holes were understood
to be in a quantum mechanically mixed state, and 't Hooft attempted
to describe them as pure states resembling ordinary particles. Thus
black holes inhabit an extension of Hilbert space with an according
Hamiltonian. This system is sensitive to observer dependence: the
free falling observer perceives matter, and 't Hooft writes that it
is this matter which he considers in this paper. The distinction between
vacuum and matter is assumed to be observer dependent when considering
coordinate transformations with a horizon. The brick wall model is
intended to show that the horizon itself rather than the black hole
as a whole determines its quantum properties. This seems to have a
relation to entanglement entropy, considered in the following section,
and that is not surprising since it is 't Hooft's explicit intention
to reconcile gravity and quantum mechanics.

The number of black hole microstates was computed much later using
string theory, beginning in 1996 when Strominger and Vafa computed
the entropy of an extremal supersymmetric black hole in string theory
and obtained the Bekenstein Hawking entropy \cite{Strominger}; this
approach was extended to a wide variety of black holes. Another approach
in loop quantum gravity considers black hole states as spin networks
on the horizon. These treatments are interesting and productive, but
they are beyond the scope of this thesis.

\subsection{Entanglement entropy}

Several hints lead us to consider a relationship between quantum entanglement
entropy and the geometry of space-time. First, the fact that entanglement
entropy arises as a consequence of a partitioning of subsystems. Such
a partition is located in space in the case of a black hole horizon,
for example: if there were no such barrier the entanglement entropy
would be zero, and it is possible that barrier width or extent determine
the amount of entanglement entropy. Second, the fact that entropy
can be related to energy. Einstein's equations point at a relationship
between geometry and energy, and it is clearly of interest whether
and how this relationship can be extended to entropy. A relation has
been shown when the entropy is thermodynamic (see Sec.\ref{sub:Jacobson etc})
but not for entropy as a measure of the amount of entanglement. Third,
it has been shown that entanglement affects the speed of evolution
\cite{Giovanetti,me}, indicating that there is a relationship between
entanglement and time. In a relativistic context this too points to
a relationship between entanglement and geometry.

Entanglement entropy differs by definition and by characteristics
from the entropy defined in statistical physics. The latter is the
logarithm of the number of states, and is extensive. Entanglement
entropy is a measure of quantum correlations between states which
are part of a composite system, and has been found to be proportional
to area \cite{Bombelli,Srednicki}. Since Bekenstein showed that black
hole entropy is proportional to area, many people view this as a hint
that BH entropy is entanglement entropy \cite{Ramy string theory,Callen Wilczek,Holzhey}.
Entanglement entropy is formally defined as $-Tr\left(\rho ln\rho\right)$
where $\rho$ is the partial trace of the system, that is, the reduced
density matrix of part of the composite system. Tractable calculation
of the entropy for bipartite states is with eigenvalues ($-\sum\lambda ln\lambda)$
in the Schmidt basis, in which the reduced density matrices are diagonalized,
and their eigenvalues are found to be identical. 

The first major papers on entanglement entropy were by Bombelli et
al. in 1986 \cite{Bombelli} and Srednicki in 1993 \cite{Srednicki}%
\footnote{These widely cited papers generated much further work, but the idea
and calculations were first given three years earlier in a talk by
Sorkin\cite{Sorkin talk}.%
}. Both showed entanglement entropy is proportional to area by treating
discrete coupled harmonic oscillators and then taking the continuum
limit. Srednicki's result is more powerful for two reasons: first,
unlike Bombelli he explicitly calculates it for a sphere and so it
can be related to a black hole. Second, and this is probably a reason
for its greater impact, he presents a convincing justification for
the area law: his logic is that surface area is the only thing the
inside and outside of the black hole have in common, so their entanglement
entropy, which is the same for inside and outside, must depend on
that. He assumes entropy depends on area and since entropy is dimensionless,
looks for a dimensionful parameter to cancel out area. He uses the
square of the lattice spacing. The area law is arrived at by numerical
means.

Srednicki first calculates the entropy for a pair of discrete harmonic
oscillators in the ground state. This is
\begin{equation}
S=-log(1-\xi)-\frac{\xi}{1-\xi}log\xi
\end{equation}
where $\xi=f(k_{1}/k_{0})$ (the interaction potential of each oscillator
is $-k_{i}x_{i}$). When summed over $n$, a number of haromonic oscillators,
the entropy becomes a function of $n$ and thus of lattice length.
He defines the radius of the region he has specified as $R=(n+\frac{1}{2})a$,
where $a$ denotes the lattice spacing. He obtains the entropy as
$S=0.30\, M^{2}R^{2}$. This looks neatly proportional to area. However
he has defined $M=\frac{1}{a}$ . When writing the entropy explicitly
with $M,R$ in terms of $a,$ the spatial component cancels out and
we are left with a function only of the number of oscillators, $S=0.30(n+\frac{1}{2})^{2}$.
That is really where the ``area law'' comes from. It is obtained
numerically.

Srednicki's entropy diverges because it sums over n, the number of
lattice sites, which is then taken to continuum. But there is no particular
divergence at the boundary of the sphere which he examines. This and
other general treatments of entanglement entropy \cite{Plenio} all
calculate it for discrete lattices and then take it to continuum.
In all these cases the entropy does not diverge at the boundary unless
- since it is proportional to area - the boundary area is infinite.
Divergence of entanglement entropy in the general treatment is a result
of having an infinite number of modes, so that a UV cutoff is necessary.

A different treatment of black hole entropy as entanglement entropy
was that of Ryu and Taskayanagi \cite{ryu}, who calculate entropy
in a $d+1$ dimensional conformal field theory from the area of a
$d$ dimensional minimal surface in $AdS_{d+2}$, and show that the
result reproduces the Bekenstein Hawking formula. This will not be
discussed in the thesis.

\subsection{Entropy as a Noether charge}

Wald \cite{wald 1,wald and iyer} obtains black hole entropy for generalized
theories of gravity by requiring diffeomorphism invariance in conjunction
with the first law of thermodynamics. In accordance with Noether's
theorem that to every continuous symmetry a conserved current and
charge can be associated, diffeomorphism symmetry has such a current
and charge. Wald uses the abstract formalism of Hamiltonian mechanics
\cite{Arnold}, defining the phase space of a linear dynamical system
as a symplectic vector space. The solutions of the equations of motion
are mapped to points on a symplectic manifold. This is done in the
language of differential forms. The advantage of this approach is
that the equations of motion are independent of choice of a coordinate
system.

Wald sets out by writing the most general possible Lagrangian for
diffeomorphism invariant theories. This is a function of the Riemann
tensor and its derivatives, as well as the metric and any other fields.
A variation of the Lagrangian by varying the fields gives the equations
of motion as well as the exterior derivative of a symplectic potential
form $\Theta$. For $\xi^{a}$ any smooth vector field on the spacetime
manifold, that is $\xi^{a}$ the infinitesimal generator of a diffeomorphism,
and for any field configuration $\phi$ (not necessarily a solution
to the equations of motion) the Noether current is defined by
\begin{equation}
J=\Theta\left(\phi,\mathcal{L}_{\xi}\phi\right)-\xi\cdot L
\end{equation}
where the dot denotes contraction of the vector field $\xi$ into
the first index of the differential form of the Lagrangian $L$. One
finds that $dJ=-E\mathcal{L}_{\xi}\phi$ where $E$ is the equations
of motion. On shell one sees that $J$ is a closed form and thus there
is a Noether charge $Q$ such that $J=dQ$. 

Wald applies this to a stationary black hole solution with bifurcate
Killing horizon, where $\xi^{a}$ is the Killing field vanishing on
the bifurcation surface. A variation of the fields away from the background
solution provides an equation relating the surface term at infinity
to a surface term on the horizon. The infinity terms give the mass
and angular momentum. These are equated to the horizon term. This
equation has the form of the first law of thermodynamics, where
\begin{equation}
\delta\intop_{\Sigma}Q\left(\xi\right)=\frac{\kappa}{2\pi}\delta S
\end{equation}
where $\Sigma$ is the bifurcation surface. He then shows that 
\begin{equation}
S=-2\pi\intop_{\Sigma}E_{R}^{abcd}\epsilon_{ab}\epsilon_{cd}\label{eq:Wald entropy}
\end{equation}
where $E_{R}^{abcd}$ is a tensor field obtained by taking the functional
derivative of $L$ with respect to $R_{abcd}$ (viewed as independent
of $g_{ab}$) and $\epsilon_{ab}$ is the binormal to the bifurcation
surface, $\epsilon_{cd}$ is the area form. Thus the black hole entropy
is proportional to the Noether charge coming from the boundary at
the black hole horizon. 

Wald's entropy is sometimes referred to as geometric. It originates
in diffeomorphism symmetry, but also from the first law of thermodynamics.
The purely geometric derivation gives the energy at the horizon, in
an equation corresponding to the first law. The entropy is obtained
by assuming the relation between mass, angular momentum, energy and
temperature which is given by the first law of thermodynamics. (Further
details are in Sec.\ref{sec:Carlip's-scheme}.)

\subsection{Entropy from confo\label{sub:Carlp intero}rmal field theory}

Carlip \cite{THE Carlip paper,Carlip ADM} attempted to obtain black
hole entropy from considerations of symmetry. General relativity is
diffeomorphism invariant, and this symmetry is expressed in an algebra,
which, when space has a boundary, can be shown to have a central extension.
In a different context Cardy \cite{Cardy} obtained a formula for
statistical entropy as a function of the central charge of a conformal
theory, and Carlip makes use of the Cardy formula to obtain black
hole entropy. He first did this using the ADM formalism. In this Hamiltonian
form, general relativity is constrained with constraints $\mathcal{H}_{\mu}$
and these constraints generate the symmetries of the theory. The generators
obey an algebra, and if space has a boundary the algebra has been
shown to be isomorphic to a pair of Virasoro algebras, with a central
extension. The boundaries in the Schwarzschild metric, for example,
are taken, as with Wald, to be infinity at one end, and the black
hole horizon at the other. In further work, Carlip uses Wald's covariant
phase space formalism and concept of entropy as a Noether charge as
a point of departure to write the algebra for general relativity and
compute the central charge. He plugs in Cardy's formula to obtain
the entropy as a function of the central charge, and shows that in
the case of a black hole, this gives the known result. This is explained
in detail in Sections \ref{sec:Carlip} and \ref{sec:Carlip's-scheme}.

\subsection{The first law of thermodyam\label{sub:Jacobson etc}ics}

In contrast to the derivation by Bardeen et al. of the four laws of
black hole mechanics from the equations of gravity, Jacobson \cite{jacobson}
derived the Einstein equations of state from the first law of thermodynamics.
This was done by taking the energy flow across a causal horizon -
not necessarily a black hole - and arguing that the entropy of the
system beyond the horizon is proportional to horizon area. The first
law is an equilibrium relation, whereas the horizon may not be in
equilibrium but rather contracting, expanding or shearing. Therefore
Jacobson takes a small neighborhood of a point on the horizon. This
neighborhood is locally flat by the equivalence principle, and there
is a Killing field $\chi^{a}$ generating boosts orthogonal to the
point $P$ in question. The energy flow across the horizon is
\begin{equation}
\delta Q=\intop_{\mathcal{H}}T_{ab}\chi^{a}d\Sigma^{b}
\end{equation}
where the integral is over a ``pencil'' of generators of the inside
past horizon of $P$. Temperature is taken as the Unruh temperature,
$\kappa/2\pi$ where $\kappa$ is the acceleration of the Killing
orbit. Taking the entropy variation $\delta S$ as a variation of
area $\delta A$, he writes the area variation as an integral of the
expansion $\theta$ of the horizon generators. The Raychauduri equation
relates this to the Ricci tensor, thus giving $\delta S$ as an integral
over a quantity including the Ricci tensor. Plugging in the first
law of thermodynamics, $\delta Q=T\delta S,$ the Einstein equations
are obtained. This remarkable work was later generalized to higher
order theories of gravity (\cite{merav ramy,Parikh,paddy on de tds}
and see Sec.\ref{sec:Variation-of-Wald's}).

Here I have detailed only a selection of a wide variety of treatments
of black hole entropy. The question remains: when discussing the entropy
of a black hole, what are we actually talking about? In the following
chapters I examine specific aspects of black hole entropy, in the
hope that clarifying the similarities and differences between the
various treatments outlined above will shed some light on this question.
Part I of the thesis treats behavior at the boundary, observer dependence
and variations of entropy. Part II contains supplementary details,
proofs and calculations.

\chapter{The black hole boundary}

If the universe is perceived as a pure state composed of the part
within the black hole and the part outside, then either of the parts
will have entanglement entropy since by themselves each is not pure
but mixed. Therefore a black hole has entanglement entropy by definition.
A more physical understanding of entanglement entropy in the black
hole context may be entanglement of Hawking pairs. This hints that
entanglement entropy of a black hole may be the same as thermodynamic
entropy, since Hawking radiation is primarily thermal. This idea is
strengthened by Bekenstein's observation that black hole entropy is
proportional to area, just as is entanglement entropy. The question
is whether the statistical entropy of a black hole coincides with
entanglement entropy.

In order to explore this we focus on behavior of entropy at the boundary.
When t'Hooft investigated black holes in a statistical mechanics framework
he found divergence on the horizon \cite{'t Hooft}. Later work \cite{de alwys,myers}
assumed that the divergence is a direct result of the black hole redshift.
However entanglement entropy in the general case does not show divergence
at the barrier, but rather diverges as a function of high momentum
cutoff . This might indicate a difference between black hole entropy
and entanglement entropy. The question is whether this divergence
is related to the horizon itself as a causal barrier, or whether it
is due to simple quantum uncertainty, in which case the divergence
does not point to an essential difference between the two entropies. 

General calculations of entanglement entropy do not show divergence
at the barrier. In Bombelli's calculation divergence comes from high
frequency modes and so he inserts a UV cutoff, but there is no other
cause of divergence. Srednicki defines $R=(n+\frac{1}{2})a,$ where
$a$ is lattice spacing and $n$ the number of discrete oscillators.
The expression for entropy which he obtains numerically is $S=0.30M^{2}R^{2}$
which does not diverge for a particular radius. In fact he has defined
M as the inverse lattice spacing $a^{-1}$ so that the actual expression
is $S=0.30(n+\frac{1}{2})^{2}$ which diverges for an infinite number
of oscillators, but again, not at a particular location. Srednicki
also performs a perturbative approximation for $l\gg N$, where $l$
is the usual quantum number for angular momentum and N the total number
of lattice sites, it diverges for infinite lattice sites but not at
a particular location. Plenio in his review of entanglement entropy
in lattice systems \cite{Plenio} has no divergence because an upper
bound for entanglement entropy is known to be logarithmic negativity,
$E_{n}>S,\: E_{n}=ln\left\Vert \rho^{PT}\right\Vert ,$ that is, the
logarithm of the trace normed partial transpose matrix. This would
diverge if the matrix were infinite, but not with respect to any particular
location.

In \cite{ramy amos cold bosons} divergence at the barrier subdividing
a system is explored in a nonrelativistic context, rather than in
the specific context of a black hole. Taking a system and looking
at a subvolume, energy fluctuations in the subvolume are proportional
to the surface area of the subvolume. If the fluctuations were thermal
they should have been extensive (proportional to volume) but they
are not. The claim is that they are not thermal but rather caused
by the division into a subvolume, and thus by entanglement.

A high momentum cutoff is necessary to prevent divergence of $\Delta E^{2}$.
Such a cutoff is equivalent to smearing the boundary which divides
off the subvolume. In \cite{ramy amos cold bosons} this was done
explicitly for particles in a box. A more general treatment appears
in \cite{Ramy string theory} where half of Minkowski space is treated
as a thermal ensemble in Rindler space. If a pure state is subdivided
into two mixed states, $\rho_{A},\rho_{B}$, then the equation $Tr(\rho_{A}\hat{O}_{A})=<\psi_{AB}|\hat{O}_{A}|\psi_{AB}>$
allows to obtain expectation values of operators in an entangled pure
state by calculating expectation values of one of the mixed states.
In this way entanglement entropy of the pure state is found to be
the thermal entropy of the mixed state. It diverges at the horizon,
scales as area, and needs a UV cutoff.

\section{\label{sec:our-paper.}Divergence on the horizon}

\textcolor{black}{Calculations of black hole entropy on the horizon
give rise to divergence}. We will show that this divergence is not
unique to a black hole, nor is it a UV divergence found in field theory
which requires appropriate renormalization. Rather it can be seen
as a result of quantum x/p uncertainty because the horizon is defined
as a perfectly sharp boundary dividing spacetime into an observable
and an unobservable region. A similar divergence arises for any quantum
mechanical system when a sharp boundary divides the whole system into
an observed and an unobserved regions. This is also the case with
a coordinate system which truncates part of flat space, as with Rindler
coordinates. The divergence is tamed by smoothing over the boundary,
rather than by renormalizing the theory. The same is true for black
hole entropy.

In quantum mechanics we know that there are questions which can, but
should not be asked. If we insist on asking them, the theory itself
lets us know in a clear way by giving us a senseless answer. For example,
if we ask ``what is the typical momentum of a perfectly localized
particle?\textquotedbl{} the formal answer will be infinite because
of the position/momentum uncertainty relation. Of course, this just
means that the momentum fluctuations will become larger as the particle
is localized in a sharper way. Here the observer needs to change the
question to ``what is the typical momentum of a particle whose wave
function has a small finite width in space?\textquotedbl{} and treat
the concept of a sharply localized particle as a limit.

In quantum field theory we are familiar with questions involving infinity.
Some of these indicate real problems with divergence, but others are
meaningless just as with momentum of a localized particle. An example
of a real problem is to ask ``what is the charge of the electron?\textquotedbl{}
where the answer comes out infinite. In this case the infinite answer
does not mean that we should not have asked the question. Rather,
it means that we have misidentified a microscopic parameter in the
theory and that this parameter should be ``renormalized\textquotedbl{}.
After a redefinition of the ``bare\textquotedbl{} (correct) microscopic
theory we can ask the question and get a finite answer. However, in
other cases the divergence can not be corrected by modifying the theory
because the question itself does not make sense.

An example of this second type of divergence would be to look at a
non-relativistic particle in a finite box and ask: What is the energy
in the left hand side of the box. If we approach this problem using
second quantization and field operators, we will see that the problem
may then be extended to relativistic fields, and that there too the
difficulty arises from an ill posed question. We ask ``what is a typical
energy or momentum in the left half of the box.\textquotedbl{} Note
that this does not involve putting a real partition into the box;
that would simply give two smaller boxes, with finite energy, of course.
However if the partition is imposed by limiting the possibility of
observations to only half the box without imposing new boundary conditions,
then if the partition is sharp the answer will be infinite because
the fluctuations of momentum and energy are infinite.

How should we interpret the infinite answer when we know that in fact
the energy is finite? In \cite{ramy amos cold bosons} it was shown
that the reason for the senseless answer is that the question is inappropriate.
The insistence on an infinitely sharp division between the (observable)
left region and the (unobservable) right region is the cause of the
divergence. In this case the sensible question should involved a smoothed
division of the box, allowing the boundary between the observable
and unobservable domains to be smoothed. If the resolution with which
the box is divided into the observable and unobservable halves is
limited, then the answer is finite and inversely proportional to the
smoothing width, exactly as in the case of the localized particle.

The distinction between the two classes of divergences is the distinction
between an ultraviolet (UV) divergence and an ill posed question.
We would like to know to which of these two classes black hole entropy
belongs. Are its divergences inherent to the system and requiring
some knowledge of the UV properties of the theory, a theory of quantum
gravity, or both? Or are they, rather, similar to those one obtains
when dividing space into two regions, one observable and the other
unobservable, and tracing over the unobservable region?

In this chapter we will consider some generic wave function and apply
a ``window-function\textquotedbl{} to it, leaving the boundary conditions
exactly the same as they were initially. The window function will
allow us to impose a smooth division between the observable and unobservable
regions. When the width of the window function is taken to zero, a
sharp division between the regions is obtained. This set up is different
than setting up the quantum system with boundary conditions that would
have made the wave function vanish outside a certain region.

We will show that the problem of divergence at the dividing boundary
can be resolved for a quantum mechanical system by asking the right
question, namely smoothing the division between the two regions. We
will then argue that the origin of the divergences encountered for
black holes is similar. This will allow us to argue that such divergences
are not a unique black hole characteristic but rather a result of
quantum uncertainty, and the correct expression must involve smearing
out the boundary. In fact Bekenstein noted in 1994 that if the boundary
of the region being traced out were absolutely sharp, the energy would
be very large due to the uncertainty principle, and so the boundary
must be thought of as \textquotedbl{}slightly fuzzy\textquotedbl{}
\cite{Bekenstein our paper}, and we will show in detail that this
is the case.

This chapter is organized as follows. First we briefly review the
behavior of black hole entropy at the horizon. Then we show the relationship
between entropy and energy near the horizon of a black hole. Next,
we clarify the concept of partitioning and define an operator which
may smooth a partition. This is then used to examine behavior of energy
at a boundary between two subsystems, first for the non-relativistic
and then for the relativistic case, and to show that in both cases
energy diverges as the boundary becomes sharp. We extend this to Rindler
space, as a partitioning of Minkowski space. Finally, we examine 't
Hooft's calculation of black hole entropy, and find that his relocation
of the boundary to avoid divergence is equivalent to smearing out
the boundary. Therefore here too the divergence is related to sharpness
of the boundary and is not unique to a black hole.

\section{Thermal and entanglement\label{sec:Thermal-and-entanglement} entropy}

't Hooft calculated thermodynamic characteristics of a black hole,
among them entropy, and in doing so found a divergence of the density
of states and hence of the entropy density at the horizon. He overcame
the problem by adjusting the limits of integration to a ``brick wall''
a finite infinitesimal distance from the horizon. Entanglement entropy
also diverges, but the divergence appears to be an ultraviolet divergence
that does not seem to diverge at any particular location.

For a BH in equilibrium, the space just outside the hole near the
horizon can be treated as a thermal state in Rindler space \cite{Ramy string theory,kabat strassler}..
In this case entanglement entropy coincides with thermal entropy,
as follows. To find entanglement entropy we take the trace of part
of the system. If that part of the system is a thermal state, the
partial trace is a thermal density matrix,
\begin{eqnarray}
\rho_{part} & = & \frac{1}{Z}\sum_{i}e^{-\beta E_{i}}|E_{i}\rangle\langle E_{i}|.\label{eq:po for bh thermal state}
\end{eqnarray}
 Entanglement entropy is given by 
\begin{eqnarray}
S & = & -Tr\left(\rho_{part}\ln\rho_{part}\right)
\end{eqnarray}
 and the energy is given by 
\begin{eqnarray}
\left\langle E\right\rangle  & = & \frac{1}{Z}\sum_{i}E_{i}e^{-\beta E_{i}}
\end{eqnarray}
 It follows that 
\begin{eqnarray}
S & = & -\frac{1}{Z}\sum_{i}e^{-\beta E_{i}}\times\left(-\beta\sum_{i}E_{i}-\ln Z\right)\nonumber \\
 & = & \beta\langle E\rangle+\ln Z.
\end{eqnarray}
 For a scalar field at a finite temperature $\ln\: Z$ is a constant,
so the entropy is linear to the expectation value of the energy. Therefore
in the case of a black hole the entanglement entropy behaves as does
the energy. Thus instead of examining entropy at a barrier dividing
the two subsystems, which is a complicated non-local quantity, we
can calculate the reduced density matrix of a subsystem and look at
the behavior of its energy which is a simpler local quantity.

\section{M\label{sec:fluctuations}omentum fluctuations, energy and entropy
for smooth partitions}

\subsection{Partitioning a subvolume}

We will examine various examples of partitioning: first we take a
single non relativistic particle in a box, then a relativistic field,
then an entire region of Minkowski space and finally a black hole.
The first case is clearest but our claim is that the others are essentially
the same.

It is crucial to clarify that the partitioning corresponds to limiting
the observability to a subvolume. If we were to take a box and place
an actual physical partition in the middle, this would impose new
boundary conditions and we would simply have two smaller boxes with
observables appropriate to the new boundary conditions. Instead we
leave the particle in the original box, but consider only a subvolume
of the box. An example of this would be to work out the probability
of finding the particle. Had we actually partitioned the box and looked
for the particle in the left half, we would find a probability of
one or zero to find it there. But we do not actually do this; rather
than making the actual observation, we just calculate the probability
to find the particle on the left, and then we will obtain a probability
of one half. Similarly in what follows we will calculate expectation
values for part of a system without actually imposing a partition
with new boundary values.

This kind of partitioning is equivalent to tracing out part of the
system. The mathematical operation of tracing defines in a clear way
the kind of partitioning of the quantum system that we have in mind.
We do not impose new boundary conditions, but rather we restrict the
domain of observability to a limited region of the total volume. This
will be implemented by a window operator, as described below. If the
partitioning is done at a sharply localized point we will see divergence
of momentum and energy, even though in fact obviously the particle
itself has the same finite energy it had initially. If the partition
is not sharply localized we will no longer see a divergence.

We are interested in the expectation value for the reduced energy
in the case where we look at a subvolume of the entire system. This
can be expressed in two ways. We can rewrite the state so that it
is multiplied by a window function: $\left|\psi\right\rangle _{window}=f(\vec{r},w)\left|\psi\right\rangle $.
Thus the expectation value for the reduced energy in this restricted
system will be $\left\langle \psi\right|fHf\left|\psi\right\rangle $.
Rather than regarding the window function as part of the state, we
can treat as part of the operator, so that we define the restricted
Hamiltonian as $H^{V}=fHf$. A striking equation relates quantum expectation
values of operators that act on part of a system to the statistical
averages for a reduced density matrix of the subsystem. Writing the
density matrix for the subsystem as $\rho^{V}$, 
\begin{equation}
\langle\psi|H^{V}|\psi\rangle=Tr\left(\rho^{V}H^{V}\right).
\end{equation}
 Therefore we can calculate the reduced energy in the subsystem by
taking the expectation value of the restricted Hamiltonian in the
entire system.

The restricted Hamiltonian may be smoothed so that the partition into
subsystems is not completely sharp. This is equivalent to giving the
window function varying width. For details see Sec.\ref{sec:appA}.
It is possible to define a smoothing function that is strictly zero
on the left and continuous at any fixed desired order at the boundary.
When we discuss the horizon in Rindler and Schwarzschild metrics,
we will see that this is the form which can be given to a smoothing
function operating on the redshift.

The function $f(\vec{r},w)$ behaves as a window enclosing part of
space, and thus it mimics the horizon by ``truncating'' part of space
for the field. We provide it with a varying width, and examine energy
as a function of its width. Our aim is to see how sharp localization
affects the reduced energy divergence.

\subsection{Energy and momentum fluctuations in a restricted non relativistic
system}

We now write the reduced density matrix for nonrelativistic bosons
restricted (in the sense defined above) to one part of space. We emphasize
again: this restriction is related to limiting the region in which
observations can be made, without imposing new boundary conditions.
In practical terms it could mean adding a Heaviside step function
as our window operator, thus integrating only up to a defined point.
We calculate the energy we as observers will measure. We take free
spinless bosons and consider states that are created by the field
operator $\Psi$ acting on the vacuum:  
\begin{eqnarray}
\left|\Psi\right\rangle  & = & \Psi^{\dagger}(\vec{r})\left|0\right\rangle =\sum_{\vec{p}}\frac{e^{-i\vec{p}\vec{r}}}{\sqrt{\Omega}}g(\vec{p})\, a_{\vec{p}}^{\dagger}\left|0\right\rangle .\label{fieldop}
\end{eqnarray}
 The function $g(\vec{p})$ is the wave function of the state in momentum
space.%
\footnote{The function $g$ will not be particularly relevant for us and in
most cases we will ignore it by setting $g(\vec{p})=1$. All our results
can be easily generalized for the case $g(\vec{p})\ne1$.%
}

The Hamiltonian is given by 
\begin{eqnarray}
H & = & \sum_{\vec{p}}\frac{p^{2}}{2m}a_{\vec{p}}^{\dagger}a_{\vec{p}}.\label{eq:H}
\end{eqnarray}

The energy of a state $|\psi\rangle$ is given by
\begin{eqnarray}
E=\left\langle \psi|H|\psi\right\rangle  & = & \left\langle 0|\Psi H\Psi^{\dagger}|0\right\rangle .\label{eq:psiTpsi}
\end{eqnarray}
 In configuration space the energy is given by 
\begin{equation}
E=\intop_{-\infty}^{\infty}d^{3}r\frac{1}{2m}\left\langle 0|\nabla_{r}\Psi\left(\vec{r}\right)\nabla_{r}\Psi^{\dagger}\left(\vec{r}\right)|0\right\rangle .
\end{equation}

We calculate the energy corresponding to the restricted Hamiltonian
$E_{\psi}^{V}=\left\langle \psi|H^{V}|\psi\right\rangle =Tr(\rho^{V}H^{V})$.
We replace the restricted Hamiltonian $H^{V}$ by its smoothed counterpart
with the help of a window function $f(\vec{r},w)$, as discussed above.
Alternatively, we can use a restricted smoothed field operator (here
we set $g=1$) 
\begin{equation}
\Psi_{\text{smoothed}}^{V}=\int d^{3}r\, f\left(\vec{r}\right)\Psi^{\dagger}\left(\vec{r}\right)=\int d^{3}r\, f\left(\vec{r},w\right)\sum_{\vec{p}}\frac{e^{-i\vec{p}\vec{r}}}{\sqrt{V}}\, a_{\vec{p}}^{\dagger}=\sum_{\vec{p}}f\left(\vec{p},w\right)a_{\vec{p}}^{\dagger},\label{psiv}
\end{equation}
 with $f\left(\vec{p},w\right)$ being the Fourier transform of $f\left(\vec{r},w\right)$.
Because $f\left(\vec{r},w\right)$ is a smooth function its Fourier
transform suppresses large momenta and acts effectively as a high
momentum cutoff. The result of Eq.(\ref{psiv}) is substituted into
Eq.(\ref{eq:psiTpsi}). The creation operators on the vacuum give
delta functions, resulting in 
\begin{eqnarray}
E_{\text{smoothed}}^{V} & = & \frac{1}{2m}\intop_{-\infty}^{\infty}d^{3}r\vec{\nabla}f\left(\vec{r},w\right)\cdot\vec{\nabla}f(\vec{r},w)\label{eq:nonrelativistic_energy_eqs-2}
\end{eqnarray}
 In Sec.\ref{sec:appB} we evaluate explicitly a related case, the
restricted smoothed momentum squared $\left\langle \psi|(P_{\text{smooth}}^{2})^{V}|\psi\right\rangle $.

\begin{figure}[h]
\centering$\:\:\:\:\:\:\:\:\:$\includegraphics[scale=0.8]{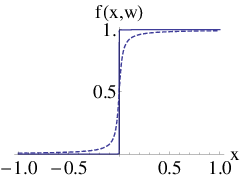}$\:\:\:\:\:\:\:\:\:\:$\includegraphics[scale=0.8]{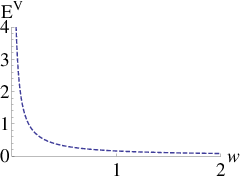}

\caption{Shown\label{fig:smoothed} is a one dimensional example of a smooth
window function (left) and the corresponding restricted energy as
a function of barrier width (right).}
\end{figure}

For specific window functions the smoothed restricted energy can be
evaluated explicitly. Consider, for example, a one dimensional case
with 
\begin{equation}
f(x,w)=\frac{1}{2}+\frac{1}{\pi}\arctan\left(\frac{x}{w}\right).
\end{equation}
 The function is depicted by the dashed line in Fig.\ref{fig:smoothed}.
In momentum space (ignoring the singularity at $p=0$) 
\begin{equation}
f(p,w)=\frac{1}{\sqrt{2\pi}}\frac{1}{p}e^{-|p|w}.
\end{equation}
 So $1/w$ acts as a high momentum cutoff suppressing any momentum
components of the smoothed wavefunction with $|p|>\frac{1}{w}$.

The value of the restricted energy, the restriction being the positive
half of the x-axis, can be calculated analytically in this case
\begin{equation}
E_{\text{smoothed}}^{V}=\frac{1}{2m}\frac{1}{2\pi w}.
\end{equation}
 This is also shown in Fig.\ref{fig:smoothed}, taking $m=1/2$. As
$w\to0$, so that the partition becomes sharper, the energy increases,
and it diverges for an infinitely sharp partition.%
\footnote{We note that this term represents the contribution of the partitioning
to the energy. A full calculation would include the wave function
for the particle, $g(\vec{p})$ as explained in the previous note.%
}

Other smoothing functions yield very similar results. The restricted
energy is inversely proportional to the smoothing width $w$ and diverges
in the limit $w\to0$.

For the nonrelativistic case, $E_{\text{smoothed}}^{V}=\frac{1}{2m}\left\langle \psi|(P_{\text{smooth}}^{2})^{V}|\psi\right\rangle $.
Since $\left\langle \psi|(\vec{P}_{\text{smooth}})^{V}|\psi\right\rangle =0$
it follows that $\left\langle \psi|(P_{\text{smooth}}^{2})^{V}|\psi\right\rangle =(\Delta P_{\text{smooth}}^{V})^{2}$
so the divergence of the energy is equal to the divergence of the
momentum fluctuations. The divergence should not be confused with
a UV divergence; the two are unrelated. The boundary behaves as if
it is a localized particle. Given a function describing barrier slope,
energy increases as the barrier grows sharper. That is, the more sharply
the position of the dividing barrier is specified, the larger the
energy. In the limit that the width tends to zero $w\to0$ the energy
diverges. This is the same phenomenon found in quantum mechanical
uncertainty, where the more sharply we specify the position of a particle,
the greater the uncertainty of its momentum. The energy in this case
is a simple function of momentum and linearly related to momentum
uncertainty, so that as the momentum fluctuations diverge so will
the energy. Thus the energy divergence here is an indication of position/momentum
uncertainty.

\subsection{Relativistic smoothed restricted energy}

We extend the previous computation from the case of non-relativistic
fields to the case of relativistic fields. It is not immediately clear
what position uncertainty means in the case of a relativistic field
because the position operator is not defined in a clear way for this
case.

The momentum operator, on the other hand, can be defined in a straightforward
way from the energy-momentum tensor $P_{j}=T_{0j}=\int\frac{d^{3}k}{\left(2\pi\right)^{3}}k_{j}a_{\vec{k}}^{\dagger}a_{\vec{k}}$
taking $c,\hbar=1$. Using the momentum operator we can have a practical
definition of the uncertainty relations based on evaluation of the
momentum fluctuations in a localized state corresponding to excitation
of the field in a limited region of space. This is what we will use
in the following, leaving the formal definitions and the deeper meaning
of this definition for more philosophical discussions. Let us consider
a single particle state $\int d^{3}xg(\vec{x}-\vec{x}_{0},w)\Psi^{\dagger}(\vec{x})|0\rangle$.
The wavefunction of this state $g(\vec{x}-\vec{x}_{0},w)$ is localized
at $x=x_{0}$ with $w$ being the scale on which the state is spread.
For example, we can take $g(x-x_{0},w)\propto e^{-\frac{(x-x_{0})^{2}}{2w^{2}}}$.
We then evaluate the momentum fluctuations in this state. They will
grow in an inverse proportionality to the localization scale $w$
of the state. Similarly, if we have an $n$-particle state $\int\prod\limits _{1}^{n}d^{3}x_{i}g(\vec{x_{i}-x_{0},w})\prod\limits _{1}^{n}\Psi_{j}^{\dagger}(\vec{x_{j}})|0\rangle$and
we evaluate the fluctuations of the total momentum of the state, they
will grow in an inverse proportionality to the localization scale
$w$. Obviously, the state can have several localization scales. In
that case the smallest one will be the most significant. The generalization
to an arbitrary state should be clear by now.

In this context, formally, the only difference between a relativistic
field and the non-relativistic field treated with second quantization
is that both creation and destruction operators appear in the field
operator. The formal analogy between the relativistic case and the
non-relativistic one is clear and must point to a real correspondence
between the two cases when considering the position/momentum uncertainty
relation despite the inability to define a covariant position operator
for the relativistic case.

In a relativistic system the energy operator is taken from the energy
momentum tensor: $H=T_{00}=\int\frac{d^{3}k}{\left(2\pi\right)^{3}}k_{0}a_{\vec{k}}^{\dagger}a_{\vec{k}}$.

In order to look for the various expectation values we recall the
relativistic scalar product: 
\begin{equation}
\left\langle \varphi|\phi\right\rangle =-i\int d^{3}x\left[\varphi\partial_{t}\phi^{*}-\left(\partial_{t}\:\varphi\right)\phi^{*}\right]
\end{equation}
 and the expression for the expectation value of the Hamiltonian in
a state $|\varphi\rangle$ 
\begin{equation}
\left\langle \varphi\left|H\right|\varphi\right\rangle =-i\int d^{3}x\left[\varphi\partial_{t}\left(H\varphi\right)^{*}-\left(\partial_{t}\:\varphi\right)(H\varphi)^{*}\right].
\end{equation}

A smoothed state with window function, as before, can be defined as
before $\int d^{3}r\, f\left(\vec{r}\right)\Psi^{\dagger}\left(\vec{r}\right)\left|0\right\rangle $
where the field operator here is the relativistic one. The resulting
smoothed restricted energy is 
\begin{eqnarray}
E_{\text{smooth}}^{V}=\left\langle \psi\left|(H_{\text{smooth}})^{V}\right|\psi\right\rangle  & = & \int d^{3}p\ f(\vec{p},w)\ p\ f(-\vec{p},w)\nonumber \\
 & = & \int d^{3}r\ f(\vec{r},w)\ \sqrt{\vec{\nabla}^{2}}\ f(\vec{r},w)
\end{eqnarray}
 The details of the derivation are given in Sec.\ref{sec:appC}. The
result clearly has the same behavior as in the non relativistic case.
Alternately, since $E^{2}\sim P^{2}$ we may calculate $\left\langle P^{2}\right\rangle $
and obtain
\begin{eqnarray}
\frac{1}{2}\intop_{-\infty}^{\infty}d^{3}r\vec{\nabla}f\left(\vec{r},w\right)\cdot\vec{\nabla}f(\vec{r},w)\label{eq:relativistic_energy sq_eqs-2}
\end{eqnarray}
 This is identical to the non relativistic result, and equals $(\Delta P_{\text{smooth}}^{V})^{2}$.

We saw that in the nonrelativistic treatment energy tends to diverge
the more sharply the boundary between the different parts of space
is specified. The relativistic case shows the same phenomenon. Here
too, the smoothing function $f(\vec{r},w)$ acts as a momentum cutoff.
In both cases the energy increases as the barrier width becomes narrower,
and diverges for a completely sharp barrier with zero width. In the
relativistic case $E^{2}\sim P^{2}$ rather than $E\sim P^{2}$ but
we still obtain $E\sim\Delta p$ . As before, the energy is proportional
to the momentum uncertainty, and just as in the previous section,
it diverges when the barrier is made sharp. This can be seen as an
example of position/momentum uncertainty.

\section{Restric\label{sec:BH}ted energy and statistical entropy of the black
hole}

So far we have discussed restricted operators in flat spacetime. The
restriction was implemented in an ad-hoc way by a choice of a (smoothed)
theta function. In the case of the BH, spacetime is restricted in
a different way. For example, in the Schwarzschild geometry, the metric
$ds^{2}=-(1-\frac{r_{s}}{r})dt^{2}+\frac{1}{1-\frac{r_{s}}{r}}dr^{2}+r^{2}d\Omega^{2}$
\textcolor{black}{is used to treat the region of space outside the
horizon $r>r_{s}$ }. So all the operators used in Schwarzschild geometry
are restricted operators. One can view the redshift factor $\frac{1}{1-\frac{r_{s}}{r}}$
as implementing the restriction by becoming infinite at the horizon
$r=r_{s}$.

We will try to explain how the redshift, acting as a restriction,
creates an infinitely sharp boundary that results in divergence of
the reduced energy and reduced entropy. We begin with the simpler
case of Rindler spacetime, that is the spacetime of an accelerated
observer in Minkowski space. Rindler space has the advantage that
it is equivalent to a restriction to half of Minkowski space so this
example allows us to explicitly compare the two restriction mechanisms.
We will explain how we can implement the ideas of smoothing the boundary
by restricting the maximal value of the redshift, and show that when
smoothing is implemented all quantities are rendered finite with magnitude
inversely proportional to the smoothing parameter, exactly as in the
cases that we have encountered before. This will allow us to show
that a similar phenomenon occurs for BH's.

\subsection{The uncertainty principle in Rindler spacetime}

We use the Minkowski space metric: 
\begin{equation}
ds^{2}=-dt^{2}+dz^{2}+d\vec{x}_{\bot}^{2},
\end{equation}
 where $z$ is the coordinate that will be used to separate space
into the left and right halves $z<0$ and $z>0$ and $\vec{x}_{\bot}$
stands for the transverse coordinates. An accelerated observer whose
acceleration is $a/2\pi$ lives in Rindler space whose metric is 
\begin{equation}
ds^{2}=-e^{2a\xi}d\eta^{2}+e^{2a\xi}d\xi^{2}+{d\vec{x}_{\bot}}^{2}.\label{E:rindler}
\end{equation}
The Minkowski coordinates and Rindler coordinates are related by:
\begin{align}
t(\xi,\eta) & =\frac{1}{a}e^{a\xi}\sinh a\eta\\
z(\xi,\eta) & =\frac{1}{a}e^{a\xi}\cosh a\eta\\
\vec{x}_{\bot} & =\vec{x}_{\bot}.
\end{align}

Choosing a fixed Rindler time, for example, $\eta=0$, we see that
the $\xi$ coordinate only covers the $z>0$ half of space. The restriction
is implemented by the redshift factor $e^{-a\xi}$ which diverges
for $\xi\to-\infty$, corresponding to $z=0$.

As it stands, the restriction implemented by the redshift is infinitely
sharp. \textcolor{black}{The Rindler observer does not see the region}
$z<0$ . We wish to understand how to implement a smoothed restriction
rather than an infinitely sharp one. So we analyze just how the redshift
leads to divergence of $(\Delta p)^{2}$ and vanishing of $(\Delta z)^{2}$,
in order to consider how the divergence may be tamed. We consider
a non-relativistic particle whose wave function has some spread $\Delta z$
in Minkowski space. For example, 
\begin{equation}
\psi(z)=\frac{1}{\sqrt{2\pi(\Delta z)^{2}}}e^{-\frac{1}{2}\ensuremath{\frac{z^{2}}{(\Delta z)^{2}}}}.
\end{equation}
 In momentum space the spread of the wave function is inversely proportional
to $\Delta z$, $(\Delta p)^{2}\sim1/(\Delta z)^{2}$. Viewed by an
accelerated observer, the wave function at the origin $z=0$ corresponding
to $\xi\to-\infty$ would be squeezed in the $\xi$ direction: $\Delta\xi=e^{a\xi}\Delta z$.
As required by the uncertainty principle the spread in momentum would
increase, $\Delta p_{\xi}=e^{-a\xi}\Delta p_{z}$. Thus finite $\Delta z$
and $\Delta p$ in Minkowski space are adjusted by the Rindler metric,
so that to the Rindler observer the position fluctuations at the origin
will vanish and momentum fluctuations will diverge.

By our choice the particle is localized at the origin (any other choice
would simply require a shift in the Rindler time $\eta$), so in the
limit $\xi\to-\infty$ the momentum fluctuations diverge because the
the wave function has been squeezed in space. This divergence obviously
does not signal a breakdown of physics. It just means that considering
the classical Rindler geometry when viewing a quantum particle requires
closer thought. Rindler geometry imposes a restriction on Minkowski
space. When the restriction is sharp, equivalent to localizing a particle
at the origin, the momentum fluctuations diverge. Limiting the Rindler
redshift factor tames the divergence and increases position fluctuations,
thus softening the localization, and smoothing the restriction.

\subsection{Momentum fluctuations and r\label{sub:Momentum-fluctuation Rindler}edshift
in Rindler spacetime}

In view of the previous discussion, and in preparation for the reinterpration
of the 't Hooft calculation, let us consider a (massless) scalar field
$\phi$ that satisfies the Klein-Gordon equation
\begin{equation}
\frac{1}{\sqrt{-g}}\left(\partial_{\mu}\sqrt{-g}g^{\mu\nu}\partial_{\nu}\right)\phi=0.
\end{equation}

In Minkowski spacetime there is an exact solution to the Klein-Gordon
equation. The $z$ dependent part of the solution is given by 
\begin{equation}
\phi(z)=e^{\pm ipz}.
\end{equation}
 However, for the purpose of making the calculation more similar to
the 't Hooft calculation we can rewrite the solution in a WKB form,
where the WKB solution is 
\begin{equation}
\phi_{_{WKB}}(z)=e^{\pm\int\limits ^{z}p(z)dz}.
\end{equation}
 Obviously, in Minkowski space $p(z)$ is a constant and the WKB solution
reduces to the exact solution. The WKB momentum can be expressed as
\begin{equation}
p^{2}(z)=E^{2}-p_{\bot}^{2}.
\end{equation}

In Rindler spacetime the WKB wave function is
\begin{equation}
\phi_{_{WKB}}(\xi)=e^{\pm i\int\limits ^{\xi}d\xi\sqrt{g_{\xi\xi}}p(\xi)}
\end{equation}
 with 
\begin{equation}
p^{2}(\xi)=g^{\eta\eta}E^{2}-p_{\bot}^{2}
\end{equation}
 which is space varying. So the WKB wave function is 
\begin{equation}
\phi_{_{WKB}}(\xi)=e^{\pm i\int\limits ^{\xi}d\xi\sqrt{g_{\xi\xi}}\sqrt{g^{\eta\eta}E^{2}-p_{\bot}^{2}}}.
\end{equation}
 Near the horizon $p(\xi)$ diverges as $\sqrt{g^{\eta\eta}E^{2}}=e^{-a\xi}E$
and the proper length $\widetilde{d\xi}=d\xi\sqrt{g_{\xi\xi}}=d\xi e^{a\xi}$
vanishes. This is a manifestation of the position/momentum uncertainty
relation caused by the redshift.

Rindler space implements a sharp division of Minkowski space. That
is, the Rindler observer sees a sharp cutoff at the horizon $\xi\to-\infty$.
Smoothing this cutoff in momentum space means restricting the momentum
$p(\xi)$ near the horizon. We saw in the previous section that restricting
the redshift widens $\Delta x$ and shrinks $\Delta p$. Therefore
restricting the redshift $g^{\eta\eta}$, $g^{\xi\xi}$ will smooth
the cutoff.

In 't Hooft's black hole calculation the energy and entropy diverge
due to a diverging density of states. In Rindler space too the density
of states diverges, and we will see that that this divergence is due
to the uncertainty principle. We define the density of states near
energy $E$ in Rindler space and evaluate it by counting the number
of WKB solutions

\begin{eqnarray}
\pi n & = & \int d\xi e^{a\xi}\int\frac{d^{2}p_{\bot}}{\left(2a\right)^{2}}p(\xi,E,p_{\bot})\nonumber \\
 & = & 2\pi\int d\xi e^{a\xi}\int\frac{dp_{\bot}}{\left(2a\right)^{2}}\ p_{\bot}\sqrt{e^{-2a\xi}E^{2}-p_{\bot}^{2}}\nonumber \\
 & = & -\frac{2}{3}\frac{\pi}{\left(2a\right)^{2}}E^{3}\int d\xi e^{-2a\xi}
\end{eqnarray}
 where we have performed first the angular integral of $p_{\bot}$
and then the radial part. This integral diverges because of the diverging
redshift factor at the horizon. So the density of states, the entropy
and energy are divergent for the same reason and if the redshift factor
is restricted, they all become finite.

We can smooth the partition by limiting the redshift, or alternately,
by implementing a smoothing function on states of the system. This
equivalent procedure will also tame the divergence. The smoothed functions
that we need to count are obtained by multiplying the original unsmoothed
function by the smoothing function, $\psi(\xi)\to\psi(\xi)f(\xi,w)$,
or in Fourier space $\phi(p)\to\phi(p)f(p,w)$. Recall that in momentum
space the function $f(p,w)$ acted as a high momentum cutoff for $p>1/w$.
Then for wavefunctions with energy $E$ we need to effectively restrict
the Rindler momentum $p(\xi)=e^{-a\xi}\sqrt{E}$ to be $p(\xi)<1/w$.
In this context it simply means that the redshift factor is limited
to some maximal value which can always be expressed as $e^{-a\xi_{min}}$.
The ``brick wall\textquotedbl{} model of 't Hooft in this context
amounts to a sharp cutoff on the momentum $p(\xi)$. However, clearly,
any other cutoff schemes will do the same job. The density of states
of smoothed wavefunctions is of course finite, 
\begin{eqnarray}
\pi n & = & \int_{\xi_{min}}d\xi e^{a\xi}\int\frac{d^{2}p_{\bot}}{\left(2a\right)^{2}}p(\xi,E,p_{\bot})\nonumber \\
 & = & 2\pi\int_{\xi_{min}}d\xi e^{a\xi}\int\frac{dp_{\bot}}{\left(2a\right)^{2}}\ p_{\bot}\sqrt{e^{-2a\xi}E^{2}-p_{\bot}^{2}}\nonumber \\
 & = & \frac{2}{3}\frac{\pi}{(2a)^{3}}E^{3}e^{-2a\xi_{min}}.
\end{eqnarray}
 This makes the energy and entropy finite and inversely proportional
to the maximal redshift which determines the smoothing width of the
division in Rindler space.

\subsection{Momentum fluctuations and entanglement entropy in Schwarzschild spacetime}

't Hooft solves the wave equation in the Schwarzschild metric, identifies
$p$, the wave number, and using a WKB approximation he obtains the
density of states. However the redshift leads this to diverge at the
horizon. The region near the black hole horizon is a thermal state
in Rindler space, and indeed just as in Rindler space, limiting the
redshift will prevent the divergence.

We recall the calculation in Schwarzschild coordinates. For simplicity
we have chosen the scalar field to be massless. The Klein-Gordon equation
in these coordinates is 
\begin{equation}
\left(1-\frac{2M}{r}\right)^{-1}E^{2}\phi+\frac{1}{r^{2}}\partial_{r}\left(r\left(r-2M\right)\partial_{r}\right)\phi-\left(\frac{l\left(l+1\right)}{r^{2}}\right)\phi=0.\label{eq:thooft wave eq}
\end{equation}
 The wave number can be defined as%
\footnote{This differs by a a factor $g_{rr}$ from 't Hooft's original defintion.%
} 
\begin{equation}
p^{2}=g^{tt}E^{2}-\left(\frac{l\left(l+1\right)}{r^{2}}\right)
\end{equation}
 Using a WKB approximation the density of states for a massless scalar
field is given by 
\begin{eqnarray}
\pi n & = & \sum\limits _{l,m}\intop_{2M}dr\sqrt{g_{rr}}\, p\left(r,l,m\right)\label{eq:t'Hooft integral}\\
 & = & \intop_{2M}dr\sqrt{g_{rr}}\int(2l+1)dl\sqrt{g^{tt}E^{2}-\frac{l\left(l+1\right)}{r^{2}}}\nonumber 
\end{eqnarray}
 where $l,m$ are the angular parameters. Evaluating the integral
over $l$ we find 
\begin{eqnarray}
\pi n & = & -\frac{2}{3}\intop_{2M}dr\sqrt{g_{rr}}r^{2}\left(g^{tt}E^{2}\right)^{3/2}\nonumber \\
 & = & -\frac{2}{3}E^{3}\intop_{2M}dr\frac{r^{2}}{\left(1-\frac{2M}{r}\right)^{2}}
\end{eqnarray}
 This integral diverges at the horizon. If we were to limit the redshift,
as we did with Rindler space, there would be no divergence. Apparently
't Hooft does otherwise: he takes the lower limit a slight distance
away from the horizon, his well known ``brick wall,'' so that the
lower limit becomes $2M+h$. From this expression he obtains the energy
and entropy, which diverge as $h\rightarrow0$.

In fact 't Hooft's adjustment of the lower limit of the integral from
$2M$ to $2M+h$ is equivalent to a change of variable which leaves
the lower limit at $2M$ but changes the redshift: 
\begin{eqnarray}
\intop_{2M+h}dr\left(1-\frac{2M}{r}\right)^{-2} & = & \intop_{2M}d\widetilde{r}\left(1-\frac{2M}{\widetilde{r}+h}\right)^{-2}\label{orint}
\end{eqnarray}
 This clearly does not diverge at the horizon. The new expression
is always finite and is limited by $\left(1-\frac{2M}{2M+h}\right)^{-2}\lesssim(2M/h)^{2}$
for $h\ll M$.

The altered redshift is equivalent to multiplication of the original
redshift in the $\widetilde{r}$ system by a smoothing function: 
\begin{eqnarray}
\left(1-\frac{2M}{\widetilde{r}+h}\right)^{-1} & = & \left(1-\frac{2M}{\widetilde{r}}\right)^{-1}\: f(\widetilde{r},h)
\end{eqnarray}
 with 
\begin{eqnarray}
f\left(\widetilde{r},h\right) & = & \frac{\left(\widetilde{r}+h\right)\left(\widetilde{r}-2M\right)}{\widetilde{r}\left(\widetilde{r}-2M+h\right)}.
\end{eqnarray}
 Thus the change of variable implemented by the brick wall has the
effect of multiplying the redshift by a smoothing function.

The original divergent integral in eq.\ref{orint} can be expressed
in terms of a sharp step function $\intop_{2M}dr\left(1-\frac{2M}{{r}}\right)^{-2}=\intop_{0}dr\ \Theta(r-2M)\left(1-\frac{2M}{{r}}\right)^{-2}$.
The altered integral can be expressed in terms of a smoothed step
function 
\begin{eqnarray}
\intop_{0}dr\left(1-\frac{2M}{r}\right)^{-2}f^{2}(r,h)\ \Theta(r-2M) & = & \intop_{0}dr\left(1-\frac{2M}{r}\right)^{-2}\widetilde{\Theta}(r-2M,h)\label{eq:soft step}
\end{eqnarray}
 Thus we see that 't Hooft's changed lower limit is exactly equivalent
to smoothing the step function to a new one $\widetilde{\Theta}(r-2M,h)=f^{2}(r,h)\ \Theta(r-2M)$
with width $h$. Formally the brick wall can be seen as either changing
the redshift or smoothing the step function \textcolor{black}{and
thus modifying the sharp partitioning of the region.} Obviously, any
other limiting procedure of the maximal redshift will render the integral
finite and make the energy and entropy finite. 
\begin{figure}[H]
\centering

\includegraphics[scale=0.8]{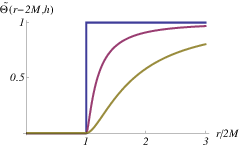}

\caption{Smoothed step function as function of $r/2M$. Curves have $h=0$
(sharp step), $h=0.1$ and $h=0.9$ (lowest). }

\end{figure}

\section{Summary of\label{sec:summary and conclusions} results}

Energy has been shown to diverge as the boundary between two quantum
subsystems, an observable subsystem and an unobservable subsystem,
becomes sharp. The divergence is due to the fact that the energy is
a simple function of the momentum fluctuations. These diverge in the
presence of a sharp boundary because of the uncertainty principle,
much in the same way that they diverge for a sharply localized particle.
For the nonrelativistic case $\left\langle E\right\rangle =\frac{1}{2m}(\Delta P)^{2}$.
In the relativistic case $\left\langle E\right\rangle =\Delta P$
so in both cases energy divergence at an infinitely sharp boundary
is clearly a consequence of position/momentum uncertainty.

In a coordinate system which implements a sharply localized boundary,
the density of states and thus energy and entropy diverge at the boundary.
Limiting the redshift tames this divergence. We have shown that limiting
the redshift smoothes the boundary by widening $\Delta x$ and limiting
$\Delta p$. Therefore the smoothing cutoff prevents the energy from
diverging. This implies that the divergence of the energy and entropy
was a result of the sharp localization of the boundary, and was due
to the uncertainty principle.

The region near the boundary of a black hole is a thermal state, where
the entropy is linear to energy. Therefore black hole entropy will
diverge at the boundary as well. We have shown that regardless of
any other cause, there would be divergence at the infinitely sharp
boundary as a result of the uncertainty principle. We have also shown
that 't Hooft's divergence at the black hole is an example of momentum/position
uncertainty, as seen by the fact that the ``brick wall'' which corrects
it in fact smoothes the sharp boundary.

This result raises the question whether the entanglement and statistical
mechanics definitions of black hole entropy might refer to the same
quantity. Both are proportional to area. The UV divergence may be
tamed with a UV cutoff, and the boundary divergence by smearing out
the boundary (both procedures might turn out to be equivalent). So
the two expressions could be expressing the same quantity. If this
is the case, then the microscopic counting of the number of states
becomes tantamount to counting the correlations between the observed
and unobserved regions of spacetime. Black hole entropy has also been
shown, from thermodynamic considerations as well as explicit calculations
in string theory, to equal one fourth of the horizon area. An open
problem is to obtain the factor of $1/4$ in either of these definitions
of black hole entropy.

\section{\label{sec:Carlip}Entropy from conformal field theory}

In the previous section we showed that the divergence of entropy at
the horizon may be due to quantum uncertainty. We explicitly refrained
from imposing boundary conditions at the horizon. The boundary served
to trace out part of the system by limiting the possibility of observations,
but did not imposing new boundary conditions on the system.

Carlip \cite{THE Carlip paper} derived black hole entropy from arguments
of symmetry. This was motivated by earlier work of Brown and Henneaux
\cite{Brown Henneaux} showing that 2+1 dimensional gravity with a
negative cosmological constant has an asymptotic symmetry consisting
of a pair of Virasoro algebras, so that a microscopic quantum black
hole theory should be a conformal field theory. In conformal theory
entropy can be obtained with the Cardy formula. Strominger \cite{Strominger Carlip}
made use of this to obtain entropy of a 2+1 dimensional black hole.

Carlip proposed a scheme applicable in any dimension. When the horizon
is treated as a boundary, the algebra of constraints in general relativity
acquires a central extension. With appropriate boundary conditions
this extended algebra contains a Virasoro subalgebra and the Cardy
formula can be used to obtain entropy. In \cite{Carlip ADM} this
was done using the ADM formula, and then in \cite{THE Carlip paper}
using covariant phase space methods. We focus on the second approach
here. We note that Carlip's entropy does NOT diverge at the boundary,
and discuss the reason.

Below we outline Carlip's scheme and focus on his treatment of the
boundary. He works with a stretched horizon, such that $\chi^{2}=\epsilon$
where $\chi$ is a Killing vector at the boundary, and at the end
of the calculation he takes $\epsilon\rightarrow0$. He does so in
order to vary the Noether charge \emph{at} the boundary. To do this
he must go a short distance away from the exact boundary. There he
decomposes $\chi$ (no longer a Killing vector at that point), into
two orthogonal vectors, as the ``r,t plane.'' It's as if in order
to wiggle something one needs to stretch it a bit. If one sits exactly
on the boundary there is no wiggle room. The question we examine is
- what happens to Carlip's entropy when taking $\chi^{2}$ to 0, that
is, exactly at the boundary? Does it diverge as in the previous section?
We find that the scheme breaks down at the boundary itself, and discuss
the implications.

\subsection{Background}

In 1986 Brown and Henneaux \cite{Brown Henneaux} showed that in the
canonical (ADM) formalism general relativity has a central extension,
so that $\left\{ H\left[\xi_{1}\right],H_{2}\left[\xi_{2}\right]\right\} =H\left\{ \xi_{1},\xi_{2}\right\} +K\left\{ \xi_{1},\xi_{2}\right\} $
where $\xi$ is a surface deformation vector, $H\left[\xi\right]$
is its symmetry generator and $K$ is the central charge. In the ADM
formalism the vector fields $\xi$ which preserve the spacetime metric
under Lie transport become deformations of a spacelike surface described
by the canonical variables $g_{ij},\pi^{ij}$ . Brown and Henneaux
show this symmetry group is isomorphic to the \textcolor{black}{two
dimensional} conformal group, its central charge is non trivial and
its algebra a direct sum of two Virasoro algebras. This was done for
$2+1$ dimensional gravity with a negative cosmological constant,
but they write that it can easily be generalized to higher dimensions.
In 1991 Barnich, Henneaux and Schomblond showed that the Hamiltonian
Poisson bracket structure is equivalent to the covariant phase space
formalism which deals with the Lagrangian \cite{barnich henneaux}.
Therefore the latter too can be used in the calculation of a central
charge for the algebra of general relativity. 

Cardy \cite{Cardy}\textsf{\textbf{ }}studied statistical systems
at a critical point, and the consequences of their conformal invariance.
He related the free energy of such a system to the central charge
of the Virasoro algebra. He thus obtained the partition function as
a function of the central charge. The number of states can be extracted
from the partition function \cite{Carlip BTZ}, and the entropy is
the logarithm of the number of states:
\begin{equation}
S\left(\mbox{\ensuremath{\Delta}}\right)=2\pi\sqrt{\frac{c_{eff}\Delta}{6}}
\end{equation}
where $\Delta$ is the eigenvalue of Virasoro generator $L_{0}$,
and the effective central charge takes different forms depending on
the particular conformal field theory under discussion. In cases where
the lowest eigenvalue of $L_{0}$ does not vanish, $\Delta_{0}\neq0$,
$c_{eff}=\left(c-24\Delta_{0}\right)$. For explicit examples see
\cite{Carlip BTZ}.

In \cite{THE Carlip paper} Carlip uses the covariant phase space
formalism, and obtains the central charge for general relativity.
The scheme is detailed in Sec.\ref{sec:Carlip's-scheme}, but here
we give a brief outline. Since he is interested in the entropy associated
with the black hole horizon, he attempts to specify boundary conditions
that will reflect the presence of the horizon. He works with a stretched
horizon. In that case the spacetime has no Killing vector, so he requires
boundary conditions which preserve the asymptotic horizon structure,
that is, as $\chi^{2}\rightarrow0$ they ensure that the boundary
will be a null surface. He focuses on vector fields $\xi^{a}$ which
generate a diffeomorphism, and decomposes them into a linear combination
of orthogonal vectors in the ``r-t plane.'' $\xi^{a}=R\rho^{a}+T\chi^{a}.$
At the horizon $\rho^{a}$ and $\chi^{a}$ coincide. One finds that
$R\sim\chi^{a}\nabla_{a}T$, so that $\xi^{a}=const\cdot\chi^{b}\nabla_{b}T\rho^{a}+T\chi^{a},$
and that closure of the algebra requires $\rho^{a}\nabla_{a}T=0$
at the horizon.

Writing $\chi^{a}\nabla_{a}T\equiv DT,$ the central charge is shown
to be
\begin{equation}
K\left[\xi_{1},\xi_{2}\right]=\frac{1}{16\pi G}\intop_{\mathcal{H}}\hat{\epsilon}_{a_{1}...a_{n-2}}\frac{1}{\kappa}\left(DT_{1}D^{2}T_{2}-DT_{2}D^{2}T_{1}\right).\label{eq:carlip's central charge}
\end{equation}
The $T$ are writen as periodic functions 
\begin{equation}
T_{n}\left(\nu,\theta^{i}\right)=\frac{1}{\kappa}e^{ink\nu}f_{n}\left(\theta^{i}\right)
\end{equation}
where $\nu$ is a parameter along the orbits of the Killing vector
$\chi$, and $\theta^{i}$ are angular coordinates. Implementing previously
derived orthogonality constraints, the central term is
\begin{equation}
K\left[T_{m},T_{n}\right]=-\frac{iA}{8\pi G}m^{3}\delta_{m+n.0}
\end{equation}
and plugging this into the Cardy formula, with $\Delta$ as a given
eigenvalue of $J\left[T_{0}\right],$ gives the black hole entropy
with the known value of $\frac{A}{4G}$.

\subsection{The horizon}

Carlip's derivation was done using the stretched horizon. To find
the entropy at the horizon one then takes $\chi^{2}\rightarrow0$.
However this is problematic.

First, at the horizon the two orthogonal vectors to which the diffeomorphism
generator was decomposed coincide. Since $\rho=\chi$ at the horizon
this means that on the horizon the central charge vanishes. A key
requirement for closure of the algebra was $\rho^{a}\nabla_{a}T=0$.
When $\rho=\chi$ then also $\chi^{a}\nabla_{a}T=0$ and clearly eq.(\ref{eq:carlip's central charge})
equals zero and the central charge vanishes, as does the entropy.
Therefore either the entropy at the horizon vanishes or the entire
scheme is not valid exactly at the horizon. This could be meshed with
the previous section of this chapter, as showing that it is not possible
to calculate entropy at a defined point in space. Cardy's entropy
is in fact that of statistical mechanics and is proportional to energy.
The problem with this idea is that Carlip's entropy does not diverge
on the boundary if the central charge vanishes, but rather the entropy
vanishes as well. 

To further investigate this we change to horizon crossing coordinates.
Since away from the horizon, $\chi,\rho$ are the ``r,t plane,''
then $\chi$ is in fact the timelike vector $\partial_{t}$ and $\rho$
is the radial vector $\partial_{r}$ . To enforce closure of the algebra,
the constraint is $\rho^{a}\nabla_{a}T=0$ . So $T$ is only a function
of $\chi$ (that is, of $t$ not of $r$) . Writing them as Kruskal
coordinates we will see that for $T$ a function of $t,$ 
\begin{equation}
T=f(t)\rightarrow T=f\left(\frac{u}{v}\right)=f\left(e^{t/2M}\right).
\end{equation}
Then $T=f\left(\frac{u}{v}\right)$ at the horizon will be either
diverge or be trivial, depending whether $u$ or $v$ goes to 0 faster.
In detail: We write the vector $\xi=ADT\hat{\rho}+T\hat{\chi}$ where
$A\equiv\frac{1}{\kappa}\frac{\chi^{2}}{\rho^{2}}.$ We use tortoise
coordinates:
\begin{equation}
r^{*}=r+2Mlog|\frac{r-2M}{2M}|\approx2Mlog|\frac{r-2M}{2M}|
\end{equation}
at the horizon. 
\begin{eqnarray}
U & = & -exp\left\{ \frac{r^{*}-t}{4M}\right\} =-exp\left\{ \frac{2Mlog\left|\frac{ADT-2M}{2M}\right|-T}{4M}\right\} =-\sqrt{\left|\frac{ADT-2M}{2M}\right|}e^{-T/4M}\nonumber \\
V & = & exp\left\{ \frac{r^{*}-t}{4M}\right\} =\sqrt{\left|\frac{ADT-2M}{2M}\right|}e^{T/4M}.
\end{eqnarray}
 So
\begin{equation}
\xi=-\sqrt{\left|\frac{ADT-2M}{2M}\right|}e^{-T/4M}\hat{U}+\sqrt{\left|\frac{ADT-2M}{2M}\right|}e^{T/4M}\hat{V}.
\end{equation}
or in a basis of $(U,V,x_{\bot})$
\begin{equation}
\xi=\left(-\sqrt{\left|\frac{ADT-2M}{2M}\right|}e^{-T/4M},\sqrt{\left|\frac{ADT-2M}{2M}\right|}e^{T/4M},0\right).
\end{equation}
To work out the central charge we write the modes

\begin{eqnarray*}
T_{n} & = & \frac{1}{k}e^{iknt}f_{n}(\theta),\:\frac{dT}{dt}=-iknT,\,\frac{d^{2}T}{dt^{2}}=-k^{2}n^{2}T\\
-\frac{U}{V} & = & e^{-t/(2M)}\rightarrow t=log\left[-\frac{U}{V}\right]^{-2M}\\
T & = & \frac{1}{k}\left[-\frac{U}{V}\right]^{-ikn2M}
\end{eqnarray*}
Then the central charge works out to be
\begin{eqnarray}
K{\color{red}} & = & \frac{1}{16\pi G}\intop_{\mathcal{H}}\hat{\epsilon}f_{n}f_{m}\left(-inm^{2}+imn^{2}\right)\left[\frac{U}{V}\right]^{-i4kM(n+m)}\nonumber \\
 & = & \frac{A}{16\pi G}\delta_{m+n,0}\left(-inm^{2}+imn^{2}\right)\left[\frac{U}{V}\right]^{-i4kM(n+m)}\nonumber \\
 & = & \frac{A}{8\pi G}\delta_{m+n,0}\left(in^{3}\right)\left[\frac{U}{V}\right]^{-i4kM(n+m)}
\end{eqnarray}
\textcolor{black}{If the integration is done before going to the horizon,
this term is finite thanks to the delta function. However the calculation
cannot be performed on the horizon itself, where if $V=0,\: U\neq0$,
the expression will vanish or diverge at the horizon. The limiting
process must be done at the end of the calculation.}

The divergence seems to recall the divergence we found in the previous
section, in particular because it arises as the stretched horizon
is taken back to the precise location of the horizon. However it results
from the constraint for closure of the algebra rather than from quantum
uncertainty. 

We see that the very constraint which ensures the existence of the
algebra of diffeomorphism invariance prevents calculation of entropy
at the horizon. This makes intuitive sense, since the horizon limits
diffeomorphism invariance by defining a border beyond which no continuous
translation is possible. In Chapter 4 we vary the entropy using a
Lie derivative of the partition function, and we see that in a space
without a horizon it would vanish, as a result of diffeomorphism invariance,
but once the space is bounded it does not vanish. In Carlip's scheme
the existence of a boundary of space gives rise to the central charge
from which the entropy is calculated; the weak point in this scheme
is that the algebra itself does not hold at the boundary, and thus
the central charge itself cannot be calculated there.

This differs from statistical and entanglement entropy in two aspects.
First, in statistical and entanglement entropy, the boundary is perceived
as a ``virtual wall'' allowing us to look at part of the system,
rather than an actual wall imposing boundary conditions. Second, problematic
behavior at the boundary in statistical entropy was shown to be a
possible artifact of quantum uncertainty, whereas in Carlip's scheme
it results from constraints on the classical algebra.

\section{Discussion}

Entanglement entropy does not necessarily diverge at the horizon,
whereas thermodynamic entropy has been shown to do so. However this
divergence can be seen as a result of quantum uncertainty rather than
related to some new unknown black hole physics. 

In this chapter I have looked at boundary behavior of different treatments
of black hole entropy. In the first part, I claimed that boundary
conditions must not be changed when tracing out part of the system:
imposing new boundary conditions would mean changing the subsystem
itself. In contrast, Carlip's treatment imposes boundary conditions
and his derivation of the entropy is a direct result of the existence
of the boundary, and of his imposition of boundary conditions. 

Carlip's entropy does not result from the state being thermal but
rather from conformal symmetry. His imposition of boundary conditions
is related to this symmetry. One effect of this is that there is no
horizon divergence. But the lack of divergence is essentially related
to the entropy's source in symmetry, as opposed to thermodynamic entropy
which is proportional to energy and thus diverges from x/p uncertainty.
On the face of it horizon divergence appears to be related to treatment
of boundary conditions, but in fact it is related to the question
whether we deal with a thermal state or not, because in that case
entropy is proportional to energy, which was shown to diverge at a
sharply localized boundary.

There is a possibility that a stretched horizon bears resemblance
to 't Hooft's brick wall, which we have here shown may be a method
of delocalizing the horizon. In neither case can the entropy be calculated
exactly at the horizon. However the source of the problem in these
two cases is different: in statistical mechanics it results from quantum
uncertainty and in Carlip's scheme it results from the classical algebraic
constraint on diffeomorphism invariance. 

Entanglement entropy and statistical entropy of a black hole may refer
to the same phenomenon. While entanglement entropy does not diverge
at the horizon, the divergence of statistical entropy may be due to
x/p uncertainty. If they are the same, the degrees of freedom that
are entangled would be the Hawking pairs, since it is they that lead
to a thermal state at the horizon. In both cases boundary conditions
are not imposed at the border. This is essentially different from
Carlip's entropy from conformal symmetry. It is clear from all this
that  when considering black hole entropy one must specify the treatment
of boundary conditions in the scheme under discussion.

\chapter{Curvature independence}

In this chapter I focus on statistical entropy. I examine the number
of states, as given by the volume of phase space. Statistical mechanical
entropy is then calculated from the number of states, following 't
Hooft. Inspired by 't Hooft's treatment of a single particle in a
black hole background, I consider whether statistical entropy of a
particle in a black hole background has any unique dependence on curvature,
for the following reason: 

Wald entropy is defined in terms of the curvature tensor. Since Wald
used the first law of thermodynamics to obtain the entropy, it should
be related to the entropy defined in statistical mechanics. However
if one calculates statistical entropy of a test particle in a region
of Minkowski space and then Rindler space one obtains two different
results (Sec.\ref{sub:Momentum-fluctuation Rindler} and see also
\cite{paddy kolekar}). Thus though the curvature in both Minkowski
and Rindler spaces is identical (and vanishes), the statistical entropy
will be different. In a space with non vanishing curvature, could
statistical entropy be uniquely dependent on curvature?

For a stationary black hole solution in Einstein gravity, Wald entropy
coincides with the original Bekenstein/Hawking entropy and is proportional
to area\cite{wald 1,wald and iyer}. The statistical entropy of a
particle in a black hole background was also shown to be dependent
on area \cite{paddy kolekar,paddy ideal gas}. This too leads one
to wonder about a connection between the two formulations of entropy.

We will show that the number of states is an explicit function of
the metric. Since the curvature derives from the metric, it would
seem that the number of states is related to curvature. However we
find that for certain transformations of the metric, the number of
states is preserved. These transformations do not preserve curvature.
This is shown only for a diagonal metric, but it serves as a counter
example showing that in the most general case the number of states
is not uniquely dependent on the spacetime curvature scalar, although
it may depend on the curvature together with other factors that at
times cancel the effect of the curvature.

This chapter is organized as follows. First we establish the definition
of the number of states. We then give an example that indicates observer
dependence. This is followed by a more thorough exploration of the
number of states, and of different methods of calculating the volume
of phase space. We then ask under what conditions transformation of
the metric will leave this volume invariant. We obtain a general transformation
of any metric which displays a clear constraint on the preservation
of the number of states. We examine characteristics of this transformation
and look for a possible relationship to curvature. We find that in
general it need not preserve curvature. That is, the number of states
and thus the entropy will remain the same for systems with different
curvature.

The constraint in our proof holds for a conformal transformation,
but in general other transformations of the metric will not preserve
the number of states. This proves that the number of states, and the
statistical entropy derived from it, are observer dependent quantities.

In \cite{Katya} numerical computation showed that for specific examples
of curved space entanglement entropy does not depend on curvature.
This secton of the thesis we prove analytically that statistical entropy
is not uniquely dependent on curvature. This gives further weight
to the idea that statistical and entanglement entropy represent the
same entity.

\section{Definition of number of states}

\textcolor{black}{The system we treat here is that of a classical,
massless free particle, confined to a spatial volume V. We want to
calculate the number of states up to a given energy E}. In statistical
mechanics the number of states of a single particle in a nonrelativistic
system is defined as follows: Take an integral over the volume of
phase space ($d^{3}xd^{3}p$), restrict it to values of momenta which
fit the energy eigenvalues of the system and \textcolor{black}{to
obtain}
\begin{eqnarray}
N & = & g\int d^{3}x\int\frac{d^{3}p}{\left(2\pi\hbar\right)^{3}}\nonumber \\
 & = & gV\int\frac{d^{3}p}{\left(2\pi\hbar\right)^{3}}
\end{eqnarray}
where $g$ is a numerical factor related to the degeneracy (eg., for
spins with Dirichlet boundary conditions $g=2\left(1/8\right)$ for
the positive octant and spin degeneracy). This will be taken as 1
is the system under consideration, and in what follows g will refer
to the determinant of the metric. Integration over the volume in phase
space gives $V=\left(\pi/L\right)^{3}$ for a potential well, $\left(2\pi/L\right)^{3}$for
periodic boundary conditions. Dividing by a unit of volume in momentum
space, $2\pi\hbar,$ gives the number of states in phase space with
the given energy, per unit volume of phase space. Since we do not
limit ourselves to nonrelativistic thermodynamics, nor to three space
dimensions, a more general definition is necessary.

The number of states, that is, the phase space volume of a single
particle, is then defined as
\begin{equation}
N(E,V,1)=\int d^{d}x\frac{d^{d}p}{(2\pi\hbar)^{d}}dE\delta(E-E(p)).
\end{equation}
where $d$ denotes the number of space dimensions%
\footnote{This integral is actually $\int d^{d}x\delta(t-t_{0})\frac{d^{d}p}{(2\pi\hbar)^{d+1}}dE\left[2\pi\hbar\delta(E-E(p))\right]$.%
}. Without loss of generality we are taking a constant time hypersurface
(Sec.\ref{sub:Special-relativity:-phases-pace}). For n particles
\[
N(E,V,n)=\prod_{i}^{n}N_{i}
\]
where $N_{i}=N(E,V,1)$, \textcolor{black}{and the particles are bosons
(this would be more complicated for fermions). For $n$ identical
particles this is $\left(N_{i}\right)^{n}/n!$, where the $\frac{1}{n!}$
comes from Bose statistics,}\textcolor{red}{{} }and the entropy for
n particles is
\[
S(E,V,n)=klogN(E,V,n)
\]
 where $k$ is Boltzmann's constant, and from here on we take $k=1.$
The number of states is Lorentz invariant. For a proof see Sec.\ref{sec:Lorentz-invariance-of}. 

In order to apply this definition to curved space, we need to clarify
what momentum and energy refer to for a matter field or gas of particles
in curved space. The discussion of the number of states assumes equilibrium
or near equilibrium thermodynamics. Introducing time dependence involves
complications on two levels: the spacetime may not be stationary,
and the number of states may be time dependent. Here we dispense with
these complications: we assume a stationary spacetime and take a constant
time slice. This is because our argument will depend on showing counter
examples to the claim that the number of states depends on the scalar
curvature. If a simpler case provides the counter example, it should
suffice. 

There are (at least) two possible ways to approach the issue. One
is that of \cite{'t Hooft}, who took $\psi(x)$ a scalar wave function
for a light spinless particle of mass $m$ in the Schwarzschild metric,
$m\ll1\ll M$ where $M$ is the BH mass, used a WKB approximation,
wrote the wave equation and defined the spatial momentum $k(r)$ in
terms of the eigenvalues of the Laplacian operator while taking energy
as the eigenvalue of the time component of the Laplacian. He obtained
the number of states by calculating $\int k(r)dr$ and then summing
over angular degrees of freedom. Another possibility is that of \cite{paddy kolekar,paddy ideal gas}
who treated a relativistic gas of particles, and rather than the wave
equation, used the scalar invariance of the squared momentum four-vector
of the particle, while the covariant energy of a particle is the projection
of the timelike Killing vector on the four momentum, $\xi^{a}p_{a}$.
Both approaches give the same relationship between energy and momentum,
which for a static diagonal metric is 
\begin{equation}
g^{00}E^{2}=\sum_{i}g^{ii}\left(p_{i}\right)^{2}\label{eq:energy momentum eq}
\end{equation}
 where we are taking a massless particle for simplicity. 

Remark on notation: for simplicity of notation in this section, $g_{00}$
refers to the positive value of the \textcolor{black}{time-time component}
of the metric, except where explicitly stated otherwise. The minus
sign appears in the form of the equation.

\section{Dependence of N on time-time component of metric}

As a warm up, we show heuristically that for constant volume and energy
(for an observer at infinity) the number of states $N$ depends on
the time-time component of the metric. For 3+1 dimensions we show
that 
\begin{equation}
N\sim V_{\bot}E^{3}\int dr\sqrt{g_{rr}}\left(g^{00}\right)^{3/2}
\end{equation}
where $V_{\bot}$ refers to volume transverse to the radial coordinate.
Here $E$ refers to energy perceived by an observer at infinity, while
$\sqrt{g^{00}}E$ is just the energy of a local observer. Thus $N$
is found to be observer dependent.

Proof of this is as follows: In general $N\sim\int d^{3}xd^{3}p\equiv V_{x}V_{p},$
the product of the volume of configuration and momentum space for
a given energy. We may write this as $\left|X\right|^{3}\left|P\right|^{3}$
where we take space as a box with sides of equal length $\left|X\right|$
, and analogously for momentum with length $\left|P\right|$. (This
is just handwaving; a precise treatment will be given later.)

Now
\begin{eqnarray}
V_{x} & = & \int\sqrt{g_{3}}d^{3}x=\left|X\right|^{3}
\end{eqnarray}
where $g_{3}$ is the determinant of the spatial metric, and similarly
\begin{equation}
V_{p}=\int\sqrt{\frac{1}{g_{3}}}d^{3}p=\left|P\right|^{3}
\end{equation}
since the metric in a cotangent space is given by the inverse matrix
of that for configuration space. 

To find $\left|P\right|^{3}$ we use the wave equation: $\Box\phi=0.$
Then
\begin{eqnarray}
g^{00}E^{2}-g^{xx}p_{x}^{2}-g^{yy}p_{y}^{2}-g^{zz}p_{z}^{2} & = & 0
\end{eqnarray}
and transferring the momenta to the RHS and taking the root of both
sides
\begin{equation}
\sqrt{g_{00}E^{2}}=\sqrt{g^{xx}p_{x}^{2}+g^{yy}p_{y}^{2}+g^{zz}p_{z}^{2}}=\left|P\right|
\end{equation}
so that 
\begin{equation}
\left(\sqrt{g_{00}E^{2}}\right)^{3}=\left|P\right|^{3}=V_{p}.
\end{equation}
Since we are interested in black holes, we assume that $g_{00}$ depends
on one space coordinate $x$. This has no impact on the argument up
till now. In that case 
\begin{eqnarray}
N & \sim & V_{x}V_{p}=\int\sqrt{g_{ii}}d^{3}x\int\sqrt{g^{ii}}d^{3}p=\int\sqrt{g_{ii}}d^{3}x\left(\sqrt{g_{00}E^{2}}\right)^{3}\nonumber \\
 & = & V_{\bot}E^{3}\int\sqrt{g_{xx}}dx\left(\sqrt{g_{00}}\right)^{3}.
\end{eqnarray}
This shows that the volume of phase space in this simple case is determined
by the energy as perceived by the observer in the specific metric
background under discussion.

\section{Calculation of N }

\subsection{Calculation by the definition:}

The number of states given by the volume of phase space is the product
of the volume of position and momentum space. The momentum component
of the number of states belongs to a constrained region in the cotangent
space of the region of configuration space in question, obtained from
the wave equation.For example, in Cartesian coordinates in flat space
a Fourier transform of the wave equation gives
\begin{equation}
E^{2}-p_{x}^{2}-p_{y}^{2}-p_{z}^{2}=0
\end{equation}
and this defines a sphere of radius E:
\begin{equation}
1=\frac{p_{x}^{2}}{E^{2}}+\frac{p_{y}^{2}}{E^{2}}+\frac{p_{z}^{2}}{E^{2}}.
\end{equation}
In statistical physics we take all energies up to a given energy,
and so we look for the volume enclosed by this sphere, $\frac{4}{3}\pi E^{3}$.
If the metric is not flat, the volume will be an ellipsoid.

The wave equation for a general diagonal metric is
\begin{eqnarray}
g^{00}E^{2} & = & \sum_{i}g^{ii}\left(p_{i}\right)^{2}\nonumber \\
1 & = & \sum_{i}\frac{g^{ii}\left(p_{i}\right)^{2}}{g^{00}E^{2}}=\sum_{i}\frac{p_{i}^{2}}{g_{ii}g^{00}E^{2}}
\end{eqnarray}
where $p_{i}$ the spatial momenta are summed in all space directions.
This is the formula for volume of an ellipsoid with axes $\sqrt{g_{ii}g^{00}}E,$
which encloses a region whose volume in three space dimensions would
be $\frac{4}{3}\pi\sqrt{g_{xx}g_{yy}g_{zz}}\left(\sqrt{g^{00}}E\right)^{3}$
. In $d+1$ spacetime dimensions this becomes 
\begin{equation}
C_{d}\sqrt{g_{d}}\left(\sqrt{g^{00}}E\right)^{d}\label{eq:ellipsoid}
\end{equation}
where $g_{d}$ denotes the determinant of the spatial part of the
metric and $C_{d}$ is the volume enclosed by the d-dimensional unit
ball. One then integrates over all momentum space. Since the measure
in the momentum integral includes the root of inverse metric $g^{d}$,
that is,
\begin{equation}
\int\frac{d^{d}p}{\sqrt{g_{d}}}
\end{equation}
then the space determinant in eq.(\ref{eq:ellipsoid}) cancels out,
and the integral over momentum space gives the volume of a sphere
with radius $\sqrt{g^{00}}E.$ 

Therefore the number of states of a test particle in $d+1$ dimensions
(d space dimensions) for a diagonal metric in a given volume $V$
is
\begin{eqnarray}
N(E,V,d,1) & = & CE^{d}\intop_{V}d^{d}x\sqrt{g_{d}}\left(g^{00}\right)^{\frac{d}{2}}\nonumber \\
C & = & \frac{\pi^{\frac{d}{2}}}{\Gamma\left(\frac{d}{2}+1\right)}.
\end{eqnarray}
and given n identical particles $N(E,V,d,n)=(N(E,V,d,1))^{n}.$ An
explicit proof for $3+1$ and $4+1$ dimensions appears in Sec.\ref{sub:Proof-in-3+1dimensions}.For
explicit calculations of this quantity for a gas of particles in various
spacetimes please see \cite{paddy kolekar}.

\subsection{Calculation with the WKB approximation:}

Another method of calculating the number of states is with the WKB
approximation, which for a free particle in one dimension is exact.
WKB quantization gives the number of modes in one dimension $n$ as%
\footnote{Note that for periodic BCs $k=2n\pi/L$ whereas WKB used for a potential
well has $k=n\pi/L$.%
}
\begin{equation}
n=\frac{1}{\pi\hbar}\int dxp(x).
\end{equation}
In flat space the WKB term can be extended to $d$ dimensions thus:
\begin{eqnarray}
n_{i} & = & \frac{1}{\pi\hbar}\int dx_{i}p(x)\nonumber \\
n & \equiv & \left[\sum_{i=1}^{d}n_{i}^{2}\right]^{1/2}
\end{eqnarray}
 and one then continues the calculation by relating $n$ to the energy
(see Sec.\ref{sec:Equivalence-of-WKB}). In curved space problems
arise.

In a spherical coordinate system relating $n$ to the energy is problematic.
For a particle within sphere of radius $R$, $k=\beta_{nl}/R$ where
$\beta_{nl}$ are the zeros of the Bessel function: $j_{l}(kR)=0$.
But unlike the Cartesian case one cannot extract $E(n)$ in neat analytical
form, because the $\beta_{n,l}$ must be calculated numerically, and
each $l$ has its own $\beta_{n,l}$.%
\footnote{For example see \cite{Sakurai}.%
} Instead one follows 't Hooft and obtains the radial modes from the
wave equation. Essentially this means incorporating energy of transverse
degrees of freedom into a radial potential. One then sums the radial
modes over angular degrees of freedom.

\section{Invariance of number of states under transformation of metrics}

We wish to examine a general transformation which changes the metric
while leaving the number of states invariant. We find that such a
transformation exists, but does not preserve curvature. We give details
of the transformation, followed by examples of the relation to curvature.

We begin with conformal rescaling. If a $3-$dimensional metric changes
by $\tilde{g}=a(x)Ig,$ then the number of states is
\begin{eqnarray}
N_{0} & = & \int\sqrt{g_{3}}d^{3}xd^{3}p=\int\sqrt{g_{3}}\frac{4\pi E^{3}}{3}\left(g^{00}\right)^{3/2}d^{3}x.\nonumber \\
\tilde{N} & = & \int\sqrt{\tilde{g}_{3}}d^{3}xd^{3}p\nonumber \\
 & = & \int a^{3/2}\sqrt{g}\frac{4\pi E^{3}}{3}\left(\frac{1}{a}g^{00}\right)^{3/2}d^{3}x\\
 & = & \int\sqrt{g}\frac{4\pi E^{3}}{3}\left(g^{00}\right)^{3/2}d^{3}x=N
\end{eqnarray}
since $\tilde{g}_{00}=a(x)g_{00}$ and so $\tilde{g}^{00}=\frac{1}{a(x)}g^{00}$.
This only works if the metric is uniformly rescaled, so that $a_{0}=a_{i}.$
Thus conformal rescaling preserves the number of states. We conclude
that preservation of the number of states requires a constraint on
the relationship between the time and space components of the metric.

In search of a more general transformation we take a general diagonal
metric in $1+3$ dimensions. Generalization to more space dimensions
will be simple.

\begin{equation}
\left(\begin{array}{cccc}
g_{00}\\
 & g_{xx}\\
 &  & g_{yy}\\
 &  &  & g_{zz}
\end{array}\right)
\end{equation}
The volume of space in this metric:
\begin{equation}
\intop_{V}\sqrt{g_{xx}g_{yy}g_{zz}}d^{3}x
\end{equation}
where the integral is over a given volume V. The volume of momentum
space is

\begin{equation}
\intop_{V_{p}}\frac{d^{3}p}{\sqrt{g_{xx}g_{yy}g_{zz}}}
\end{equation}
where $V_{p}$ is the volume in momentum space. As explained above,
from the wave equation
\begin{equation}
1=\frac{1}{g_{xx}g^{00}E^{2}}p_{x}^{2}+\frac{1}{g_{yy}g^{00}E^{2}}p_{y}^{2}+\frac{1}{g_{zz}g^{00}E^{2}}p_{z}^{2}\label{eq:wave eq}
\end{equation}
which is the equation for volume of ellipsoid with axes $\sqrt{g_{xx}g^{00}}E,\sqrt{g_{yy}g^{00}}E,\sqrt{g_{zz}g^{00}E}$.
The momentum volume is obtained by integration, or more simply by
just plugging in the formula for volume of ellipsoid in 3 dimensions:
$\frac{4}{3}\pi abc=\frac{4}{3}\pi\sqrt{g_{xx}g_{yy}g_{zz}}\left(g^{00}E^{2}\right)^{3/2}.$

Phase space is given as%
\footnote{This can have a prefactor of $\left(2\pi\right)^{-3}$ when calculating
the density of modes per unit of phase space. %
}:
\begin{equation}
N=\intop_{V}d^{3}x\intop_{V_{p}}d^{3}p.
\end{equation}
We now transform the metric in arbitrary way but keeping it diagonal:
\begin{equation}
\left(\begin{array}{cccc}
a(\vec{x})g_{00}\\
 & b(\vec{x})g_{xx}\\
 &  & c(\vec{x})g_{yy}\\
 &  &  & d(\vec{x})g_{zz}
\end{array}\right)
\end{equation}
We plug this into the term for phase space. First we calculate the
volume of momentum space for the transformed metric. The wave equation
is now
\begin{eqnarray}
\frac{1}{a(\vec{x})}g^{00}E^{2} & = & \frac{1}{b(\vec{x})}g^{xx}p_{x}^{2}+\frac{1}{c(\vec{x})}g^{yy}p_{y}^{2}+\frac{1}{d(\vec{x})}g^{zz}p_{z}^{2}
\end{eqnarray}
 and using eq.(\ref{eq:wave eq})
\begin{equation}
1=\frac{a(\vec{x})}{b(\vec{x})g_{xx}g^{00}E^{2}}p_{x}^{2}+\frac{a(\vec{x})}{c(\vec{x})g_{yy}g^{00}E^{2}}p_{y}^{2}+\frac{a(\vec{x})}{d(\vec{x})g_{zz}g^{00}E^{2}}p_{z}^{2}\label{transf wave eq}
\end{equation}
so that the volume becomes
\begin{equation}
V_{p}=\frac{4}{3}\pi\sqrt{b(\vec{x})c(\vec{x})d(\vec{x})g_{xx}g_{yy}g_{zz}}\left(\frac{g^{00}}{a(\vec{x})}E^{2}\right)^{3/2}.
\end{equation}
This will equal the volume before the transformation if
\begin{equation}
b(\vec{x})c(\vec{x})d(\vec{x})=a(\vec{x})^{3}.\label{eq:THE CONSTRAINT}
\end{equation}
Thus we have identified the constraint for an arbitrary transformation
to preserve the volume of phase space.

\subsection{Characteristics of the transformation}

We looked for some kind of general algebraic characterization for
this kind of matrix, but found none. It belongs to $GL(n,R)$ but
does not represent a particular symmetry. Conformal transformations
are a subgroup of our transformation, as shown in Sec.\ref{sub:Examples-in-various}.
Certain non conformal transformations also preserve the number of
states. This holds if the determinants cancel out: That is, for $d$
space dimensions, the time part $a(x)$ when raised to the $d^{th}$
power, has to equal the determinant of the space part. Take
\begin{equation}
A=\left(\begin{array}{cccc}
a(x) & 0 & 0 & 0\\
0 & a(x)^{2} & 0 & 0\\
0 & 0 & a(x) & 0\\
0 & 0 & 0 & 1
\end{array}\right)
\end{equation}
As with the conformal transformation, we still have $\sqrt{\tilde{g}_{3}}=a^{3/2}\sqrt{g_{3}}$
and and so $\tilde{N}=N_{0}$. 

So in general our constraint is:
\begin{equation}
\left(g_{00}\right)^{d}=det\: g_{space}\label{eq:constraint in GENERAL}
\end{equation}
where $d$ is the number of space dimensions and $g_{space}$ is the
determinant of the spatial part of the metric.

We can regard the transformation matrix, labeled $A$, as two blocks,
separating the time and space components:
\begin{equation}
A=\left(\begin{array}{cc}
T\\
 & S
\end{array}\right)
\end{equation}
where $T$ is a 1x1 matrix, and S is a diagonal matric of rank $d$
where $d$ is the dimension of space (rank of $A$ is the dimension
of space-time, $d+1$). Then the constraint requires 
\begin{eqnarray}
det(S) & = & det(T)^{d}\nonumber \\
det(A) & = & det(T)^{2d}.
\end{eqnarray}

\subsection{Relation to curvature}

In $d+1$ dimensions 
\begin{eqnarray}
N & \sim & \int d^{3}x\sqrt{\frac{g_{d}}{\left(g_{00}\right)^{d}}}
\end{eqnarray}
where $g_{d}$ denotes the determinant of the space part of metric.
To preserve the number of states we have to preserve the ratio $g_{d}/g_{00}^{d}$
, which entails the constraint on the determinant as detailed above.
The question becomes: given a change of metric for which this constraint
holds, will such a constraint ensure preservation of scalar curvature?
If so preservation of the number of states would entail preservation
of curvature, which is an observer independent characteristic.

We take two matrices representing two possible transformations of
a given $3$- dimensional metric:
\begin{eqnarray}
A & = & \frac{1}{L}\left(\begin{array}{ccc}
x\\
 & x\\
 &  & x
\end{array}\right)\nonumber \\
B & = & \left(\begin{array}{ccc}
\frac{\sqrt{2}x}{L}\\
 & 2\\
 &  & \frac{x^{2}}{L^{2}}
\end{array}\right)
\end{eqnarray}
where $L$ is a constant with dimension of length. Both transformation
matrices preserve the constraint given above, while their intrinsic
curvature differs: the first has $R=\frac{3L}{2x^{3}}$ , the second
has $R=\frac{L}{\sqrt{2}x^{3}}.$ This is because the second one has
\textcolor{black}{fewer nonzero components of the Christoffel connection},
since the derivative must be $\partial_{x}$, and $\partial_{x}g_{yy}=0$
\footnote{This reason is purely formal, in terms of the coordinates chosen here;
R is not dependent on coordinate system of course. Just for that reason,
a different choice of coordinates would also involve different scalar
curvature for the two cases.%
}. Therefore clearly imposing the constraint on a metric transformation
will not necessarily preserve the curvature of the original metric.

Curvature in these examples is affected by the number of terms with
an $x$ derivative, while the determinant is not. Thus the constraint
on the determinant does NOT preserve curvature. This is intuitively
understandable: the determinant indicates volume but gives no information
as to the spatial distribution of the volume.

\subsection{Examples in vario\label{sub:Examples-in-various}us dimensions}

In $1+1$ dimensions a transformation that preserves $N$ \emph{must}
be conformal: $g_{00}=g_{xx}$ since $det\, g_{space}=g_{xx.}$ In
2+1 dimensions we give two examples of transformations that preserve
N. One is conformal, the other non conformal but symmetric: We give
the matrices for the transformations, and the scalar curvature when
they are applied to a flat Lorentzian metric:

Conformal: {\small{}
\begin{equation}
A=\frac{1}{L}\left(\begin{array}{ccc}
-x\\
 & x\\
 &  & x
\end{array}\right),R=-\frac{3L}{2x^{3}}
\end{equation}
} 

Symmetric:{\small{}
\begin{equation}
B=\frac{1}{L}\left(\begin{array}{ccc}
-\sqrt{xy}\\
 & x\\
 &  & y
\end{array}\right),R=L\left(\frac{-6y^{2}+5x\sqrt{xy}}{8\left[xy\right]^{5/2}}\right).
\end{equation}
\[
\]
}An asymmetric example is like the one given in the previous section
for a Euclidean metric.

Note that plugging in the value $x=y$ \emph{after} deriving $R$
for matrix $B$ does not give the curvature of matrix $A.$ This is
because the derivation of $R$ takes into account the direction of
each component as well as its numerical value. If one plugs in $y=x$
before deriving $R$ all the derivatives $\partial_{y}$ vanish, giving
the different result.

We next look at 3+1 dimensions. The constraint requires $\left|\left(g_{00}\right)^{3}\right|=detg_{3}$.
Comparing several matrices that obey this constraint and inspecting
their curvature:{\footnotesize{}
\begin{eqnarray}
A=\left(\begin{array}{cccc}
-\frac{x}{L}\\
 & \frac{x^{3}}{L^{3}}\\
 &  & 1\\
 &  &  & 1
\end{array}\right),\: R & = & \frac{2L^{3}}{x^{5}}\nonumber \\
B=\frac{1}{L}\left(\begin{array}{cccc}
-\left(xyz\right)^{\frac{1}{3}}\\
 & x\\
 &  & y\\
 &  &  & z
\end{array}\right),\: R & = & \frac{4L}{9}\left(\frac{1}{x^{3}}+\frac{1}{y^{3}}+\frac{1}{z^{3}}\right)\nonumber \\
C=\frac{1}{L}\left(\begin{array}{cccc}
-x\\
 & x\\
 &  & x\\
 &  &  & x
\end{array}\right),\: R & = & \frac{3L}{2x^{3}}
\end{eqnarray}
}A few comments: 1) The curvature for the third transformation is
the same as for the conformal matrix in 1+2 dimensions. 2) As before,
setting $x=y=z$ after calculating the curvature for matrix $B$ does
not give the same result as the curvature for matrix $C$. Again,
this is because the direction of the variable contributes in calculating
$R,$ and not just its numeric value. This sheds light on the fact
that the number of states, which is proportional to the volume of
phase space, is different from curvature, which incorporates information
on the distribution of that volume. A constraint on the determinant,
representing Euclidean volume, is not the same as that on Ricci curvature,
which in fact represents the amount by which the volume of a geodesic
ball in a curved Riemannian manifold deviates from that of the standard
ball in Euclidean space\emph{.}

\subsubsection{Rindler vs Schwarzschild: }

The transformation from Minkowski metric to Rindler metric is not
diagonal. \textcolor{black}{Rindler coordinates mix time and space
coordinates of the Minkowski metric,} and that is why N in the Rindler
calculation is different from flat space. We cannot conclude from
this that curvature is irrelevant to statistical entropy. That conclusion
can only be drawn from the general proof given above.

The Schwarzschild calculation gives divergence at the boundary and
we found that the number of states (and thus the entropy) is different
from that of Minkowski space as shown in the previous chapter. This
is not the same as the difference between the number of states in
Rindler and Minkowski spaces. \textcolor{black}{The basic difference
is that Schwarzschild and Minkowski spaces have different geometry,
whereas the transformation to Rindler space is a coordinate change
from Minkowski space. In terms of the calculation given above,} in
the Schwarzschild case the argument here given does apply, since the
transformation metric from Minkowski to Schwarzschild metric is diagonal.
The Schwarzschild number of states differs from that of Minkowski
because of the redshift on energy: $g_{00}(r)$.

\subsection{Discussion}

Our transformation leaves N invariant because it preserves the relationship
between the volume of momentum space and of position space. $\left(g^{00}\right)^{3/2}$
is the variable part of momentum space, and $\sqrt{g_{d}}$ is the
variable part of position space. $N$ is invariant so long as the
relation between the two is preserved, so that if position space shrinks,
momentum space grows and vice verse: $a(x)^{d}$ multiplying $\sqrt{g_{space}}$
equals $1/a(x)^{d}$ multiplying momentum space.

It would seem that a proof of curvature independence must show that
there are no cases at all where curvature is preserved under a transformation
that preserved the number of states. In fact it is quite possible
that in some case curvature might be preserved. We claim that this
must be seen as a coincidence because the constraint on preservation
of the number of states relates to the determinant. By definition,
there is a difference between the determinant, which represents Euclidean
volume and does not depend on directions in space, and curvature which
does depend on directions in space. The number of states does not
depend on directions in space and so it can be preserved even if the
directional characteristics and thus the curvature are changed. There
will be a subgroup where transformation of the number of states will
indeed preserve curvature. But one cannot assume that any given number
of states, and the entropy derived from it, relate in a unique manner
to a spacetime with a given curvature. Curvature may affect the number
of states but there is no one to one correspondence between the two.

We examined the question whether in curved space the number of states,
and the statistical entropy derived from this, is observer dependent
or is related to a physical quantity such as curvature. We found that
- as with any measurement of length - the result just depends on your
ruler. Finding $N$ is actually just a generalization of finding a
length: it involves finding a volume in space and momentum space.
Obviously people with different rulers will give you a different answer
for the length. The statistical number of states is essentially like
a scaled ruler and thus is observer dependent and not related to the
intrinsic geometry. This reflects the scheme in the previous chapter,
in that boundary conditions are imposed on the observer rather than
on the state. This differs from the work of Wald and Carlip.

In addition, in \cite{Katya} it was shown for explicit examples that
entanglement entropy does not depend on curvature. For a discretized
region in curved space it was found that even when the space is large
enough for the effects of curvature to be noticeable, entropy remains
proportional to area and is not affected by the curvature of the background.
This qualitative similarity to our result reinforces the idea that
entanglement and statistical entropy are one and the same thing, and
that they possibly differ from Noether charge entropy.

The results in this chapter apply to a diagonal metric only. A general
metric could only be diagonalized locally, and that would not be relevant
to a discussion of curvature. However since a diagonal metric is seen
to be curvature independent, and there is certainly at least one diagonal
metric, this serves as a counter example to the claim that the number
of states is uniquely dependent on curvature.

In conclusion, we have shown that the number of states is a function
of the metric and is preserved under specific transformations of the
metric, which do not preserve curvature. Therefore the number of states
calculated with the accepted definition of phase space does not depend
uniquely on curvature. In this it appears to differ from Wald entropy.
However, as discussed in Sec.\ref{sec:summary and conclusions}, for
general theories of gravity a wider definition of phase space may
be necessary.

\chapter{Entropy variation}

\section{Introduction}

In this chapter we write the variation of the statistical entropy
of matter fields outside a black hole along a Killing vector, and
we compare this to the variation of Wald's entropy. The calculation
applies to any generalized theory of gravity. If the variations are
equivalent, this would imply a natural connection between the two
entropies, where one is derived from the presence of matter in the
region of a black hole and the other derived exclusively from spacetime
geometry. Such a connection would relate statistical mechanics to
the gravity field equations, with profound implications for the idea
of gravity as an emergent phenomenon. We find that under certain conditions
for a stationary black hole the two entropy variations do in fact
coincide.

Wald \cite{wald 1,wald and iyer} has studied black holes in generalized
theories of gravity and proposed that the correct dynamical entropy
of stationary black hole solutions with bifurcate Killing horizons
is a Noether charge entropy. The microphysical understanding of Wald's
entropy is unknown.

Another approach to the calculation of black hole entropy is that
of statistical mechanics, as discussed above. Since statistical entropy
is calculated using the partition function of the fields outside the
horizon, its microphysical origin is understood. However, as statistical
entropy is known to depend on the number of fields in the theory,
whereas Wald's entropy does not, it seems that the two entropies may
be essentially different.

We note that Wald's calculation was for general (higher derivative)
theories of gravity. Therefore we attempt to relate these two entropies
for generalized theories of gravity. A connection between the two
entropies could indicate a relationship between the entropy of microstates
within the black hole and that of matter outside. This would be similar
to the concept of black hole entropy as entanglement entropy between
the states inside and outside the black hole. 

The variation of Wald's entropy was calculated in \cite{merav ramy}
(see also \cite{Parikh}). In \cite{merav ramy} the authors differentiate
Wald's entropy along a Killing vector to obtain $\delta S$. They
use the generalized gravitational field equations to derive a term
for the energy flow across the black hole horizon, and find that the
first law $\delta Q=T\delta S$ is fulfilled. Thus the entropy variation
along a Killing vector is:
\begin{equation}
\delta S_{Wald}=2\pi\int T_{ab}\chi^{a}\epsilon^{b}
\end{equation}
 where $\chi^{a}$ is the Killing field and $\epsilon^{b}$ is a $(D-1)$
volume form. (Details of the derivation are in Section \ref{sec:Variation-of-Wald's}).
We here obtain a similar relationship for the variation of statistical
entropy and $T_{ab}$, where our variation will be along an arbitrary
vector, which on the horizon we will take to be a Killing vector.
Thus we will \textcolor{red}{compare} the variation of the two different
entropies along a Killing vector on the black hole horizon. 

\emph{Note:} This chapter, unlike the rest of this thesis, is written
using differential forms because it follows the variation of Wald
entropy \cite{merav ramy}\textsf{\textbf{\textit{ }}}which was calculated
in that language.

\section{Preliminaries}

For the following treatment we will need the partition function. In
the path integral approach the amplitude to go from a field configuration
$\phi_{1}$ at a time $t_{1}$ to a field configuration $\phi_{2}$
at time $t_{2}$ is given 
\begin{equation}
\left\langle \phi_{2},t_{2}|\phi_{1},t_{1}\right\rangle =\int D\phi e^{iI\left[\phi\right]}\label{eq:amplitude and pi}
\end{equation}
where the path integral is over all field configurations which take
the values $\phi_{1}$at time $t_{1}$ and $\phi_{2}$ at time $t_{2}$.
However if $H$ is the Hamiltonian, 
\begin{equation}
\left\langle \phi_{2},t_{2}|\phi_{1},t_{1}\right\rangle =\left\langle \phi_{2}|e^{-iH(t_{2}-t_{1})}|\phi_{1}\right\rangle .
\end{equation}
One defines the time period $t_{2}-t_{1}=-i\tau$, and in going to
Euclidean time one takes the time as periodic with period $\beta,$
and one identifies the fields $\phi_{1}=\phi_{2}$ , sums over all
$\phi_{1}$ and thus obtains the partition function for the canonical
ensemble of field $\phi$ at inverse temperature $\beta$, 
\begin{equation}
Z=Tr(e^{-\beta H})=\intop D\phi e^{-I_{E}\left[\phi\right]}\label{eq:Z as PI}
\end{equation}
and $I_{E}\left[\phi\right]$ is the matter action as defined above,
but now with Euclidean time. (see for instance \cite{Peskin}, Chap.9).

\section{Variation of statistical entropy}

The entropy is given by 
\begin{eqnarray}
S_{st}= & \beta\langle E\rangle+\ln Z.\label{eq:entagldentropy2}
\end{eqnarray}
since, as discussed in Chapter 2, the region near a black hole is
a thermal state. We now perform a variation of the entropy. Since
we are interested in comparing to Wald entropy, we look for the effect
on the statistical entropy of a change in the geometry. A natural
way to vary the geometry is to push the metric forward with the flow
of a vector field $\xi$. The corresponding variation of the metric
is given by its Lie derivative along $\xi$, therefore it is this
derivative that will appear in the entropy variation below.

Since it has been shown \cite{merav ramy,Parikh} that the variation
of Wald's entropy is related to the energy momentum tensor, we would
like to relate the variation of the statistical entropy to $T_{ab}$
as well. We label by $\delta_{g}S_{st}$ the variation of the entropy
we have just described. From eq.(\ref{eq:entagldentropy2}) we have
\begin{eqnarray}
\delta_{g}S_{st}=\delta_{g}\left(\beta\langle E\rangle\right)+\delta_{g}\ln Z.\label{variation entropy}
\end{eqnarray}
We want to examine the effect of a change only in the metric on the
expression for the entropy. To do this, we take the variation as described
above, where the metric is varied by taking its Lie derivative along
a vector field $\xi$, as follows.

We begin with the second term on the RHS. 
\begin{eqnarray}
\delta_{g}lnZ & = & \int\epsilon\frac{\delta lnZ}{\delta g^{ab}}\mathcal{L}_{\xi}g^{ab}\label{eq:second term}\\
 & = & \frac{1}{Z}\int\epsilon\frac{\delta Z}{\delta g^{ab}}\mathcal{L}_{\xi}g^{ab}\\
 & = & \frac{1}{Z}\int\frac{\epsilon}{2}\left\langle T_{ab}\right\rangle \mathcal{L}_{\xi}g^{ab}
\end{eqnarray}
 where $\epsilon$ is a D-dimensional volume form, and the last line
was obtained as in \cite{birrell and davies}, Sec.6.1.%
\footnote{In that text one obtains $i\left\langle T_{ab}\right\rangle $ which
here for Euclidean space time should become $-\left\langle T_{ab}\right\rangle $.
We omit the minus sign in what follows, and note that in \cite{wald and iyer}
the Noether charge has a minus sign.%
} Note that $\left\langle T_{ab}\right\rangle $ is not normalized.

We express the Lie derivative of the metric in terms of $\xi_{a}$:
$\mathcal{L}_{\xi}g_{ab}=\nabla_{a}\xi_{b}+\nabla_{b}\xi_{a}$ and
the variation of the action becomes: 
\begin{eqnarray}
\delta_{g}lnZ & = & \frac{1}{Z}\intop_{V}\frac{\epsilon}{2}\left\langle T_{ab}\right\rangle \left[\nabla^{a}\xi^{b}+\nabla^{b}\xi^{a}\right]\nonumber \\
 & = & \frac{1}{Z}\intop_{V}\epsilon\left\langle T_{ab}\right\rangle \nabla^{a}\xi^{b}\nonumber \\
 & = & \frac{1}{Z}\left[\intop_{V}\epsilon\nabla^{a}\left[\left\langle T_{ab}\right\rangle \xi^{b}\right]-\intop_{V}\epsilon\nabla^{a}\left\langle T_{ab}\right\rangle \xi^{b}\right]\label{eq:Lie derivative of I_m}
\end{eqnarray}
 where we have used symmetry of $T_{ab}$ and integration by parts.
Since $\nabla^{a}\left\langle T_{ab}\right\rangle =0$ we obtain 
\begin{equation}
\delta_{g}lnZ=\frac{1}{Z}\intop_{V}\epsilon\nabla^{a}\left[\left\langle T_{ab}\right\rangle \xi^{b}\right]\label{delta St v}
\end{equation}
 From the generalized Stokes law we have $\intop_{V}\left(\nabla^{c}\omega_{c}\right)\epsilon=\intop_{H}\omega^{c}\epsilon_{c}$
and using $\omega_{c}=\left\langle T_{ab}\right\rangle \xi^{b}$ we
have 
\begin{equation}
\delta_{g}lnZ=\frac{1}{Z}\intop_{H}\left\langle T_{ab}\right\rangle \xi^{a}\epsilon^{b}\label{delta St}
\end{equation}
 where $\epsilon^{b}$ is a (D-1) volume form and $H$ denotes the
boundary of $V$. The volume is bounded by the black hole horizon
and at infinity. We assume space is asymptotically flat and so $T_{ab}$
vanishes at infinity, and the contribution in eq.(\ref{delta St})
comes only from the black hole horizon.%
\footnote{Here we have used an IR cutoff at infinity, but this is so far away
that $T_{ab}$ vanishes there.%
}

Eq.(\ref{delta St}) gives part of the variation of the matter action
of the fields found in the volume outside the black hole when the
metric changes along some arbitrary vector field. We now specialize
to the horizon. At the horizon we require $\xi\rightarrow\chi$ to
be a Killing vector which fulfills $\chi^{a}\nabla_{a}\chi^{b}=\chi^{b}$.
Since $\chi$ vanishes on the horizon, we take a region just outside
the horizon and take the limit to the horizon, as Carlip does (Chap.2,
and see \cite{jacobson,merav ramy,Parikh}). The second term in the
variation of the matter action along the Killing vector is thus 
\begin{equation}
\delta_{g}lnZ=\frac{1}{Z}\intop_{H}\left\langle T_{ab}\right\rangle \chi^{a}\epsilon^{b}.\label{eq:ln Z variation}
\end{equation}
 If we had taken $\xi$ to be a Killing vector throughout the volume
in eq.(\ref{eq:second term}), eq\emph{.}(\ref{eq:ln Z variation})
would vanish as $\mathcal{L}_{\xi}g_{ab}$ would vanish. However eq.(\ref{eq:second term})
is an integral over volume, and $\xi$ is a Killing vector only on
the boundary of this volume, not in the bulk. So eq.(\ref{eq:second term})
does not vanish, and from Stokes' law eq\emph{.}(\ref{eq:ln Z variation})
does not vanish.%
\footnote{For example, take the following integral over a volume with radius
R: $\int(\frac{r}{R}-1)dV$. The integrand vanishes at the surface,
but clearly the integral over the volume does not vanish, and so the
surface integral obtained by Stokes law will not vanish either.%
}

We now turn to the first term in eq.\ref{variation entropy}. $\langle E\rangle=-\partial lnZ/\partial\beta.$
In this case $\beta$ is constant, as the period of Euclidean time
(and as is the temperature for a stationary black hole) and will not
itself be varied. Thus using eq.(\ref{eq:ln Z variation}) we find
\begin{eqnarray}
\delta_{g}\left(\beta\langle E\rangle\right) & = & -\beta\delta_{g}\frac{\partial}{\partial\beta}lnZ=-\beta\frac{\partial}{\partial\beta}\delta_{g}lnZ\nonumber \\
 & = & -\beta\frac{\partial}{\partial\beta}\left(\frac{1}{Z}\intop_{H}\left\langle T_{ab}\right\rangle \chi^{a}\epsilon^{b}\right).\label{eq:energy variation}
\end{eqnarray}
Eqs.(\ref{eq:ln Z variation}),(\ref{eq:energy variation}) give 

\begin{equation}
\delta_{g}S_{st}=-\beta\frac{\partial}{\partial\beta}\left(\frac{1}{Z}\intop_{H}\left\langle T_{ab}\right\rangle \chi^{a}\epsilon^{b}\right)+\frac{1}{Z}\intop_{H}\left\langle T_{ab}\right\rangle \chi^{a}\epsilon^{b}.\label{eq:big entropy variation term}
\end{equation}
First we take $\left\langle T_{ab}\right\rangle $ as well as $Z$
as independent of Euclidean time, in this case the first term on the
RHS vanishes. We define $\epsilon^{b}=k^{b}d\tau dA$ where $k^{b}$
is the tangent to the vectors generating the horizon for parameter
$\tau$, and $dA$ is the area element of a cross section of the horizon.
Then 
\begin{eqnarray}
\delta_{g}S_{st} & = & \frac{1}{Z}\intop_{H}\left\langle T_{ab}\right\rangle \chi^{a}\epsilon^{b}\nonumber \\
 & = & \frac{1}{Z}\intop_{\tau_{2}}^{\tau_{1}}d\tau\intop_{A}\left\langle T_{ab}\right\rangle \chi^{a}k^{b}dA.\label{eq:lorentzian delta s}
\end{eqnarray}
 Integration over Euclidean time gives
\begin{equation}
\delta_{g}S_{st}=\frac{\beta}{Z}\intop_{A}\left\langle T_{ab}\right\rangle \chi^{a}k^{b}dA
\end{equation}
We normalize $\left\langle T_{ab}\right\rangle $ with the partition
function, obtaining
\begin{equation}
\delta_{g}S_{st}=\beta\intop_{A}\left\langle \tilde{T}_{ab}\right\rangle \chi^{a}k^{b}dA
\end{equation}
where the tilde represents normalization, and taking $\beta=2\pi$
this is equal to the variation of Wald entropy. If we do not integrate
over the entire time period but rather take a finite interval we obtain
a term proportional to Wald entropy. In either case it appears that
for time independent $\left\langle \tilde{T}_{ab}\right\rangle $
the variation of Wald's entropy is equivalent up to a constant to
a purely geometric variation of statistical entropy. The factor of
$2\pi$ in the variation of Wald's entropy comes from the original
term for Wald entropy, eq.\ref{eq:Wald entropy} and not from the
variation, and represents the temperature in the first law of thermodynamics,
rather than being inherent to the derivation of the Noether charge
for diffeomorphism invariance. 

If $\left\langle \tilde{T}_{ab}\right\rangle $ is not independent
of time the first term of eq.(\ref{eq:big entropy variation term})
does not vanish and becomes
\begin{equation}
\delta_{g}\left(\beta\langle E\rangle\right)=-\beta\frac{\partial}{\partial\beta}\left(\frac{1}{Z}\intop_{H}\left\langle T_{ab}\right\rangle \chi^{a}\epsilon^{b}\right).
\end{equation}
where we have abandoned normalization in order to explore the role
of the partition function. The next question is whether to take $Z$
as a function of $\beta$. If not, we obtain
\begin{equation}
\delta_{g}\left(\beta\langle E\rangle\right)=-\frac{\beta}{Z}\intop_{H}\frac{\partial}{\partial\beta}\left\langle T_{ab}\right\rangle \chi^{a}\epsilon^{b}
\end{equation}
for which one needs to know the precise dependence of the stress energy
tensor on the Euclidean time period. If $Z$ is a function of $\beta$
we obtain two terms, 
\begin{equation}
\delta_{g}\left(\beta\langle E\rangle\right)=-\frac{\beta}{Z}\intop_{H}\frac{\partial}{\partial\beta}\left\langle T_{ab}\right\rangle \chi^{a}\epsilon^{b}+\frac{\beta}{Z^{2}}\int D\phi(-L)e^{-I_{E}}\intop_{H}\left\langle T_{ab}\right\rangle \chi^{a}\epsilon^{b}
\end{equation}
This may also be written
\begin{equation}
\delta_{g}\left(\beta\langle E\rangle\right)=-\frac{\beta}{Z}\intop_{H}\frac{\partial}{\partial\beta}\left\langle T_{ab}\right\rangle \chi^{a}\epsilon^{b}-\frac{\beta}{Z^{2}}\left\langle L\right\rangle \intop_{H}\left\langle T_{ab}\right\rangle \chi^{a}\epsilon^{b}
\end{equation}
The expectation value of the Lagrangian is not normalized just as
$\left\langle T_{ab}\right\rangle $ is not (see \cite{birrell and davies}
Chap. 6). Normalization gives
\begin{equation}
\delta_{g}\left(\beta\langle E\rangle\right)=-\beta\intop_{H}\frac{\partial}{\partial\beta}\left\langle \tilde{T}_{ab}\right\rangle \chi^{a}\epsilon^{b}-\beta\left\langle \tilde{L}\right\rangle \intop_{H}\left\langle \tilde{T}_{ab}\right\rangle \chi^{a}\epsilon^{b}.
\end{equation}

\section{Summary}

We have three possibilities for this variation: one is just the same
as the variation of Wald entropy, resulting from a Lie derivative
of the metric, and the other two contain additional terms which derive
from time dependence of $T_{ab}$ and $Z$ for the matter fields in
question. Thus the existence of matter fields are responsible for
the difference between even a purely geometric variation as above,
and the variation of Wald entropy.

The first possibility is this: If the expectation value of the stress
energy tensor of the matter fields is time independent then we obtain
a term proportional to the variation of Wald's entropy. Integration
over all Euclidean time gives the exact same term as the variation
of Wald entropy. We are treating a stationary black hole, so that
it seems reasonable to assume time independence. (We recall that $\beta$
appears as inverse temperature in the partition function, but is also
the period of Euclidean time.)

If the expectation value of the stress energy tensor is time dependent,
the question then arises whether $Z$ is as well. If not, then the
entropy variation is
\[
\delta_{g}S=\beta\intop_{A}\left\langle \tilde{T}_{ab}\right\rangle \chi^{a}k^{b}dA-\beta\intop_{H}\frac{\partial}{\partial\beta}\left\langle \tilde{T}_{ab}\right\rangle \chi^{a}\epsilon^{b}
\]
which includes one term which is equal to the variation of Wald entropy,
and a second which is not. If $Z$ is taken as dependent on the time
period,
\[
\delta_{g}S=\beta\intop_{A}\left\langle \tilde{T}_{ab}\right\rangle \chi^{a}k^{b}dA-\beta\intop_{H}\frac{\partial}{\partial\beta}\left\langle \tilde{T}_{ab}\right\rangle \chi^{a}\epsilon^{b}-\frac{\beta}{Z^{2}}\left\langle \tilde{L}\right\rangle \intop_{H}\left\langle \tilde{T}_{ab}\right\rangle \chi^{a}\epsilon^{b}.
\]

In conclusion, if one assumes a stationary black hole with time independent
$\left\langle T_{ab}\right\rangle $ and $Z$, one obtains a variation
according to a change in the geometry which is just that of the variation
of Wald's entropy. It is not surprising that the variation of statistical
entropy due to a slight change in the geometry should be similar to
the variation of the Noether charge for diffeomorphism invariance.
If we were to perform a complete variation of the statistical entropy
including variation by the fields, and not just by the geometry, the
two terms would differ. 

If we do not make the above assumptions, the variation due to a change
in the geometry does not coincide with the variation of Wald entropy,
but is affected by the nature of the stress energy tensor and the
partition function, that is, by the nature of the matter fields in
the black hole metric. 

Thus we have shown that along a Killing vector at the horizon, the
variation of the statistical entropy of the matter fields caused by
variation of the matter Lagrangian density due to a change of the
metric differs from the variation of Wald's entropy, which is derived
 from the diffeomorphism invariance of spacetime. This is true for
any generalized theory of gravity. In previous work, it is not clear
what Wald's entropy counts: it could be quanta of Planck length, or
Planck area, some geometrical or other quantity. The statistical entropy
term in our work is explicitly taken only from the matter Lagrangian.
In all the above possibilities one finds that the two variations coincide
in the purely geometrical aspect. The variation itself must be only
according to a change in the geometry and not in the matter fields.
In addition, for the variations to coincide one must ignore dependence
of the matter fields and partition function on Euclidean time. This
makes it seem that the Wald's entropy isolates the geometric characteristics
of space time at the black hole horizon, which also affect the matter
fields in its vicinity, but do not consitute their statistical entropy.

\chapter{Conclusions and d\label{cha:Conclusions-and-discussion}iscussion}

In this thesis we discussed several issues which differentiate between
different conceptions of black hole entropy: divergence at the horizon,
imposition of boundary conditions, observer dependence and behavior
under variation.

We saw that divergence at the boundary as in 't Hooft's model may
result from simple x/p uncertainty, since thermodynamic entropy is
linear to energy and energy diverges as $\Delta x$ vanishes. Therefore
statistical entropy may be identified in the case of the black hole
with entanglement entropy even though entanglement entropy does not
diverge at the boundary, because this divergence may simply be a result
of quantum uncertainty. Entanglement entropy expresses correlations
between two different parts of a system, while statistical entropy
counts degrees of freedom of one of those two parts. Counting correlations
is thus the same as counting the number of degrees of freedom in one
of the subsystems, since each degree of freedom in one subsystem is
correlated with one in the other subsystem.

Another criterion for distinguishing between entropies is the imposition
of boundary conditions. Carlip's treatment of entropy as resulting
from symmetry necessitates the imposition of boundary conditions and
must be distinguished from statistical/entanglement entropy for this
reason. This is also true of Wald entropy, which is derived from surface
terms at the boundaries at infinity and at the horizon. In contrast,
entanglement entropy traces out one part of a system, that is, looks
at only part of a system without imposing boundary conditions. In
the example given of a potential well, the probability to find a particle
in the left side of the well would be one half, whereas if we imposed
boundary conditions, it would be either 1 or 0 because we would have
two smaller wells. However we have not divided the system into two
smaller wells, but rather we look only at one half of the entire well. 

It is possible that a stretched horizon, like 't Hooft's brick wall,
is just a method of delocalizing the horizon. In that case Carlip's
central charge would not vanish despite the problem raised by imposing
boundary conditions. Further work is needed to test this possibility. 

A third characteristic that must be taken into account is curvature
independence of the entropy. It has been shown that statistical entropy
is not uniquely related to the curvature scalar, although the curvature
may be one of a number of factors affecting it. It is possible that
the particular method used to explore statistical entropy is not sufficiently
general. In particular it holds only for a diagonal metric. However
since this method provides a counter example, it is difficult to see
how in a more general framework statistical entropy would not be observer
dependent. Since statistical entropy is derived from the number of
states, whereas Wald's entropy is defined using the curvature tensor,
this seems to indicate a difference between these two concepts of
entropy. However the issue may not be so simple. Phase space is defined
as the product of spatial volume and its canonical conjugate. This
definition arose in a context where the canonical conjugate of the
variable in the Lagrangian was its time derivative. However the gravitational
Lagrangian includes the Ricci scalar, and Ricci tensors as well in
the generalized theories of gravity with which Wald dealt. The Lagrangian
of a particle in a gravitational background will include at least
two terms, the matter Lagrangian and the graviational term. Each will
have a generalized momentum conjugate to the dynamical variable in
the Lagrangian. Therefore it may be necessary to redefine statistical
entropy to take into account a more general formulation of phase space.

In conclusion, we have shown that the number of states is a function
of the metric and is preserved under specific transformations of the
metric, which do not preserve curvature. Therefore the number of states
calculated with the accepted definition of phase space does not depend
on curvature. For general theories of gravity it may be necessary
to redefine statistical entropy taking into account a more general
concept of phase space.

Variation of statistical entropy resulting from a change in the metric
was shown to differ from the variation of Wald's entropy. The two
variations coincide only in the purely geometrical aspect. To coincide,
the variation of statistical entropy must be only according to a change
in the geometry and not a change in the matter fields. In addition,
one must ignore dependence of the matter fields and partition function
on Euclidean time. This makes it seem that the Wald's entropy isolates
the geometric characteristics space time at the black hole horizon,
which also affect the matter fields in its vicinity, but do not consitute
their statistical entropy.

It appears from all the above that statistical entropy and entanglement
entropy may refer to the same thing, whereas Wald's geometric entropy,
as well as Carlip's attempt to obtain entropy from conformal symmetry,
refer to a different entity. This does not explain what the entropy
actually means, or what degrees of freedom statistical entropy represents.
But it does draw a significant distinction between these two basic
types of entropy, and that is a necessary first step in understanding
what black hole entropy refers to.

Further investigation is necessary to clarify these points. In particular
it is necessary to explore higher theories of gravity. The main objective
of this thesis was to implement the idea that in trying to understand
what the various concepts of black hole entropy refer to, and which
of them coincide, one should focus on specific aspects of this entropy.
Here I have pointed out several crucial aspects: behavior at the boundary,
observer dependence, imposition of boundary conditions, and behavior
of a variation of this entropy. This investigation appears to lead
to the conclusion that statistical entropy and entanglement entropy
may coincide, and differ from the entropy of Wald and Carlip which
derives from spacetime symmetry.

\part{Supplementary Details and Calculations}

\chapter{Supplement to background}

\section{\label{sec:'t-Hooft's-calculation}'t Hooft's calculation of statistical
entropy}

To find the number of modes for a particle in the black hole metric,
t'Hooft uses a one dimensional WKB approximation. He takes the contribution
to energy of the transverse momenta as an effective radial potential
$V_{eff}=l(l+1)/r^{2}$ (since it behaves as a centrifugal potential).

The wave equation for a free massive particle in a diagonal metric
is
\begin{equation}
\left(g^{00}E^{2}-g^{rr}k_{r}^{2}-g^{\theta\theta}k_{\theta}^{2}-g^{\phi\phi}k_{\phi}^{2}-m^{2}\right)\psi=0.
\end{equation}
The one-dimensional WKB approximation has
\begin{eqnarray}
n\pi & = & \intop_{0}^{R}dr\sqrt{g_{rr}}k(r).
\end{eqnarray}
From the wave equation we obtain the radial eigenfunction 
\begin{equation}
k_{l}^{2}(r)=g_{rr}\left(g^{00}E^{2}-g^{\theta\theta}k(\theta)^{2}-g^{\phi\phi}k(\phi)^{2}-m^{2}\right)
\end{equation}
for each eigenfunction $k_{l}$. In the Schwarzschild metric this
becomes
\begin{equation}
k_{l}^{2}(r)=\frac{1}{1-\frac{2M}{r}}\left(\frac{1}{1-\frac{2M}{r}}E^{2}-\frac{1}{r^{2}}\left(l\left(l+1\right)\right)-m^{2}\right).
\end{equation}
The number of radial modes is then summed over the angular degrees
of freedom,
\begin{equation}
N\pi=\intop_{0}^{R}dr\frac{1}{1-\frac{2M}{r}}\intop_{0}^{E^{2}r^{2}}dl(2l+1)\sqrt{E^{2}-\left(1-\frac{2M}{r}\right)\left(\frac{l(l+1)}{r^{2}}+m^{2}\right)}
\end{equation}
where the upper limit of the second integral is in order to ensure
a positive root. 

The free energy is obtained from N as follows:
\begin{eqnarray}
F & = & -TlnZ=\left\langle E\right\rangle -TS\nonumber \\
S & = & =\beta\left\langle E\right\rangle +lnZ=-\frac{\partial F}{\partial T}\nonumber \\
e^{-\beta F} & = & \sum e^{-\beta E}=\Pi_{n,l,l_{z}}\frac{1}{1-e^{-\beta E}}
\end{eqnarray}
\begin{eqnarray}
\beta F & = & \sum log(1-e^{-\beta E}).
\end{eqnarray}
Now, taking the sum to an integral
\begin{equation}
\pi\beta F=\pi\int dN\, log(1-e^{-\beta E})
\end{equation}
and $\pi N=g(E)$ so integrating by parts
\begin{equation}
\pi\beta F=-\intop_{0}^{\infty}dE\frac{\beta g(E)}{e^{\beta E}-1}
\end{equation}
because $dN=dg(E),$$\frac{d}{dE}log(1-e^{-\beta E})=\frac{\beta}{1-e^{-\beta E}},$and
plugging in $\pi N=g(E)$ gives
\begin{equation}
\pi\beta F=-\beta\intop_{0}^{\infty}\frac{dE}{e^{\beta E}-1}\intop_{2M+h}^{L}dr\frac{1}{1-\frac{2M}{r}}\int(2l+1)dl\sqrt{E^{2}-(1-\frac{2M}{r})(m^{2}+\frac{l(l+1)}{r^{2}}}
\end{equation}
so
\begin{equation}
F=-\frac{1}{\pi}\intop_{0}^{\infty}\frac{dE}{1-e^{-\beta E}}N.
\end{equation}
Integration gives two terms, one of which is the contribution from
the vacuum surrounding the system at large distances and 't Hooft
discards it. The second term is the horizon contribution:
\begin{equation}
F_{horizon}=-\frac{2\pi^{3}}{45h}\left(\frac{2M}{\beta}\right)^{4}
\end{equation}
which diverges as $h\rightarrow0.$ Then taking $U=\frac{\partial}{\partial\beta}\left(\beta F\right)$
\begin{equation}
S=\beta\left(U-F\right)=\frac{8\pi^{3}}{45h}2M\left(\frac{2M}{\beta}\right)^{3}Z
\end{equation}
where Z is the total number of particle types. 't Hooft then adjusts
parameters of the model, such that $S=4M^{2}/\lambda$, in accordance
with area dependence derived previously. The main point of this derivation
is that the entropy diverges on approaching the horizon.

\section{\label{sec:Carlip's-scheme}Wald and Carlip}

Carlip has worked on obtaining black hole entropy from conformal theory
over a period of years, in the ADM formalism as well as Wald's canonical
phase space formalism. This section is taken from \cite{THE Carlip paper}.

\subsection{Canonical phase space formalism}

Wald \cite{Wald and Lee} based his work on the abstract formalism
of Hamiltonian mechanics where states of a system with $n$ degrees
of freedom are represented by points in a $2n$ dimensional manifold
referred to as phase space, on which a symplectic form is defined.
Wald mapped the space of the solutions of the field equations of motion
to phase space. Then variation of the Lagrangian is

\begin{eqnarray}
\delta\mathbf{L} & = & \mathbf{E}\cdot\delta\phi+d\Theta
\end{eqnarray}
where $\mathbf{L}$ is an n form Lagrangian and $\phi$ are fields
(taken as scalar in this treatment, but could be otherwise), $\mathbf{E}=0$
are the Euler-Lagrange equations, and $\Theta$ is an $n-1$ form
given by surface terms. The symplectic current obtained from symplectic
variation of $\Theta$ is
\begin{eqnarray}
\omega\left[\phi,\delta_{1}\phi,\delta_{2}\phi\right] & = & \delta_{1}\Theta\left[\phi,\delta_{2}\phi\right]-\delta_{2}\Theta\left[\phi,\delta_{1}\phi\right]\nonumber \\
\Omega & = & \intop_{C}\omega
\end{eqnarray}
where $C$ is a Cauchy surface.%
\footnote{$\Theta$ is like $\frac{\partial\mathcal{L}}{\partial\nabla_{\mu}\phi^{a}}\delta\phi^{a}$,
which is a vector quantity. So one then needs to integrate over all
the vectors.%
}

Given a diffeomorphism generated by vector field $\xi$ we can define
a Noether current (n-1 form) 
\begin{eqnarray}
J\left[\xi\right] & = & \Theta\left[\phi,\mathcal{L}_{\xi}\phi\right]-\xi\cdot\mathbf{L}\nonumber \\
dJ & = & d\Theta-d\left[\xi\cdot\mathbf{L}\right].
\end{eqnarray}
We have $d\Theta=\delta\mathbf{L}-\mathbf{E}\cdot\delta\phi$ but
$\mathbf{E}=0$ on shell.

Note that $\delta\mathbf{L}=\mathcal{L}_{\xi}\mathbf{L}$ and that
the Lie derivative of a form $\Lambda$ is given by $\mathcal{L}_{\xi}\Lambda=\xi\cdot d\Lambda+d\left(\xi\cdot\Lambda\right).$
Plugging this in:
\begin{eqnarray}
dJ & = & \delta\mathbf{L}-d\left(\xi\cdot\mathbf{L}\right)\nonumber \\
 & = & \xi\cdot d\mathbf{L}+d\left(\xi\cdot\mathbf{L}\right)-d\left(\xi\cdot\mathbf{L}\right).
\end{eqnarray}
$\xi\cdot d\mathbf{L}=0$ since $\mathbf{L}$ is an n form, and so
\begin{equation}
dJ=0,\rightarrow J=dQ.
\end{equation}
Given $\xi^{a}$ and a generator of diffeomorphism $H\left[\xi\right]$,
Hamilton's equations are:
\begin{equation}
\delta H\left[\xi\right]=\Omega=\intop_{C}\omega\left[\phi,\delta\phi,\mathcal{L}_{\xi}\phi\right].
\end{equation}
By definition, $\omega\left[\phi,\delta\phi,\mathcal{L}_{\xi}\phi\right]=\delta_{\phi}\Theta-\mathcal{L}_{\xi}\Theta$.
Plug that in and we find
\begin{eqnarray}
\omega & = & \delta J\left[\xi\right]-d\left(\xi\cdot\Theta\left[\phi,\delta\phi\right]\right)\nonumber \\
\delta H & = & \intop_{C}\delta J-d\left(\xi\cdot\Theta\right)
\end{eqnarray}
and since $J=dQ$
\begin{eqnarray}
H\left[\xi\right] & = & \intop_{\partial C}\left(Q\left[\xi\right]-"undelta"\: of\: d\left(\xi\cdot\Theta\right)\right)\nonumber \\
 & = & \intop_{\partial C}\left(Q\left[\xi\right]-\xi\cdot B\right)
\end{eqnarray}
where $B$ is defined so that $\delta\intop_{\partial C}\xi\cdot B=\intop_{\partial C}\xi\cdot\Theta$,
thus explaining the unprofessional use of the term ``undelta.''

\subsection{Carlip's implementation of this }

Using this formalism as his basis, Carlip writes out $\mathbf{L}$,
$\Theta,$ $Q$ specifically for general relativity.

The brackets of generators $H\left[\xi\right]$ form an algebra. If
the spacetime has a boundary we get a central term because a generator
is unique up to a constant, and if space has a boundary the constants
of the $H_{i}$ might not match (and won't be absorbed by infinity
if there is a defined boundary). So
\begin{equation}
\left\{ H\left[\xi_{1}\right],H\left[\xi_{2}\right]\right\} =H\left[\left\{ \xi_{1},\xi_{2}\right\} \right]+K\left[\left\{ \xi_{1},\xi_{2}\right\} \right].
\end{equation}
Take 2 vector fields, $\xi_{1},\xi_{2}$ and a field $\phi$ that
fulfills the equations of motion.: The diffeomorphism along $\xi_{2}$
of $J\left[\xi_{1}\right]$ (of the Noether current generated by $\xi_{1}$)
is given by the Lie derivative 
\begin{equation}
\delta_{\xi_{2}}J\left[\xi_{1}\right]=\mathcal{L}_{\xi_{2}}J\left[\xi_{1}\right]=\xi_{2}\cdot dJ\left[\xi_{1}\right]+d\left(\xi_{2}\cdot J\left[\xi_{1}\right]\right).
\end{equation}
The first term on the RHS vanishes so
\begin{eqnarray}
\mathcal{L}_{\xi_{2}}J\left[\xi_{1}\right] & = & d\left(\xi_{2}\cdot J\left[\xi_{1}\right]\right)\nonumber \\
 & = & d\left(\xi_{2}\cdot\left(\Theta-\xi_{1}\cdot\mathbf{L}\right)\right).
\end{eqnarray}
Plug into $\delta H$:
\begin{eqnarray}
\delta_{\xi_{2}}H\left[\xi_{1}\right] & = & \intop_{C}\delta_{\xi_{2}}J\left[\xi_{1}\right]-d\left(\xi_{1}\cdot\Theta\left(\phi,\mathcal{L}_{\xi_{2}}\phi\right)\right)\nonumber \\
 & = & \intop_{c}d\left(\xi_{2}\cdot\left(\Theta\left(\phi,\mathcal{L}_{\xi_{1}}\phi\right)-\xi_{1}\cdot\mathbf{L}\right)\right)-\left(\xi_{1}\cdot\Theta\left(\phi,\mathcal{L}_{\xi_{1}}\phi\right)\right)\nonumber \\
 & = & \intop_{\partial C}\xi_{2}\cdot\Theta\left(\phi,\mathcal{L}_{\xi_{1}}\phi\right)-\xi_{2}\xi_{1}\mathbf{L}-\xi_{1}\cdot\Theta\left(\phi,\mathcal{L}_{\xi_{2}}\phi\right)
\end{eqnarray}
(and $\mathbf{L}=0$ on shell). In the bulk $\delta_{\xi_{2}}H=0$
because constraints hold on shell. Then $\delta_{\xi_{2}}H$ is the
variation $\delta_{\xi_{2}}J\left[\xi_{1}\right]$ where $J$ is the
boundary term (as in the $\delta L$ variation it's the boundary term,
and also Carlip shows this explicitly in Appendix B of the paper).

By definition 
\begin{equation}
\delta_{\xi_{2}}J\left[\xi_{1}\right]=\left\{ J\left[\xi_{1}\right],J\left[\xi_{2}\right]\right\} ,
\end{equation}
the change in $J\left[\xi_{1}\right]$ under surface deformation generated
by $J\left[\xi_{2}\right]$.

Using the algebra we get
\begin{equation}
\left\{ J\left[\xi_{1}\right],J\left[\xi_{2}\right]\right\} =J\left\{ \xi_{1},\xi_{2}\right\} +K\left\{ \xi_{1},\xi_{2}\right\} 
\end{equation}
 where $K$ the central term can be found. Carlip then plugs the general
relativity expressions into this.

\subsection{Boundary conditions}

For a Killing vector $\chi^{a}$, $\chi^{2}=0$ \textcolor{black}{on
a Killing horizon}. Carlip works with a stretched horizon such that
$\chi^{2}=\epsilon,$ and at the end of the calculation takes $\epsilon\rightarrow0$.
Near the boundary he defines an orthogonal vector $\rho:$
\begin{equation}
\nabla_{a}\chi^{2}=-2\kappa\rho_{a},\:\chi^{a}\rho_{a}=0.
\end{equation}
At the horizon $\chi$ is a Killing vector:
\begin{eqnarray}
\chi^{a}\nabla_{a}\chi_{b} & = & \kappa\chi_{b}
\end{eqnarray}
and then 
\begin{equation}
\kappa\rho_{a}=-\frac{1}{2}\nabla_{a}\chi^{2}=-\frac{1}{2}2\chi^{b}\nabla_{a}\chi_{b}=\chi^{b}\nabla_{b}\chi_{a}=\kappa\chi_{b}
\end{equation}
so at the horizon $\rho_{a}\rightarrow\chi_{a}$ (and thus the normal
and tangent vectors are the same at the horizon). \emph{Away} from
horizon they are orthogonal.

If one varies the metric there is no Killing vector. But the scheme
requires boundary conditions such that $\chi^{2}=0$. He therefore
imposes boundary conditions :
\begin{eqnarray}
\delta\chi^{2} & = & 0,\:\delta\rho_{a}=0,\:\chi^{a}t^{b}\delta g_{ab}=0\: for\:\chi^{2}=0\nonumber \\
(\frac{\chi^{a}\chi^{b}}{\chi^{2}}\delta g_{ab} & = & 0).
\end{eqnarray}
This guarantees a boundary where $\chi^{2}=0$ remains null, $\chi^{a}$
is a null normal vector.

\subsection{Decomposition of $\xi$:}

A crucial step in this scheme is the decomposition of $\xi^{a}$ (which
becomes $\chi^{a}$ at the boundary) into components 

\begin{equation}
\xi^{a}=R\rho^{a}+T\chi^{a}
\end{equation}
as Carlip says ``deformations in r-t plane.'' One finds that $R\sim\chi^{a}\nabla_{a}T\equiv DT$.
In order for diffeomorphisms along such a $\xi^{a}$ to form a closed
algebra one requires $\rho^{a}\nabla_{a}T=0.$ (To see this work out:
$\left\{ \xi_{1},\xi_{2}\right\} =\xi_{3}$ , substituting the decomposition
into orthogonal components, and require that $\xi_{3}$ too has the
form $DT\rho+T\chi$ just like $\xi_{1},\xi_{2}$.)

There is an additional subtlety: these boundary conditions don't guarantee
the existence of $H\left[\xi\right]$. You need another. Define 
\begin{equation}
\tilde{\kappa}^{2}=-\frac{a^{2}}{\chi^{2}},\: a^{a}=\chi^{b}\nabla_{b}\chi^{a}.
\end{equation}
At horizon $\tilde{\kappa}\rightarrow\kappa,$ as $\chi^{2}\rightarrow0$.
But away from the horizon $\tilde{\kappa}=\frac{\kappa\rho}{\left|\chi\right|}.$ 

If you vary the metric $\tilde{\kappa}$ is not constant. We require
an average value over the cross section of the horizon:
\begin{equation}
\delta\intop_{\partial C}\hat{\epsilon}\left(\tilde{\kappa}-\frac{\kappa\rho}{\left|\chi\right|}\right)=0.
\end{equation}
This guarantees existence of $H$ , AND in order to obtain it we require:
\begin{equation}
\intop_{\partial C}\epsilon D^{3}T=0
\end{equation}
(this results from calculation of variation of the metric),
\begin{eqnarray}
\delta & = & \frac{\partial}{\partial g_{ab}}\delta g_{ab}\nonumber \\
\delta g_{ab} & = & \nabla_{(a}\xi_{b)}=\nabla_{(a}(DT\rho+T\chi)_{b)}
\end{eqnarray}
If $DT=\lambda T$ then $T_{a}\sim e^{i\lambda_{a}T}$and $\int e^{i(a+b)k}dk=\delta_{a+b}$.

\subsubsection*{Result:}

For such a $\xi$
\begin{eqnarray}
\left\{ \xi_{1},\xi_{2}\right\} ^{a} & = & \left(T_{1}DT_{2}-T_{2}DT_{1}\right)\chi^{a}+\frac{1}{\kappa}\frac{\chi^{2}}{\rho^{2}}D\left(T_{1}DT_{2}-T_{2}DT_{1}\right)\rho^{a}
\end{eqnarray}
or labelling the term in parenthesis $A$, this is $A\chi^{a}+const\cdot D\left(A\right)\rho^{a},$
that is, some other vector $\xi_{3}.$ This is isomorphic to the algebra
of diffeomorphisms of $S^{1}$ or $R$. 

Next we want to compute $K$. 
\begin{equation}
K\left\{ \xi_{1},\xi_{2}\right\} =\left\{ J\left[\xi_{1}\right],J\left[\xi_{2}\right]\right\} -J\left[\left\{ \xi_{1},\xi_{2}\right\} \right].
\end{equation}
For general relativity Carlip had calculated:
\begin{equation}
\left\{ J\left[\xi_{1}\right],J\left[\xi_{2}\right]\right\} ^{*}=\frac{1}{16\pi G}\intop_{\partial C}\epsilon_{bca_{1}...a_{n-2}}\left[\xi_{2}^{b}\nabla_{d}\left(\nabla^{d}\xi_{1}^{c}-\nabla^{c}\xi_{1}^{2}\right)-\left[1\leftrightarrow2\right]\right].\label{eq:J brackets}
\end{equation}

\subsection{Central charge as function of T,DT:}

First he deals with the measure in eq.(\ref{eq:J brackets}), $\epsilon_{bca_{1}....a_{n-2}}$.
The integral itself is over the horizon ($n-2$ dimensions) and the
$b,c$ contract with the integrand. The aim is to obtain an expression
written in terms of $\chi,\rho$ (that is, $T,DT.$) This is done
by writing the $bc$ part as $\chi_{[b}N_{c]}$ where $N$ is the
other null normal to the horizon besides $\chi$ , $N^{a}\chi_{a}=-1$
. The measure becomes $\hat{\epsilon}\left(\chi_{[b}N_{c]}\right)$
where $\hat{\epsilon}$ is the measure over the horizon ($\epsilon_{a_{1}....a_{n-2}}$).
So the integral is $\int\hat{\epsilon}\chi_{[b}N_{c]}A^{b}B^{c}$
where$A,B$ are the other indexed components in the integral. 

After similar manipulations he obtains the term $\xi^{b}\epsilon_{bc}$
decomposed into $\chi$ and $\rho$ components. That is, the integrand
includes $\xi^{b}\epsilon_{bca_{1}....a_{n-2}}$ expressed as the
induced metric $\hat{e}_{a_{1}.....a_{n-2}}$ multiplied by a $T\rho_{c}$
and an $R\chi_{c}$ component. He plugs this into the term for $\left\{ J\left[\xi_{1}\right],J\left[\xi_{2}\right]\right\} $
and gets

\begin{equation}
\left\{ J\left[\xi_{1}\right],J\left[\xi_{2}\right]\right\} =\frac{1}{16\pi G}\intop_{\partial C}\hat{\epsilon}_{a_{1}...a_{n-2}}\left[-\frac{1}{\kappa}\left(T_{1}D^{3}T_{2}\right)+2\kappa\left(T_{1}DT_{2}\right)-\left(1\leftrightarrow2\right)\right]
\end{equation}
where he omitted terms of $\mathcal{O}\left(\chi^{2}\right)$. Then
\begin{equation}
Q_{a_{1}...a_{n-2}}=\frac{1}{16\pi G}\hat{\epsilon}_{a_{1}...a_{n-2}}\left(2\kappa T-\frac{1}{\kappa}D^{2}T\right)
\end{equation}
(also up to$\mathcal{O}\left(\chi^{2}\right)$). 

He needs $J[\{\xi_{1},\xi_{2}\}],$ the ''surface term of H'' where
$H=\intop_{\partial C}Q[\xi]-\xi\cdot B$, but the B term doesn't
contribute (as shown in his appropriately named Appendix B). Recall
that $J=\intop_{\partial C}Q$. Thus the central term is

\begin{equation}
K\left\{ \xi_{1},\xi_{2}\right\} =\frac{1}{16\pi G}\intop_{\partial C}\hat{\epsilon}_{a_{1}...a_{n-2}}\left(DT_{1}D^{2}T_{2}-\left(1\leftrightarrow2\right)\right).
\end{equation}

\subsubsection*{Measures}

A Virasoro algebra has $K=\frac{c}{24}\int\frac{dz}{2\pi i}\left(\xi_{1}^{'}\xi_{2}^{"}-\xi_{2}^{'}\xi_{1}^{"}\right).$
Here we have $K\sim\int\hat{\epsilon}\left(T_{1}^{'}T_{2}^{"}-T_{2}^{'}T_{1}^{"}\right)$.
It's the same form of integrand as Virasoro algebra, but the measures
are different. $\hat{\epsilon}$ that appears in Carlip's $K$ has
dimension $d-2$ (or $2d$ in $R^{4}$). Integration is over cross
section of $H$ (that is you integrate over all dimensions \emph{except}
the radial and time directions.) Whereas the complex integral is only
over one complex variable. Carlip needs to reduce his integral to
one variable. 

Taking $v$ as a parameter along $\chi^{a}$, $Dv=1$ , we have $T_{i}$
as functions of $v$ and of horizon angular variable $\theta$, we
require
\begin{equation}
\intop_{\partial C}\hat{\epsilon}T_{1}\left(v,\theta\right)T_{2}\left(v,\theta\right)=const\int dvT_{1}\left(v,\theta\right)T_{2}\left(v,\theta\right).\label{eq:carlips eq.5.2}
\end{equation}
But there is a mismatch: LHS is over horizon. RHS is over parameter
of Killing orbits, which are NOT one of the angles of horizon. Since
he can't fix this, he adds variables to $T$ so that he can integrate
over horizon, and then use a Kronecker delta to adjust the result.

Since $\chi^{a}$ are over the horion, he takes $v$, their parameter,
as periodic with period $2\pi/\kappa.$ (Writing it as an exponent
with discrete coefficient $n$ \emph{means} it's periodic. The only
notable thing here is that as a period, he takes $\kappa/2\pi$ as
natural to a BH horizon.) 

He uses the orthogonality condition obtained earlier. The $T_{i}$
are orthogonal\emph{ }from before (the eq. $DT=\lambda T$ gives $T_{i}$
as exponents) and are now given the specific form
\begin{eqnarray}
T_{n}(v,\theta^{i}) & = & \frac{1}{k}e^{inkv}f_{n}\left(\theta^{i}\right),\nonumber \\
\intop_{\partial C}\hat{\epsilon}f_{m}f_{n} & \sim & \delta_{m+n,0}.
\end{eqnarray}
Thus the angles in the horizon will give a Kronecker delta times the
horizon area. 
\[
T_{m}=\frac{1}{k}e^{imkv}f(\theta)\equiv\tilde{T}_{m}f_{m}(\theta)
\]
where $\int\hat{\epsilon}f_{m}f_{n}\sim\delta_{m+n,0}.$ (the $f$s
need a Kronecker delta because they are indeed angular.) $v$ is a
parameter along $\chi^{a}$ and so $DT=\chi^{a}\nabla_{a}T=\partial_{v}T.$
Integral (4.21) of his paper is over the horizon 
\begin{eqnarray}
K\left[\xi_{m},\xi_{n}\right] & = & \frac{1}{16\pi G}\intop_{\mathcal{H}}\hat{\epsilon}_{a_{1}....a_{n-2}}\frac{1}{k}\left(DT_{m}D^{2}T_{n}-DT_{n}D^{2}T_{m}\right)\nonumber \\
 & = & \frac{1}{16\pi G}\intop_{\mathcal{H}}\hat{\epsilon}_{a_{1}....a_{n-2}}\frac{1}{k}\left(\partial_{v}\left(f_{m}e^{imkv}\right)\partial_{vv}\left(f_{n}e^{inkv}\right)-\left[m\leftrightarrow n\right]\right)\nonumber \\
 & = & \frac{1}{16\pi G}\intop_{\mathcal{H}}\hat{\epsilon}_{a_{1}....a_{n-2}}\frac{1}{k}f_{m}f_{n}\left(\left(\frac{imk}{k}e^{imkv}\right)\left(\frac{-n^{2}k^{2}}{k}e^{inkv}\right)-\left[m\leftrightarrow n\right]\right)\nonumber \\
 & = & \frac{-iA}{16\pi G}\delta_{m+n}\left(mn^{2}-nm^{2}\right)e^{i\left(m+n\right)kv}\nonumber \\
 & = & \frac{-iA}{16\pi G}\delta_{m+n}m^{3}.
\end{eqnarray}
 \textbf{Note that the integral is only over area}. The exponents
from the $v$ dependence of $T$ vanish because the delta function
makes them equal 1.

\subsection{Counting states:}

For a rotating BH, $\chi^{a}=t^{a}+\sum\Omega_{\left(\alpha\right)}\psi_{\left(\alpha\right)}^{a}$,
where $t^{a}$ is a time Killing vector, $\Omega$ angular velocity
and $\psi^{a}$ the rotational Killing vector. Then
\begin{equation}
T_{n}=\frac{1}{k}exp\left\{ in\left(kv+\sum l_{_{a}}\left(\phi_{\left(\alpha\right)}-\Omega_{\left(\alpha\right)}v\right)\right)\right\} .
\end{equation}
Plug this into the commutation relations for $\{\xi_{m},\xi_{n}\}^{a}$
and indeed it gives a new vector $\xi_{m+n}^{a}\sim T_{m+n}\chi^{a}+DT_{m+n}\rho^{a}$.

Then:
\begin{eqnarray}
\left\{ T_{m},T_{n}\right\}  & = & -i\left(m-n\right)T_{m+n}\nonumber \\
K\left\{ T_{m},T_{n}\right\}  & = & -\frac{iA}{8\pi G}\delta_{m+n,0}\nonumber \\
i\left\{ J\left[T_{m}\right],J\left[T_{n}\right]\right\}  & = & \left(m-n\right)J\left[T_{m+n}\right]+\frac{A}{8\pi G}m^{2}\delta_{m+n,0}\nonumber \\
\frac{c}{12} & = & \frac{A}{8\pi G}.
\end{eqnarray}
Boundary term; $J\left[T_{0}\right]\equiv\Delta=\frac{A}{8\pi G}.$

Cardy formula:
\begin{equation}
\rho=exp\left\{ 2\pi\sqrt{\frac{c}{6}\left(\Delta-\frac{c}{24}\right)}\right\} .
\end{equation}
Taking the log and plugging in $c,\Delta$
\begin{equation}
S=2\pi\sqrt{\frac{A}{4\pi G}\left(\frac{A}{8\pi G}-\frac{A}{16\pi G}\right)}=\frac{A}{4G}.
\end{equation}

\subsection{Extra term}

Later work \cite{Koga,Silva,Compere} showed that an extra term was
necessary in Carlip's scheme. The bracket of two Noether fields
\begin{equation}
\left[J_{m},J_{n}\right]=\delta_{\xi_{m}}J_{n}
\end{equation}
gives the Lie derivative of $J_{n}$ along vector field $\xi_{m}^{\mu}\left(x\right)$,
where $J_{m}$ is the Noether currrent associated with the diffeomorphism
generated by $\xi_{m}$. In Carlip's work
\begin{eqnarray}
\left[J_{m},J_{n}\right] & = & \intop_{B}\frac{\sqrt{g}}{16\pi G}\{\xi_{m}^{\mu}\nabla_{\rho}\left(\nabla^{\rho}\xi_{n}^{\nu}-\nabla^{\nu}\xi_{n}^{\rho}\right)-\nonumber \\
 &  & -\xi_{n}^{\mu}\nabla_{\rho}\left(\nabla^{\rho}\xi_{m}^{\nu}-\nabla^{\nu}\xi_{m}^{\rho}\right)+\xi_{m}^{\mu}\xi_{n}^{\nu}\mathcal{L}-\left(\mu\leftrightarrow\nu\right)\}dS_{\mu\nu}
\end{eqnarray}
where the integral is over the boundary. However Silva \cite{Silva}
gives this as
\begin{eqnarray}
\left[J_{m},J_{n}\right] & = & \intop_{B}\frac{\sqrt{g}}{16\pi G}\{\nabla^{\mu}\xi_{n}^{\nu}\nabla_{\rho}\xi_{m}^{\rho}-\nabla^{\mu}\xi_{m}^{\nu}\nabla_{\rho}\xi_{n}^{\rho}-\nonumber \\
 &  & -2\nabla_{\rho}\xi_{n}^{\mu}\nabla^{\rho}\xi_{m}^{\nu}+\left(R_{\,\,\,\rho\sigma}^{\mu\nu}-2\delta_{\rho}^{\mu}R_{\,\sigma}^{\nu}\right)\xi_{n}^{\rho}\xi_{m}^{\sigma}-\left(\mu\leftrightarrow\nu\right)\}dS_{\mu\nu}.
\end{eqnarray}
The difference in these two terms up to a total derivative is
\begin{equation}
\intop_{B}\frac{\sqrt{g}}{16\pi G}\{\nabla^{\mu}\left(\xi_{m}^{\rho}\nabla_{\rho}\xi_{n}^{\nu}-\xi_{n}^{\rho}\nabla_{\rho}\xi_{m}^{\nu}\right)-\left(\mu\leftrightarrow\nu\right)\}dS_{\mu\nu}.
\end{equation}
This is actually $J\left[\left\{ \xi_{m},\xi_{n}\right\} \right]$
where 
\begin{equation}
J\left[\xi\right]=\intop_{B}\frac{\sqrt{g}}{16\pi G}\left(\nabla^{\mu}\xi^{\nu}-\nabla^{\nu}\xi^{\mu}\right)dS_{\mu\nu}.
\end{equation}
 Carlip varies the metric, not $\xi$: his variation acts on the metric
but not on the parameters. Therefore it is necessary to subtract this
from his result.

\chapter{Boundary divergence}

\section{Smooth r\label{sec:appA}estricted operators}

We will now define a smoothing function which, when applied to an
operator that is restricted to a sub-volume, will soften the sharp
partition and serve as a momentum cutoff. Let us discuss a quantum
system in a volume $\Omega$ which is initially prepared in a pure
state $|\psi\rangle$ defined in $\Omega$. We divide the total volume
into some sub-volume $V$, and its complement $\widehat{V}$ so that
$\Omega=V\oplus\widehat{V}$. The Hilbert space inherits a natural
product structure ${\cal H}_{\Omega}={\cal H}_{V}\otimes{\cal H}_{\widehat{V}}$.
We are interested in states $|\psi\rangle$ that are entangled with
respect to the Hilbert spaces of $V$ and $\widehat{V}$ so that they
can not be brought into a product form $|\psi\rangle=|\psi\rangle_{V}\otimes|\psi\rangle_{\widehat{V}}$
in terms of a pure state $|\psi\rangle_{V}$ that belongs to the Hilbert
space of $V$, and another pure state $|\psi\rangle_{\widehat{V}}$
that belongs to the Hilbert space of $\widehat{V}$.

The total density matrix is defined in terms of the total state $|\psi\rangle$
\begin{equation}
\rho=|\psi\rangle\langle\psi|.
\end{equation}
 The partition of the total volume of the system into two parts 
\begin{equation}
\Omega=V\oplus\widehat{V}
\end{equation}
 induces a product structure on the Hilbert space and allows defining
the reduced density matrix by performing a trace over part of the
Hilbert space
\begin{equation}
\rho^{V}=Tr_{\widehat{V}}\rho.
\end{equation}

Operators that act on part of the Hilbert space are defined as integrals
over densities in a part of space 
\begin{equation}
O^{V}=\int\limits _{V}d^{3}r{\cal O}(\vec{r})
\end{equation}
 or alternatively in terms of a theta function 
\begin{equation}
\Theta^{V}(\vec{r})=\begin{cases}
1 & \vec{r}\in V\\
0 & \vec{r}\in\widehat{V}
\end{cases},
\end{equation}
\begin{equation}
O^{V}=\int\limits _{\Omega}d^{3}r{\cal O}(\vec{r})\Theta^{V}(\vec{r}).
\end{equation}
 The relation between quantum expectation values of operators that
act on part of the Hilbert space to the statistical averages with
a reduced density matrix is given by 
\begin{equation}
\langle\psi|O^{V}|\psi\rangle=Tr\left(\rho^{V}O^{V}\right).
\end{equation}

We can also define a smoothed operator 
\begin{equation}
O_{\text{smooth}}^{V}=\int\limits _{\Omega}d^{3}r{\cal O}(\vec{r})\Theta_{\text{smooth}}^{V}(\vec{r},w)
\end{equation}
 where $\Theta_{\text{smooth}}^{V}(\vec{r},w)$ represents a smoothed
step function that rather than changing in a discontinuous way from
zero to unity on the boundary of $V$ changes in a smooth way over
a region of width $w$ near the boundary of $V$. Expressing $\Theta_{\text{smooth}}^{V}$
as the product of a step function and an auxiliary smoothing function
$\left(f(\vec{r},w)\right)^{2}$ (the reason for the square will become
clear in what follows): 
\begin{equation}
\Theta_{\text{smooth}}^{V}(\vec{r},w)=\left(f(\vec{r},w)\right)^{2}\Theta^{V}(\vec{r})=\begin{cases}
\to1 & \vec{r}\in V\\
0\to1 & \vec{r}\in\partial V\:\text{with width }w\\
\to0 & \vec{r}\in\widehat{V}
\end{cases}
\end{equation}
 The smooth theta function defined in this way can be made continuous
to any fixed desired order in derivatives. So if a class of operators
has at most a given order of derivatives it is possible to define
a smooth theta function that will be effectively analytic for this
class. For example, the one dimensional function 
\begin{equation}
\Theta_{\text{smooth}}^{V}(x,w)=\begin{cases}
\frac{x^{n}}{x^{n}+w^{n}} & x\ge0\\
0 & x\le0
\end{cases}
\end{equation}
 has $n-1$ continuous derivatives at $x=0$.

Rather than using the smoothed step function to modify the operators
$O^{V}$, we can view the smoothing function $f(\vec{r},w)$ as modifying
the wave function (or state) in which the operator is being evaluated
\begin{equation}
\langle\psi|O_{\text{smooth}}^{V}|\psi\rangle=\langle\psi|\left(f(\vec{r},w)\right)^{2}O^{V}|\psi\rangle=\langle f(\vec{r},w)\psi|O^{V}|f(\vec{r},w)\psi\rangle.
\end{equation}
 Defining
\begin{equation}
|\psi_{\text{smooth}}\rangle=f(\vec{r},w)|\psi\rangle
\end{equation}
 we may express the expectation value of the smoothed operator in
the original state $|\psi\rangle$ in terms of an expectation value
of the original operator in a smoothed state 
\begin{equation}
Tr\left(\rho^{V}O_{\text{smooth}}^{V}\right)=Tr\left(\rho_{\text{smooth}}^{V}O^{V}\right)
\end{equation}
 where 
\begin{equation}
\rho_{\text{smooth}}^{V}=|\psi_{\text{smooth}}\rangle\langle\psi_{\text{smooth}}|.
\end{equation}
 In momentum space 
\begin{equation}
|\psi_{\text{smooth}}\rangle=\int d^{3}pf(\vec{p},w)\psi(\vec{p})e^{-i\vec{p}\cdot\vec{r}}.
\end{equation}
 Here the smoothing function $f(\vec{p},w)$ looks as if it is a UV
cutoff suppressing the the high momentum components of the wave function.

\section{Details of nonrelativistic smoothed momentum fluct\label{sec:appB}uations}

We now calculate the expectation value of the smoothed operators $(P^{2})^{V}$
which can be used to evaluate $H^{V}$ and other smooth operators.
The partial volume $V$ is defined by a window function as described
in the text.

The operator $P^{2}$ is given by 
\begin{eqnarray}
P^{2} & = & \sum_{\vec{p}}\vec{p}a_{\vec{p}}^{\dagger}a_{\vec{p}}\cdot\sum_{\vec{k}}\vec{k}a_{\vec{k}}^{\dagger}a_{\vec{k}}\nonumber \\
 & = & \sum_{\vec{p},\vec{k}}\vec{p}\cdot\vec{k}\ a_{\vec{p}}^{\dagger}\left(a_{\vec{k}}^{\dagger}a_{\vec{p}}+\left[a_{\vec{p}},a_{\vec{k}}^{\dagger}\right]\right)a_{\vec{k}}\nonumber \\
 & = & \sum_{\vec{p},\vec{k}}\vec{p}\cdot\vec{k}\left(a_{\vec{p}}^{\dagger}a_{\vec{k}}^{\dagger}a_{\vec{p}}a_{\vec{k}}+\delta_{\vec{p}\vec{k}}a_{\vec{p}}^{\dagger}a_{\vec{k}}\right).\nonumber \\
 & = & \sum_{\vec{p},\vec{k}}\vec{p}\cdot\vec{k}\left(a_{\vec{p}}^{\dagger}a_{\vec{k}}^{\dagger}a_{\vec{p}}a_{\vec{k}}\right)+\sum_{\vec{p}}p^{2}a_{\vec{p}}^{\dagger}a_{\vec{p}}.
\end{eqnarray}
 Evaluating the expectation value: {\footnotesize{}
\begin{eqnarray}
 &  & \left\langle \psi|(P_{\text{smooth}}^{2})^{V}|\psi\right\rangle =\int d^{3}r_{1}\, d^{3}r_{2}\,\left\langle 0\right|\Psi\left(\vec{r}_{1}\right)f\left(\vec{r}_{1},w\right)\,\sum_{\vec{p},\vec{k}}\vec{p}\cdot\vec{k}\left(a_{\vec{p}}^{\dagger}a_{\vec{k}}^{\dagger}a_{\vec{p}}a_{\vec{k}}+\delta_{\vec{p}\vec{k}}a_{\vec{p}}^{\dagger}a_{\vec{k}}\right)f\left(\vec{r}_{2},w\right)\Psi^{\dagger}\left(\vec{r}_{2}\right)\left|0\right\rangle \nonumber \\
 & = & \int d^{3}r\, d^{3}r_{2}\, f\left(\vec{r}_{1},w\right)f\left(\vec{r}_{2},w\right)\sum_{\vec{q},\vec{s}}\frac{e^{i\vec{q}\vec{r}_{1}}}{\sqrt{\Omega}}\frac{e^{-i\vec{s}\vec{r_{2}}}}{\sqrt{\Omega}}\left\langle 0\right|\sum_{\vec{p},\vec{k}}\vec{p}\cdot\vec{k}\ a_{\vec{q}}a_{\vec{p}}^{\dagger}a_{\vec{k}}^{\dagger}a_{\vec{p}}a_{\vec{k}}a_{\vec{s}}^{\dagger}\,\,+\sum_{\vec{p}}p^{2}a_{\vec{q}}a_{\vec{p}}^{\dagger}a_{\vec{p}}a_{\vec{s}}^{\dagger}\left|0\right\rangle .\hspace{0.5in}
\end{eqnarray}
} Since 
\begin{eqnarray}
\left\langle 0\left|a_{\vec{q}}a_{\vec{p}}^{\dagger}a_{\vec{p}}a_{\vec{s}}^{\dagger}\right|0\right\rangle  & = & \delta_{\vec{p}\vec{q}}\delta_{\vec{p}\vec{s}}\label{eq:a operators com relation}
\end{eqnarray}
 and 
\begin{eqnarray}
\left\langle 0\right|a_{\vec{q}}a_{\vec{p}}^{\dagger}a_{\vec{k}}^{\dagger}a_{\vec{p}}a_{\vec{k}}a_{\vec{s}}^{\dagger}\,\,\left|0\right\rangle =0,\label{eq:a operators com relation1}
\end{eqnarray}
 the expectation value of the smooth operator is then 
\begin{eqnarray}
\left\langle \psi|(P_{\text{smooth}}^{2})^{V}|\psi\right\rangle  & = & \int d^{3}r\, d^{3}r_{2}\, f\left(\vec{r}_{1},w\right)f\left(\vec{r}_{2},w\right)\sum_{\vec{q},\vec{s}}\frac{e^{i\vec{q}\vec{r}_{1}}}{\sqrt{\Omega}}\frac{e^{-i\vec{s}\vec{r_{2}}}}{\sqrt{\Omega}}p^{2}\delta_{\vec{p}\vec{q}}\delta_{\vec{p}\vec{s}}\nonumber \\
 & = & \int d^{3}r\,\vec{\nabla}f\left(\vec{r},w\right)\cdot\vec{\nabla}f\left(\vec{r},w\right)\nonumber \\
 & = & \sum_{\vec{p}}p^{2}f(\vec{p},w)f(-\vec{p},w)\ .
\end{eqnarray}

\section{Details of\label{sec:appC} relativistic smoothed energy}

In a relativistic theory the hamiltonian is given by $\widehat{H}=\int\frac{d^{3}k}{\left(2\pi\right)^{3}}k_{0}a_{k}^{\dagger}a_{k}$
in momentum space. In configuration space, the expectation value of
the smoothed restricted hamiltonian is given by the relativistic scalar
product, 
\begin{eqnarray}
\left\langle \psi\left|(H_{\text{smooth}})^{V}\right|\psi\right\rangle  & = & -i\int\, d^{3}r_{1}\, d^{3}r_{2}\left[\,\,\biggl\langle0\biggr|\Psi\left(\vec{r}_{1},t_{1}\right)f\left(\vec{r}_{1},w\right)\partial_{t_{2}}\biggl(H\, f\left(\vec{r}_{2},w\right)\Psi^{\dagger}\left(\vec{r}_{2},t_{2}\right)\biggr)-\right.\nonumber \\
 &  & \left.-\partial_{t_{1}}\biggl(\Psi\left(\vec{r}_{1},t_{1}\right)f\left(\vec{r}_{1}\right)\biggr)H\, f\left(\vec{r}_{2}\right)\Psi^{\dagger}\left(\vec{r}_{2}\right)\biggl|0\biggr\rangle\right]_{\biggl|t_{1}=t_{2}}\equiv A-B
\end{eqnarray}
 The first term $A$ is given by 
\begin{eqnarray}
A & = & \int\, d^{3}r_{1}\, d^{3}r_{2}\,202f\left(\vec{r}_{1},w\right)f\left(\vec{r}_{2},w\right)\biggl\langle0\biggr|\int\frac{d^{3}p}{\sqrt{\left(2\pi\right)^{3}2p_{0}}}\left(a_{\vec{p}}e^{i\vec{p}\cdot\vec{r}_{1}-ip_{0}t_{1}}+a_{\vec{p}}^{\dagger}e^{-i\vec{p}\cdot\vec{r}_{1}+ip_{0}t_{1}}\right)\,\times\nonumber \\
 &  & \int\frac{d^{3}q}{\left(2\pi\right)^{3}}q_{0}a_{q}^{\dagger}a_{q}\ \times-i\partial_{t_{2}}\int\frac{d^{3}k}{\sqrt{\left(2\pi\right)^{3}2k_{0}}}\left(a_{\vec{k}}e^{i\vec{k}\cdot\vec{r}_{2}-ik_{0}t_{2}}+a_{\vec{k}}^{\dagger}e^{-i\vec{k}\cdot\vec{r}_{2}+ik_{0}t_{2}}\right)\,\biggl|0\biggr\rangle_{\biggl|t_{1},t_{2}=0}\label{Aeq1}
\end{eqnarray}
 where $p_{0}^{2}=\vec{p}^{2}$, $k_{0}^{2}=\vec{k}^{2}$. The second
term $B$ can be expressed in a similar straightforward manner.

We first perform the momentum integrals and evaluate the expectation
value. This integral includes the following sets of operators:
\[
a_{\vec{p}}a_{q}^{\dagger}a_{q}a_{\vec{k}}\,,\, a_{\vec{p}}a_{q}^{\dagger}a_{q}a_{\vec{k}}^{\dagger}\,,\, a_{\vec{p}}^{\dagger}a_{q}^{\dagger}a_{q}a_{\vec{k}}\:,\, a_{\vec{p}}^{\dagger}a_{q}^{\dagger}a_{q}a_{\vec{k}}^{\dagger},
\]
but only the second term yields a non-vanishing contribution, 
\begin{eqnarray}
 &  & \int\frac{d^{3}p}{\left(2\pi\right)^{3}}\frac{d^{3}k}{\left(2\pi\right)^{3}}\ d^{3}q\frac{q_{0}k_{0}}{\sqrt{2k_{0}}\sqrt{2p_{0}}}\biggl\langle0\biggr|\left(a_{\vec{p}}e^{i\vec{p}\cdot\vec{r}_{1}}+a_{\vec{p}}^{\dagger}e^{-i\vec{p}\cdot\vec{r}_{1}}\right)\, a_{q}^{\dagger}a_{q}\left(a_{\vec{k}}e^{i\vec{k}\cdot\vec{r}_{2}}+a_{\vec{k}}^{\dagger}e^{-i\vec{k}\cdot\vec{r}_{2}}\right)\biggl|0\biggr\rangle\nonumber \\
 & = & \int\frac{d^{3}p}{\left(2\pi\right)^{3}}\frac{d^{3}k}{\left(2\pi\right)^{3}}\ d^{3}q\frac{q_{0}k_{0}}{\sqrt{2k_{0}}\sqrt{2p_{0}}}e^{i\vec{p}\cdot\vec{r}_{1}-i\vec{k}\cdot\vec{r}_{2}}\biggl\langle0\biggr|a_{\vec{p}}\ a_{q}^{\dagger}\ a_{q}\ a_{\vec{k}}^{\dagger}\biggl|0\biggr\rangle\nonumber \\
 & = & \int\frac{d^{3}p}{\left(2\pi\right)^{3}}\frac{d^{3}k}{\left(2\pi\right)^{3}}\ d^{3}q\frac{q_{0}k_{0}}{\sqrt{2k_{0}}\sqrt{2p_{0}}}e^{i\vec{p}\cdot\vec{r}_{1}-i\vec{k}\cdot\vec{r}_{2}}\delta(\vec{p}-\vec{q})\ \delta(\vec{k}-\vec{q}).\label{Aeq2}
\end{eqnarray}
 Substituting the result of eq.(\ref{Aeq2}) into eq.(\ref{Aeq1})
we find 
\begin{eqnarray}
A & = & \int\, d^{3}r_{1}\, d^{3}r_{2}\, f\left(\vec{r}_{1},w\right)f\left(\vec{r}_{2},w\right)\int\frac{d^{3}p}{\left(2\pi\right)^{3}}\frac{d^{3}k}{\left(2\pi\right)^{3}}\ d^{3}q\frac{q_{0}k_{0}}{\sqrt{2k_{0}}\sqrt{2p_{0}}}e^{i\vec{p}\cdot\vec{r}_{1}-i\vec{k}\cdot\vec{r}_{2}}\delta(\vec{p}-\vec{q})\ \delta(\vec{k}-\vec{q})\nonumber \\
 & = & \int\frac{d^{3}p}{\left(2\pi\right)^{3}}\frac{d^{3}k}{\left(2\pi\right)^{3}}\ d^{3}q\frac{q_{0}k_{0}}{\sqrt{2k_{0}}\sqrt{2p_{0}}}f(\vec{p},w)\ f(-\vec{k},w)\ \delta(\vec{p}-\vec{q})\ \delta(\vec{k}-\vec{q})\nonumber \\
 & = & \frac{1}{2}\int d^{3}p\ p\ f(\vec{p},w)\ f(-\vec{p},w),
\end{eqnarray}
 where $p^{2}=\vec{p}^{2}$, and $f(\vec{p},w)$ is the Fourier transform
of $f(\vec{r},w)$. Repeating the same steps for $B$ we find $B=-A$
so that 
\begin{eqnarray}
\left\langle \psi\left|(H_{\text{smooth}})^{V}\right|\psi\right\rangle  & = & \int d^{3}p\ p\ f(\vec{p},w)\ f(-\vec{p},w)\nonumber \\
 & = & \int d^{3}r\ f(\vec{r},w)\ \sqrt{\vec{\nabla}^{2}}\ f(\vec{r},w).
\end{eqnarray}

\chapter{Number of states}

\section{\label{sec:Lorentz-invariance-of}Lorentz invariance of phase space}

For simplicity of notation we take $\hbar,c=1$. In classical thermodynamics
the density of states is defined for a non relativistic system as
follows: Take an integral over the volume of phase space ($d^{3}xd^{3}p$),
restrict it to values of p which fit the energy eigenvalues that solve
the Schroedinger equation: $p^{2}=2mE$, and divide by a unit of volume
in momentum space: This gives the number of states in phase space
with the given energy, per unit volume of phase space, and one multiplies
this by $g$, a numerical factor related to the degeneracy (eg., for
spins with Dirichlet BCs $g=2\left(1/8\right)$ for the positive octant
and spin degeneracy), and integration over the volume in phase space
would give$V=\left(\pi/L\right)^{3}$ for a potential well, $\left(2\pi/L\right)^{3}$for
periodic BCs. 

We will show that the number of states is Lorentz invariant by explicit
calculation. For simplicity we will calculate this in 1+1 dimensions,
and assume the transformed system moves in the x direction relative
to original system. Generalization to more dimensions is transparent.
We take $g=1$ and assume $E=p^{2}/2m:$
\begin{equation}
N=\int dx\frac{dp}{dE}dE.
\end{equation}

\subsection{Special relativity: phase sp\label{sub:Special-relativity:-phases-pace}ace}

The question of invariance arises because we are restricting 8 dimensional
space to the product of two 3 dimensional hypersurfaces, in both volume
and momentum, by choosing $t=0$ and $E=p^{2}/2m$ which is the energy
that solves the Schroedinger equation, or $E^{2}=p^{2}$ for the relativistic
wave equation. However we will impose the restriction afterwards and
first perform the transformation. 

If we were not restricting space to a hypersurface, the transformation
would be as follows (in 1+1 dimensions): Going to frame moving with
velocity $\beta$ relative to the original one, the transformed momenta
is
\[
\tilde{p}^{\mu}=\Lambda_{\nu}^{\mu}p^{\nu}
\]
where
\[
\Lambda_{\nu}^{\mu}=\left(\begin{array}{cc}
\gamma & \gamma\beta\\
\gamma\beta & \gamma
\end{array}\right)
\]
and so the differential of the new time component is
\[
d\tilde{p}^{0}=d\left(\Lambda_{0}^{0}p^{0}+\Lambda_{i}^{0}\beta p^{i}\right).
\]
and plugging in the explicit values of the transformation matrix elements
this is 
\begin{eqnarray*}
d\tilde{p}^{0} & = & d\left(\gamma p^{0}+\gamma\beta p^{i}\right).
\end{eqnarray*}
In what follows, for clarity, we write $p^{0}$as $E$, $p^{i}$as
$\vec{p}$, $\sqrt{p_{i}p^{i}}$ as $p$.

\subsubsection*{Volume elements}

Recall that the volume form is invariant. Since $dxdt$ actually refers
to $dx\wedge dt.$ 
\begin{eqnarray}
dx\wedge dt & = & \left(dxdt-dtdx\right).\nonumber \\
d\tilde{x}\wedge d\tilde{t} & = & \left(\left(\gamma dx+\gamma\beta dt\right)\left(\gamma dt+\gamma\beta dx\right)-\left(\gamma dt+\gamma\beta dx\right)\left(\gamma dx+\gamma\beta dt\right)\right)\nonumber \\
 & = & \gamma^{2}\left(dxdt+\beta^{2}dtdx+\beta dx^{2}+dt^{2}\right)-\left(dtdx+\beta^{2}dxdt+\beta dx^{2}+dt^{2}\right)\nonumber \\
 & = & \gamma^{2}\left(dxdt\left(1-\beta^{2}\right)+dtdx\left(\beta^{2}-1\right)\right)=\gamma^{2}(1-\beta^{2})(dxdt-dtdx)\nonumber \\
 & = & dx\wedge dt.
\end{eqnarray}

\subsubsection*{Energy momentum relation}

\begin{equation}
E^{2}=m^{2}+\vec{p}^{2},\: p=\sqrt{E^{2}-m^{2}},\;\frac{dp}{dE}=\frac{E}{p}.
\end{equation}
In this case $\vec{p}=p^{i},p=\sqrt{\vec{p}^{2}}$ .

After imposing the restriction on energy we get
\[
d\tilde{E}=d\left(\gamma E+\gamma\beta\sqrt{E^{2}-m^{2}}\right)=\gamma(1+\frac{\beta E}{p})dE.
\]
.Instead we can write
\[
d\tilde{p}=d\left(\gamma p_{x}+\gamma\beta E\right)=\gamma(1+\beta\frac{dE}{dp})dp=\gamma(1+\frac{\beta p}{E})dp.
\]

\subsubsection*{Density of states}

In original 4-dimensional frame, taking hypersurface $t=0$ and the
energy momentum relation, we have
\begin{eqnarray}
N & = & \int dx\int dt\delta(t)\int dp\int dE\delta(p^{2}-E^{2})\nonumber \\
 & = & \int dx\int dt\delta(t)\int\frac{dp}{2E}.
\end{eqnarray}
 First we deal with the space-time part. After Lorentz transformation,
and taking into account the invariance of $dx\wedge dt$, this becomes
\begin{equation}
\int d\tilde{x}d\tilde{t}\delta(\tilde{t})=\int dxdt\delta(\gamma t+\gamma\beta x)=\int dxdt\frac{\delta(t)}{\gamma}=\frac{1}{\gamma}\int dx.
\end{equation}
\begin{equation}
\frac{1}{\gamma}\intop_{0}^{\tilde{L}}dx=\frac{\tilde{L}}{\gamma}=\frac{\gamma L}{\gamma}=L
\end{equation}
and so
\begin{equation}
\intop_{0}^{\tilde{L}}d\tilde{x}\intop_{-\infty}^{\infty}d\tilde{t}\delta(\tilde{t})=\intop_{0}^{L}dx\intop_{-\infty}^{\infty}dt\delta(t)=L.
\end{equation}
For the momentum integral the delta function on the energy is scalars,
so as we know from field theory, the form of the integral is invariant:
\begin{equation}
\int\frac{d\tilde{p}}{2\tilde{E}}=\int\frac{dp}{2E}
\end{equation}
and writing the integrand explicitly:
\begin{eqnarray}
\int\frac{d\tilde{p}}{2\tilde{E}} & = & \frac{1}{2}\int\frac{\gamma dp+\gamma\beta dE}{\gamma E+\gamma\beta p}=\frac{1}{2}\int\frac{dp+\beta dE}{E+\beta p}.\nonumber \\
dE & = & \frac{p}{E}dp\nonumber \\
\frac{1}{2}\int\frac{dp+\beta dE}{E+\beta p} & = & \frac{1}{2}\int\frac{dp+\beta\frac{p}{E}dp}{E+\beta p}=\frac{1}{2}\int\frac{(E+\beta p)}{E(E+\beta p)}dp\nonumber \\
\int\frac{d\tilde{p}}{2\tilde{E}} & = & \int\frac{dp}{2E}.
\end{eqnarray}
Therefore
\begin{equation}
\int d\tilde{x}d\tilde{t}\delta(\tilde{t})d\tilde{p}d\tilde{E}\delta(\tilde{E}^{2}-\tilde{p}^{2})=\int dxdt\delta(t)dpdE\delta(E^{2}-p^{2})
\end{equation}
and the density of states is Lorentz invariant.

Extension to more space dimensions is simple but long. A general if
somewhat tricky proof can also be found in \cite{Paddy book},\emph{
}p.36.

\section{Equivalence of WKB and formal calculations \label{sec:Equivalence-of-WKB}of
the number of states}

In this section we calculate the number of states for non relativistic
massless particles in a box in several different ways and show the
different methods of calculation are equivalent. We do this for a
cube and a sphere, but one can see that the result would be the same
for other geometries.

\subsection{Photons in a cubic box}

\subsubsection{Formal method}

First we use the textbook method, obtained by solving wave equation:

Take a box with each side of length L, and Dirichlet boundary conditions.
The wave equation gives
\begin{equation}
\frac{\pi^{2}}{L^{2}}\left(n_{x}^{2}+n_{y}^{2}+n_{z}^{2}\right)=\frac{\omega^{2}}{c^{2}}
\end{equation}
where $n_{i}$ are integers that index the modes, and of course one
can generalize to more dimensions.
\begin{equation}
n\equiv\left[\sum_{i=1}^{d}n_{i}^{2}\right]^{1/2},\:\omega=\frac{n\pi c}{L}.\label{eq:dispersion-1}
\end{equation}
Taking the sum to an integral in the space of mode indices
\begin{equation}
\sum_{n}(....)\rightarrow\frac{1}{8}\intop_{0}^{\infty}4\pi n^{2}dn\label{eq:mode integral}
\end{equation}
using only the positive octant. Usually one also multiplies by 2 for
spin or polarization but we don't do so here for simplicity. Plugging
in eq.(\ref{eq:dispersion-1}) we get
\begin{eqnarray}
N & = & =\frac{\pi}{2}\intop_{0}^{\infty}\left(\frac{L}{\pi c}\right)^{3}\omega^{2}d\omega=\frac{V}{2\pi^{2}c^{3}}\intop_{0}^{\infty}\omega^{2}d\omega
\end{eqnarray}
and since $\omega=E/\hbar$, 
\begin{equation}
N=\frac{V}{2\pi^{2}c^{3}\hbar^{3}}\intop_{0}^{\infty}E^{2}dE\label{eq:N for cube}
\end{equation}
which is dimensionless as it should be. For photons in finite temperature
this will be multiplied by the Bose Einstein distribution. The density
of states is 
\begin{eqnarray}
g(E) & = & \frac{VE^{2}}{2\pi^{2}c^{3}\hbar^{3}.}\label{eq:photons in cube}
\end{eqnarray}
and number of states for fixed energy is
\begin{equation}
N=\frac{VE^{3}}{6\pi^{2}c^{3}\hbar^{3}}=\frac{1}{\left(2\pi\right)^{3}c^{3}\hbar^{3}}\frac{4}{3}\pi E^{3}V^{3}.
\end{equation}

\subsubsection{WKB method for a cubic box}

WKB is generally used in one dimension, so we begin with a one dimensional
system. For such a system the textbook method gives $n=Lk/\pi$ and
using $E=\hbar\omega=\hbar kc$ we obtain $dn=(L/\hbar\pi c)dE$ and
$g(E)=L/\hbar\pi c$. 

Using the WKB approximation $n\pi\hbar=\intop_{0}^{L}pdx=\intop_{0}^{L}\hbar kdx$
. This gives $n=kL/\pi,$ and taking $E=pc=\hbar kc,$ 
\begin{equation}
N=\frac{E}{\hbar c}\frac{L}{\pi},\:\frac{dn}{dE}=\frac{L}{\pi c\hbar}
\end{equation}
which is the same as the ``textbook'' result. 

We get $1/\pi\hbar$ rather than $2/\pi\hbar$ as in the general definition
of $N$ because WKB is defined for Dirichlet boundary conditions rather
than periodic.

Using WKB for a 3D system, solving the wave equation by separation
of variables, $\psi(xyz)=X(x)Y(y)Z(z)$, 
\begin{equation}
\psi_{WKB}\sim e^{i\vec{k}\vec{x}/\hbar}=e^{ik_{x}x/\hbar}e^{ik_{y}y/\hbar}e^{ik_{z}z/\hbar}.
\end{equation}
For each exponent we take the one dimensional calculation as above,
and so for each direction we get $k_{i}=n_{i}\pi/L$ and 
\begin{eqnarray}
n_{i} & = & \frac{k_{i}L}{\pi}.\label{eq:n_i}
\end{eqnarray}
As before,
\begin{eqnarray*}
n\equiv\left[\sum_{i=1}^{d}n_{i}^{2}\right]^{1/2},\:\omega=\frac{n\pi c}{L}
\end{eqnarray*}
and the rest of the calculation is identical to the exact one in the
previous section, eqs.(\ref{eq:N for cube}),(\ref{eq:photons in cube}).

\subsubsection{Phase space calculation for cubic box}

We have
\begin{eqnarray}
E^{2}-p_{x}^{2}-p_{y}^{2}-p_{z}^{2} & = & 0\nonumber \\
p_{x}^{2} & = & E^{2}-p_{y}^{2}-p_{z}^{2}\nonumber \\
N & = & \frac{1}{\left(2\pi\right)^{3}}\intop_{0}^{L}dx\intop_{0}^{L}dy\intop_{0}^{L}dz\intop d^{3}p=\frac{L^{3}}{\left(2\pi\right)^{3}}\int d^{3}p.
\end{eqnarray}
Dealing with the momentum integral we have 
\begin{eqnarray*}
\intop_{-E}^{E}dp_{y}\intop_{-\sqrt{E^{2}-p_{y}^{2}}}^{\sqrt{E^{2}-p_{y}^{2}}}dp_{z}\intop_{-\sqrt{E^{2}-p_{y}^{2}-p_{z}^{2}}}^{\sqrt{E^{2}-p_{y}^{2}-p_{z}^{2}}}dp_{x} & = & 2\intop_{0}^{\sqrt{E^{2}-p_{z}^{2}}}dp_{y}\sqrt{E^{2}-p_{z}^{2}-p_{y}^{2}}\int dp_{z}\\
E^{2}-p_{z}^{2} & \equiv & A^{2}\\
2\intop_{-A}^{A}dp_{y}\sqrt{A^{2}-p_{y}^{2}} & = & 2A\intop_{-A}^{A}dp_{y}\sqrt{1-\frac{p_{y}^{2}}{A^{2}}}=A^{2}\intop_{0}^{1}du\sqrt{1-u^{2}}=2A^{2}\frac{\pi}{2}
\end{eqnarray*}
where we substituted $u=cos\theta$. We need $E^{2}-p_{z}^{2}\geq0$
for the above equations to be real, so the momentum integral is now
\begin{equation}
2\int dp_{z}A^{2}\frac{\pi}{2}=\pi\intop_{-E}^{E}dp_{z}\left(E^{2}-p_{z}^{2}\right)=\pi\left[E^{2}p_{z}-\frac{p_{z}^{3}}{3}\right]_{-E}^{E}=\pi\frac{4E^{3}}{3}
\end{equation}
 and so
\begin{equation}
N=\frac{1}{\left(2\pi\right)^{3}}\frac{4\pi VE^{3}}{3}
\end{equation}
as before.

\subsection{Spherical ``box''}

Definition of number of states:
\begin{equation}
N=\frac{1}{(2\pi)^{3}}\int d^{3}xd^{3}p.
\end{equation}
This is the product of volume of two spheres, one with radius R and
the other with radius E, so it has to be
\begin{equation}
N=\frac{1}{(2\pi)^{3}}\frac{4\pi R^{3}}{3}\frac{4\pi E^{3}}{3}=\frac{2R^{3}E^{3}}{9\pi}.
\end{equation}
Solving in detail: The wave equation in its most general form is
\begin{eqnarray}
-g^{tt}E^{2}-g^{rr}p_{r}^{2}-g^{\bot\bot}p_{\bot}^{2} & = & 0.
\end{eqnarray}
For spherical coordinates we have $g^{tt}=-1,\: g^{rr}=1,\: g^{\bot\bot}=r^{-2}.$
and so the wave equation is
\[
E^{2}-p_{r}^{2}-\frac{p_{\bot}^{2}}{r^{2}}=0.
\]
\begin{eqnarray*}
p_{r} & = & \sqrt{E^{2}-\frac{p_{\bot}^{2}}{r^{2}}}\\
\int d^{3}xd^{3}p & = & 4\pi\int drr^{2}\int dp_{r}d^{2}p_{\bot}\frac{1}{r^{2}}
\end{eqnarray*}
where the space integral has a factor of $\sqrt{g_{\bot\bot}}=r^{2},$
and the momentum integral has a factor of $\sqrt{g^{\bot\bot}}=\frac{1}{r^{2}}$,
these cancel and won't appear in what follows.
\begin{eqnarray}
N & = & \frac{1}{8\pi^{3}}\intop_{0}^{R}dr\intop_{0}^{4\pi}d\Omega\intop_{-\sqrt{E^{2}-p_{\bot}^{2}/r^{2}}}^{\sqrt{E^{2}-p_{\bot}^{2}/r^{2}}}dp_{r}\intop_{0}^{\infty}d^{2}p_{\bot}\nonumber \\
 & = & \frac{1}{2\pi^{2}}\intop_{0}^{R}dr\intop_{0}^{\infty}d^{2}p_{\bot}2\sqrt{E^{2}-\frac{p_{\bot}^{2}}{r^{2}}}\nonumber \\
 & = & \frac{1}{2\pi^{2}}\intop_{0}^{R}dr\intop_{0}^{Er}2\pi p_{\bot}dp_{\bot}2\sqrt{E^{2}-\frac{p_{\bot}^{2}}{r^{2}}}\nonumber \\
 & = & \frac{2}{\pi}\intop_{0}^{R}dr\intop_{0}^{E}dp_{\bot}p_{\bot}\sqrt{E^{2}-\frac{p_{\bot}^{2}}{r^{2}}}.
\end{eqnarray}
Substituting $E^{2}-p_{\bot}^{2}=u,$ the momentum integral is
\begin{equation}
-\frac{1}{2}\int du\sqrt{u}=-\frac{1}{3}u^{3/2}
\end{equation}
Plugging in the limits of the $p_{\bot}$ integral 
\begin{equation}
N=\frac{2E^{3}}{3\pi}\intop_{0}^{R}drr^{2}=\frac{2E^{3}R^{3}}{9\pi}.
\end{equation}
In terms of volume
\begin{equation}
N=\frac{E^{3}V}{6\pi^{2}}=\frac{1}{\left(2\pi\right)^{3}}V\left(\frac{4\pi E^{3}}{3}\right)=\frac{1}{\left(2\pi\right)^{3}}V_{x}V_{p}.\label{eq:N 3d sphere}
\end{equation}

\subsubsection{WKB calculation for spherical box}

To find the number of modes we sum over the number of radial modes
$\intop_{0}^{R}drk(r,l)$ where R is an IR cutoff. This is multiplied
by the sum over angular momentum since each level is degenerate due
to the angular momentum:. 
\begin{eqnarray}
k(r,l,E) & = & \sqrt{\frac{E^{2}}{c^{2}}-\frac{l(l+1)}{r^{2}}}\nonumber \\
N\pi & = & \intop_{0}^{R}dr\intop_{0}^{\infty}d\left(l(l+1)\right)\sqrt{\frac{E^{2}}{c^{2}}-\frac{l(l+1)}{r^{2}}}
\end{eqnarray}
The root has to be positive. This determines the upper limit for $l(l+1):$
\begin{eqnarray}
\frac{E^{2}r^{2}}{c^{2}} & = & l(l+1)_{max}\nonumber \\
\intop_{0}^{E^{2}r^{2}/c^{2}}d\left(l(l+1)\right)\sqrt{\frac{E^{2}}{c^{2}}-\frac{l(l+1)}{r^{2}}} & = & -\frac{2}{3}r^{2}\left(\left[\frac{E^{2}}{c^{2}}-\frac{E^{2}r^{2}/c^{2}}{r^{2}}\right]^{3/2}-\frac{E^{3}}{c^{3}}\right)\nonumber \\
 & = & \frac{2}{3}\frac{E^{3}}{c^{3}}r^{2}.\nonumber \\
N\pi & = & \frac{2}{3}\frac{E^{3}}{c^{3}}\intop_{0}^{R}drr^{2}.\nonumber \\
N & = & \frac{2}{9}\frac{R^{3}E^{3}}{\pi c^{3}}
\end{eqnarray}
as before.

\subsection{Comparisons}

Comparing the sphere to the calculation for a cube, setting $\hbar=1$:
\begin{equation}
g(E)_{sphere}=\frac{VE^{2}}{2\pi^{2}c^{3}},,\: g(E)_{cube}=\frac{VE^{2}}{\pi^{2}c^{3}\hbar^{3}.}
\end{equation}
This is approximately the ratio of their volumes. Taking length of
side of cube as $2R,$
\begin{equation}
V_{cube}=8R^{3},V_{sphere}=\frac{\pi}{3}4R^{3}\approx\frac{1}{2}V_{cube}.
\end{equation}
and as expected, $N\sim V_{x}V_{p}.$

\section{Phase space in 3+1 and in 4+1 dimensions}

The number of states in $d+1$ dimensions (d space dimensions) for
a diagonal metric works out to be
\begin{eqnarray}
N & = & CE^{d}\intop_{V}d^{d}x\sqrt{g_{d}}\left(g^{00}\right)^{\frac{d}{2}}\nonumber \\
C & = & \frac{\pi^{\frac{d}{2}}}{\Gamma\left(\frac{d}{2}+1\right)}
\end{eqnarray}
and $g_{d}$ is the determinant of the spatial components of the metric.

\subsection{Proof in 3+1 dimensi\label{sub:Proof-in-3+1dimensions}ons:}

The wave equation:
\begin{eqnarray*}
g^{00}E^{2}-g^{xx}p_{x}^{2}-g^{yy}p_{y}^{2}-g^{zz}p_{z}^{2} & = & 0
\end{eqnarray*}
 
\[
p_{x}=\sqrt{g_{xx}}\sqrt{g^{00}E^{2}-g^{yy}p_{y}^{2}-g^{zz}p_{z}^{2}}
\]
\begin{eqnarray*}
\intop d^{3}p & = & \intop_{-\sqrt{g_{yy}g^{00}}E}^{\sqrt{g_{yy}g^{00}}E}dp_{y}\intop_{-\sqrt{g_{zz}}\sqrt{g^{00}E^{2}-g^{yy}p_{y}^{2}}}^{\sqrt{g_{zz}}\sqrt{g^{00}E^{2}-g^{yy}p_{y}^{2}}}dp_{z}\intop_{-\sqrt{g_{xx}}\sqrt{g^{00}E^{2}-g^{yy}p_{y}^{2}-g^{zz}p_{z}^{2}}}^{\sqrt{g_{xx}}\sqrt{g^{00}E^{2}-g^{yy}p_{y}^{2}-g^{zz}p_{z}^{2}}}dp_{x}\\
 & = & 2\sqrt{g_{xx}}\intop_{-\sqrt{g_{yy}g^{00}}E}^{\sqrt{g_{yy}g^{00}}E}dp_{y}\intop_{-\sqrt{g_{zz}}\sqrt{g^{00}E^{2}-g^{yy}p_{y}^{2}}}^{\sqrt{g_{zz}}\sqrt{g^{00}E^{2}-g^{yy}p_{y}^{2}}}dp_{z}\,\sqrt{g^{00}E^{2}-g^{yy}p_{y}^{2}-g^{zz}p_{z}^{2}}
\end{eqnarray*}
We label $g^{00}E^{2}-g^{yy}p_{y}^{2}\equiv A^{2}$. Then the integral
over $p_{z}$ becomes
\begin{eqnarray}
\intop_{-\sqrt{g_{zz}}A}^{\sqrt{g_{zz}}A}dp_{z}\sqrt{A^{2}-g^{zz}p_{z}^{2}} & = & A\intop_{-\sqrt{g_{zz}}A}^{\sqrt{g_{zz}}A}dp_{z}\sqrt{1-\frac{p_{z}^{2}}{g_{zz}A^{2}}}\nonumber \\
 & = & A^{2}\sqrt{g_{zz}}\intop_{-1}^{1}du\sqrt{1-u^{2}}=A^{2}\sqrt{g_{zz}}\frac{\pi}{2}.
\end{eqnarray}
Plugging this in we get
\begin{eqnarray}
\int d^{3}p & = & 2\sqrt{g_{xx}g_{zz}}\frac{\pi}{2}\intop_{-\sqrt{g_{yy}g^{00}}E}^{\sqrt{g_{yy}g^{00}}E}dp_{y}\left(g^{00}E^{2}-g^{yy}p_{y}^{2}\right)\nonumber \\
 & = & \sqrt{g_{xx}g_{zz}}\pi\left[2\sqrt{g_{yy}}\left(g^{00}E^{2}\right)^{3/2}-\frac{2}{3}g^{yy}\left(\sqrt{g_{yy}g^{00}}E\right)^{3}\right]\nonumber \\
 & = & \sqrt{g_{xx}g_{zz}g_{yy}}\frac{4}{3}\pi\left(\sqrt{g^{00}}E\right)^{3}.
\end{eqnarray}

\subsection{One more dimension:}

\begin{eqnarray*}
g^{00}E^{2}-g^{xx}p_{x}^{2}-g^{yy}p_{y}^{2}-g^{zz}p_{z}^{2}-g^{ww}p_{w}^{2} & = & 0\\
p_{w}=\sqrt{g_{ww}}\sqrt{g^{00}E^{2}-g^{xx}p_{x}^{2}-g^{yy}p_{y}^{2}-g^{zz}p_{z}^{2}}
\end{eqnarray*}

\textsf{\textbf{\textit{
\begin{eqnarray}
\intop d^{3}p & = & \intop_{-\sqrt{g_{yy}g^{00}}E}^{\sqrt{g_{yy}g^{00}}E}dp_{y}\intop_{-\sqrt{g_{zz}}\sqrt{g^{00}E^{2}-g^{yy}p_{y}^{2}}}^{\sqrt{g_{zz}}\sqrt{g^{00}E^{2}-g^{yy}p_{y}^{2}}}dp_{z}\intop_{-\sqrt{g_{xx}}\sqrt{g^{00}E^{2}-g^{yy}p_{y}^{2}-g^{zz}p_{z}^{2}}}^{\sqrt{g_{xx}}\sqrt{g^{00}E^{2}-g^{yy}p_{y}^{2}-g^{zz}p_{z}^{2}}}dp_{x}\times\nonumber \\
 &  & \times\intop_{-\sqrt{g_{ww}}\sqrt{g^{00}E^{2}-g^{xx}p_{x}^{2}-g^{yy}p_{y}^{2}-g^{zz}p_{z}^{2}}}^{\sqrt{g_{ww}}\sqrt{g^{00}E^{2}-g^{xx}p_{x}^{2}-g^{yy}p_{y}^{2}-g^{zz}p_{z}^{2}}}dp_{w}\\
 & = & 2\sqrt{g_{ww}}\intop_{-\sqrt{g_{yy}g^{00}}E}^{\sqrt{g_{yy}g^{00}}E}dp_{y}\intop_{-\sqrt{g_{zz}}\sqrt{g^{00}E^{2}-g^{yy}p_{y}^{2}}}^{\sqrt{g_{zz}}\sqrt{g^{00}E^{2}-g^{yy}p_{y}^{2}}}dp_{z}\times\nonumber \\
 &  & \times\intop_{-\sqrt{g_{xx}}\sqrt{g^{00}E^{2}-g^{yy}p_{y}^{2}-g^{zz}p_{z}^{2}}}^{\sqrt{g_{xx}}\sqrt{g^{00}E^{2}-g^{yy}p_{y}^{2}-g^{zz}p_{z}^{2}}}dp_{x}\sqrt{g^{00}E^{2}-g^{xx}p_{x}^{2}-g^{yy}p_{y}^{2}-g^{zz}p_{z}^{2}}
\end{eqnarray}
}}}We label $g^{00}E^{2}-g^{zz}p_{z}^{2}-g^{yy}p_{y}^{2}\equiv A^{2}$.
Then the integral over $p_{x}$ becomes
\begin{eqnarray}
\intop_{-\sqrt{g_{xx}}A}^{\sqrt{g_{xx}}A}dp_{x}\sqrt{A^{2}-g^{xx}p_{x}^{2}} & = & A\intop_{-\sqrt{g_{xx}}A}^{\sqrt{g_{xx}}A}dp_{x}\sqrt{1-\frac{p_{x}^{2}}{g_{xx}A^{2}}}\nonumber \\
 & = & A^{2}\sqrt{g_{xx}}\intop_{-1}^{1}du\sqrt{1-u^{2}}=A^{2}\sqrt{g_{xx}}\frac{\pi}{2}.
\end{eqnarray}
Plugging this in we get
\begin{eqnarray*}
\int d^{3}p & = & 2\sqrt{g_{xx}g_{ww}}\frac{\pi}{2}\intop_{-\sqrt{g_{yy}g^{00}}E}^{\sqrt{g_{yy}g^{00}}E}dp_{y}\intop_{-\sqrt{g_{zz}}\sqrt{g^{00}E^{2}-g^{yy}p_{y}^{2}}}^{\sqrt{g_{zz}}\sqrt{g^{00}E^{2}-g^{yy}p_{y}^{2}}}dp_{z}\left(g^{00}E^{2}-g^{yy}p_{y}^{2}-g^{zz}p_{z}^{2}\right)
\end{eqnarray*}
Let us label $g^{00}E^{2}-g^{yy}p_{y}^{2}\equiv B^{2}.$ Then the
$p_{z}$integral becomes
\begin{equation}
\intop_{-\sqrt{g_{zz}}B}^{\sqrt{g_{zz}}B}dp_{z}\left(B^{2}-g^{zz}p_{z}^{2}\right)=\frac{4}{3}\sqrt{g_{zz}}B^{3}=\frac{4}{3}\sqrt{g_{zz}}\left(g^{00}E^{2}-g^{yy}p_{y}^{2}\right)^{3/2}.
\end{equation}
Integrate over $p_{y}$:
\begin{eqnarray}
\intop_{-\sqrt{g_{yy}g^{00}}E}^{\sqrt{g_{yy}g^{00}}E}dp_{y}\left(g^{00}E^{2}-g^{yy}p_{y}^{2}\right)^{3/2} & = & \sqrt{g_{yy}}\frac{3}{8}\pi\left(\sqrt{g^{00}}E\right)^{4}.
\end{eqnarray}
(this was done with Mathematica, you get a result containing $Arctan[\infty]=\frac{\pi}{2}$).

Plugging this back in,
\begin{eqnarray*}
\int d^{3}p & = & \frac{4}{3}\left(\frac{3}{8}\right)\pi^{2}\sqrt{g_{xx}g_{ww}g_{yy}g_{zz}}\left(g^{00}\right)^{2}E^{4}\\
 & = & \frac{\pi^{2}}{2}\sqrt{g_{4}}\left(g^{00}\right)^{2}E^{4}
\end{eqnarray*}
and so
\begin{eqnarray}
N & =\frac{\pi^{2}}{2}E^{4} & \intop_{V}d^{4}x\sqrt{g_{4}}\left(g^{00}\right)^{2}\nonumber \\
 & = & CE^{4}\intop_{V}d^{4}x\sqrt{g_{4}}\left(g^{00}\right)^{\frac{4}{2}},\:\left(C=\frac{\pi^{\frac{4}{2}}}{\Gamma\left(\frac{4}{2}+1\right)}\right)
\end{eqnarray}
just as we claimed. It would be good to be able to prove by induction
that if it's true for $N_{d}$ it's true for $N_{d+1}$ but (so far)
I have not been able to generalize the integration: the integrand
becomes $\left(g^{00}E^{2}-g^{(d+1,d+1)}p_{d+1}\right)^{d/2}.$

\section{Numb\label{sec:Number-of-states Rindler}er of states in Rindler
space}

We here explicitly calculate the number of states in Rindler space.
The number of states can be obtained by writing out the wave equation
for Rindler space:
\[
-e^{-2a\xi}\partial_{tt}\psi+e^{-2a\xi}\partial_{\xi\xi}\psi+\partial_{\bot\bot}\psi=0
\]
where $\partial_{\bot\bot}$is the second derivative over transverse
degrees of freedom.
\begin{eqnarray}
e^{-2a\xi}E^{2}-e^{-2a\xi}p_{\xi}^{2}-p_{\bot}^{2} & = & 0\nonumber \\
p_{\xi} & = & \sqrt{E^{2}-e^{2a\xi}p_{\bot}^{2}}.
\end{eqnarray}

\begin{equation}
N=\frac{1}{(2\pi)^{3}}\int d^{2}x_{\bot}d\xi\sqrt{g_{\xi\xi}}\intop_{-\sqrt{E^{2}-e^{2a\xi}p_{\bot}^{2}}.}^{\sqrt{E^{2}-e^{2a\xi}p_{\bot}^{2}}.}dp_{\xi}\int dp_{\bot}2\pi p_{\bot}.
\end{equation}
Integration of $dp_{\xi}$ within these limits gives
\begin{equation}
\intop_{-\sqrt{E^{2}-e^{2a\xi}p_{\bot}^{2}}.}^{\sqrt{E^{2}-e^{2a\xi}p_{\bot}^{2}}.}dp_{\xi}=2\sqrt{E^{2}-e^{2a\xi}p_{\bot}^{2}}.
\end{equation}
\begin{eqnarray}
N & = & \frac{1}{(2\pi)^{3}}\int d^{2}x_{\bot}d\xi\sqrt{g_{\xi\xi}}\intop_{0}^{e^{-a\xi}E}dp_{\bot}2\pi p_{\bot}2\sqrt{E^{2}-e^{2a\xi}p_{\bot}^{2}}\\
 & = & \frac{2V_{\bot}}{\left(2\pi\right)^{2}}\int d\xi\sqrt{g_{\xi\xi}}\intop_{0}^{e^{-a\xi}E}dp_{\bot}p_{\bot}\sqrt{E^{2}-e^{2a\xi}p_{\bot}^{2}}
\end{eqnarray}
where we integrated over $x_{\bot}$ and limits of $p_{\bot}$are
to ensure the root is positive. (it can't be $-e^{-a\xi}E$ because
$p_{\bot}$here is like a radial coordinate). To integrate over $p_{\bot}$we
substitute $u=E^{2}-e^{2a\xi}p_{\bot}^{2}$ and obtain
\begin{equation}
-\frac{1}{2}e^{-2a\xi}\int du\sqrt{u}=-\frac{1}{3}e^{-2a\xi}u^{3/2}.
\end{equation}
Putting back $p_{\bot},$ we have
\begin{equation}
N=\frac{2V_{\bot}}{\left(2\pi\right)^{2}}\int d\xi\sqrt{g_{\xi\xi}}\left(-\frac{1}{3}e^{-2a\xi}\left[0-\left(E^{2}\right)^{3/2}\right]\right)=\frac{2V_{\bot}E^{3}}{3\left(2\pi\right)^{2}}\intop_{-\infty}^{\infty}d\xi\sqrt{g_{\xi\xi}}e^{-2a\xi}
\end{equation}
and since $\sqrt{g_{\xi\xi}}=e^{a\xi}$, this becomes
\[
N=\frac{2V_{\bot}E^{3}}{3\left(2\pi\right)^{2}}\intop_{-\infty}^{\infty}d\xi e^{-a\xi}
\]
which diverges at the lower boundary (as expected). Putting in a lower
limit of $\xi_{min}$ we have
\begin{equation}
N=\frac{V_{\bot}E^{3}}{6\pi^{2}a}e^{-a\xi_{min}}.
\end{equation}
Checking units, this is dimensionless as it should be.
\begin{equation}
\frac{dN}{dE}=\frac{V_{\bot}E^{2}}{2\pi^{2}a}e^{-a\xi_{min}}.
\end{equation}
This differs from the result in Sec.\ref{sec:our-paper.}. In that
work we took $V_{\bot}=(2a)^{-2}$ by comparing Unruh temperature
with Hawking temperature and relating it to the Schwarzschild radius.
\begin{eqnarray}
T_{H} & = & \frac{1}{8\pi M},\: T_{Rind}=\frac{a}{2\pi}\rightarrow2a=\frac{1}{2M}\nonumber \\
R_{horizon} & = & 2M=\frac{1}{2a}\nonumber \\
\frac{1}{(2a)^{2}} & = & \xi_{\bot}^{2}.
\end{eqnarray}
This was then incorporated into a WKB approximation following 't Hooft,
rather than the exact calculation presented here. However the dependence
on energy is the same, as is the divergence at the horizon.

\subsection{Comparisons }

Photons in a cube in Minkowski space are proportional to the entire
volume. Comparing the density of modes ot our result:
\begin{eqnarray}
\frac{dN}{dE}_{(Mink\, cube)} & = & \frac{VE^{2}}{\pi^{2}}\nonumber \\
\frac{dN}{dE}_{(Rind)} & = & \frac{V_{\bot}E^{2}}{\pi^{2}}\frac{e^{-a\xi_{min}}}{2a}.
\end{eqnarray}
The density of states in spherical and Schwarzschild metrics are
\begin{eqnarray}
\frac{dN}{dE}_{(sphere)} &  & \frac{2R^{3}E^{2}}{3\pi}\nonumber \\
\frac{dN}{dE}_{(Schwarz)} & = & \frac{2E^{2}}{\pi}\intop_{r_{0}}^{R}dr\frac{r^{9/2}}{(r-2M)^{5/2}}.
\end{eqnarray}

\subsection{Momentum only in $\xi$ direction:}

If there is no momentum in transverse directions we will have
\[
n=\frac{1}{\pi}\intop_{-\infty}^{\infty}d\xi e^{a\xi}\sqrt{e^{-2a\xi}E^{2}}=\frac{1}{\pi}\intop_{-\infty}^{\infty}d\xi E=\frac{E}{\pi}L
\]
where $L$ is the length of all space on the $\xi$ axis. If we had
a partition, the length factor would be shorter accordingly, but the
momentum still isn't a function of $\xi$. Thus we note that: \textbf{for
divergence at the boundary you need transverse momenta as well.}

\chapter{Variations of entropy}

\section{Variatio\label{sec:Variation-of-Wald's}n of Wald's entropy }

In \cite{merav ramy} the authors calculated the variation of Wald's
entropy. They write the variation of energy as 
\begin{equation}
\delta Q=\intop_{\mathcal{H}}T_{ab}\chi^{a}\epsilon^{d}
\end{equation}
 where $\chi^{a}$ is a Killing vector and $\epsilon^{d}$ a $(D-1)$
volume form. (See also \cite{jacobson}). The generalized gravitational
field equations give 
\begin{equation}
T^{ab}=2\left[-2\nabla_{p}\nabla_{q}\frac{\partial\mathcal{L}}{\partial R_{pabq}}+\frac{\partial\mathcal{L}}{\partial R_{pqra}}R_{pqr}^{\,\,\, b}\right]-g^{ab}\mathcal{L}.
\end{equation}
 This term for $T^{ab}$ is then substituted into $\delta Q$, thus
writing it as a function of the Riemann tensor and its derivatives.

Wald entropy is 
\begin{equation}
S_{W}=-\frac{1}{T}\ointop_{\partial\mathcal{H}}\frac{\partial\mathcal{L}}{\partial R_{abcd}}\hat{\epsilon}_{ab}\epsilon_{cd}
\end{equation}
 where $\epsilon_{cd}$ is a $(D-2)$ volume form and $\hat{\epsilon}_{ab}$
is the binormal vector to the area element. Since any $W^{cd}$ satisfies
$d\left(W^{cd}\epsilon_{cd}\right)=2\nabla_{c}W^{cd}\epsilon_{d}$
this can be written 
\begin{equation}
S_{W}=-\frac{2}{T}\intop_{\mathcal{H}}\nabla_{c}\left(\frac{\partial\mathcal{L}}{\partial R_{abcd}}\hat{\epsilon}_{ab}\right)\epsilon_{d}
\end{equation}
 and then differentiated along a Killing vector to obtain $\delta S$
\begin{equation}
\delta S_{W}=-\frac{2}{T}\intop_{\mathcal{H}}\chi_{m}\nabla^{m}\nabla_{c}\left(\frac{\partial\mathcal{L}}{\partial R_{abcd}}\hat{\epsilon}_{ab}\right)\epsilon_{d}.
\end{equation}
 Using $T=\frac{\kappa}{2\pi}$ they find that this fulfills the first
law of thermodynamics, $\delta Q=T\delta S$ and thus they show that
\begin{equation}
\delta S_{W}=2\pi\int T_{ab}\chi^{a}\epsilon^{b}.
\end{equation}

\end{document}